\documentclass[showkeys,showpacs,onecolumn,pre,11pt]{revtex4-2}
\usepackage{epsfig}
\usepackage{color}
\usepackage{amsmath}
\usepackage{amsfonts}
\usepackage{amssymb}

\normalsize

\begin{document}

\title{Compressible vortex structures and their role in the onset of hydrodynamic turbulence}
\author{D.S. Agafontsev$^{(1,2,a)}$, E.A. Kuznetsov$^{(2,3,4,b)}$,  A.A. Mailybaev$^{(5,c)}$, E.V. Sereshchenko$^{(2,6,d)}$}
\affiliation{\textit{{\small $^{(1)}$ Shirshov Institute of Oceanology, Russian Academy of Sciences, Nakhimovsky prosp. 36, Moscow, 117997, Russian Federation}}\\
\textit{{\small $^{(2)}$ Skolkovo Institute of Science and Technology, Bolshoy Boulevard 30, bld. 1, Moscow, 121205, Russian Federation}}\\
\textit{{\small $^{(3)}$ Lebedev Physical Institute, Russian Academy of Sciences, Leninsky prosp. 53, Moscow, 119991, Russian Federation}}\\
\textit{{\small $^{(4)}$ Landau Institute for Theoretical Physics, Russian Academy of Sciences, Akademika Semenova av., 1A, Moscow, 142432, Russian Federation}}\\
\textit{{\small $^{(5)}$ Instituto Nacional de Matem\'atica Pura e Aplicada, Estrada Dona Castorina 110, Rio de Janeiro, CEP 22460-320, Brasil}}\\
\textit{{\small $^{(6)}$ Khristianovich Institute of Theoretical and Applied Mechanics SB RAS, Institutskaya str., 4/1, Novosibirsk, 630090, Russian Federation}}}
\email{E-mail: $^{(a)}$ dmitrij@itp.ac.ru, $^{(b)}$ kuznetso@itp.ac.ru, $^{(c)}$ alexei@impa.br,	$^{(d)}$ s\_evgeniy@yahoo.com}

\begin{abstract}
We study formation of quasi two-dimensional (thin pancakes) vortex structures in three-dimensional flows, and quasi one-dimensional structures in two-dimensional hydrodynamics. These structures are formed at high Reynolds numbers, when their evolution is described at the leading order by the Euler equations for an ideal incompressible fluid. We show numerically and analytically that the compression of these structures and, as a consequence, the increase in their amplitudes is related to the compressibility of the frozen-in-fluid fields: the field of continuously  distributed vortex lines in the three-dimensional case and the field of vorticity rotor (divorticity) for two-dimensional flows. We find that the growth of vorticity and divorticity can be considered as a process of breaking of the corresponding fields. At high intensities, the process demonstrates a Kolmogorov-type scaling relating the maximum amplitude with the characteristic width of the structures. The possible role of these coherent structures is analyzed in the formation of the turbulent Kolmogorov spectrum, as well as the Kraichnan spectrum corresponding to a constant flux of enstrophy in the case of two-dimensional turbulence.
\end{abstract}



\maketitle
\tableofcontents

\section{Introduction}
Despite the fact that since the classical works of L.F. Richardson \cite{Richardson}, A.N. Kolmogorov \cite{kolmogorov1941dissipation} and A.M. Obukhov \cite{Obukhov} on hydrodynamic turbulence at high Reynolds numbers $Re\gg 1$, about 80 years have passed and a significant understanding of its nature has been achieved, the problem of developed hydrodynamic turbulence still remains unresolved. The main reason is that the developed hydrodynamic turbulence in no way can be studied using the perturbation theory, unlike, for example, wave turbulence \cite{ZLF}.

In wave turbulence, there are two fundamental factors that determine the dynamics of a wave system: linear wave dispersion and non-linearity. If dispersion effects prevail over nonlinear ones, then in this case each wave with frequency $\Omega$ and wave vector ${\bf k}$ moves freely for a long time and only at large distances $L \gg k^{-1} $ begins to be influenced by other waves due to non-linearity. This is the basis for applying a statistical description based on perturbation theory to such an ensemble of waves. As a result, we come to the theory of weak (wave) turbulence, which describes the behavior of an ensemble of waves in the language of kinetic equations for the wave action, which is the classical limit of occupation numbers. In this way, it is possible to advance quite far, in particular, to find turbulence spectra as exact solutions of kinetic equations, which are called Kolmogorov-Zakharov spectra (see, for example, \cite{ZLF}). 

These solutions are characterized by a constant flux of energy, number of particles, etc. It is important that these solutions cannot be considered as thermodynamically equilibrium ones - they, like the Kolmogorov-Obukhov spectrum for developed hydrodynamic turbulence, are realized in the inertial interval - the intermediate region between pumping and damping, and are characterized by a finite value of the flux of one or another integral of motion. As the amplitude increases (i.e., with an increase in nonlinearity), the main role begins to be played by coherent structures in the form of solitons, breathers, and vortices, for which the nonlinearity is compensated by dispersion effects. Such objects sometimes turn out to be stable, more often for integrable models like the Korteweg-de Vries equation (KdV) or the nonlinear Schrodinger equation (NLSE). At the same time, turbulence in integrable models, called integrable \cite{zakharov2009turbulence}, has a number of features (see, for example, \cite{suret2011wave,picozzi2014optical}, as well as recent works \cite{walczak2015optical,agafontsev2015integrable,agafontsev2016integrable,gelash2018strongly,agafontsev2020integrable} ). In the case of instability of coherent structures, a typical scenario of their nonlinear development is collapse, i.e., the formation of a singularity in a finite time (see the review \cite{ZakharovKuznetsov2012} and the literature cited there). Classical examples of collapse are the self-focusing of light in media with Kerr nonlinearity and the breaking of sound-type waves. 

If we talk about Euler's hydrodynamics, then it describes turbulence with good accuracy at high Reynolds numbers, $Re\gg 1$, in the inertial range of scales, an intermediate region between long-wavelength pumping and viscous damping. It is important that Euler's hydrodynamics cannot be studied perturbatively at any limit. Its Hamiltonian, coinciding with the total kinetic energy of the fluid, is the interaction Hamiltonian (see, for example, \cite{ZakharovKuznetsov1997}). Thus, Euler's hydrodynamics is a system with an extremely strong nonlinear interaction. Therefore, it should be expected that for developed hydrodynamic turbulence, coherent structures should play a more significant, and possibly decisive, role than for wave turbulence. It should be noted that the problem of the interaction of coherent structures and chaotic components is one of the central, still unsolved problems in the modern theory of turbulence. 

It is known that the Kolmogorov-Obukhov theory \cite{kolmogorov1941dissipation, Obukhov} describes developed hydrodynamic turbulence in the inertial range of scales. The size of this region, the ratio of the energy containing scale $L$ to the viscous scale $\ell_{vis}$, grows proportionally to $Re^{3/4}$ (see, for example, \cite{landau2013fluid}). The Kolmogorov-Obukhov theory is based on two important assumptions:\\
- turbulence in the inertial interval is homogeneous and isotropic;\\
- non-linear interaction between fluctuations is local.\\ 
The latter means that in the inertial interval the interaction between scales of the same order exceeds the interaction between fluctuations with very different scales. In the case of stationary turbulence, the behavior of the system is determined by the (constant) energy flux $\varepsilon$ from the pumping region to the damping region (essentially based on the locality property). The turbulence spectrum - the so-called Kolmogorov spectrum - has a universal behavior determined by a unique dimensional quantity $\varepsilon$. For this reason, up to a constant $C_K$ (Kolmogorov's constant), the spectrum of $E(k)$ can be obtained from dimensional considerations. In dimension, the spectrum $E(k)$ is the energy density in the phase space multiplied by $4\pi k^2$:
$$
E(k)= 4\pi k^2\frac{\rho c^2}{k^3}F\left(\frac{\varepsilon}{\rho c^2\cdot kc} \right),
$$
where $\rho$ is the density (below equal to 1), $c$ is the speed of light (introduced for convenience), and $F$ is a function of the dimensionless parameter ${\varepsilon}/(\rho c^ 2\cdot kc)$. Obviously, the spectrum $E(k)$ should not depend on the speed of light, hence the function $F(\xi)=C\xi^{2/3}$. As a result, we arrive at the Kolmogorov spectrum
$$ 
E(k)= C_K \varepsilon^{2/3} k^{-5/3}, \,\, C_K=4\pi C. 
$$
This derivation belongs to R.Z. Sagdeev; there is nothing in this derivation other than dimensional considerations. It follows from the same considerations that the time of energy transfer from energy-containing scales $L$ to the dissipation region is finite and is determined only by $L$ and $\varepsilon$: $T\sim {L^{2/3}}{\varepsilon^{ -1/3}}$. The fluctuations of the velocity and vorticity $\omega=[\nabla\times{\bf v}]$ of the scale $\ell$ are respectively given by the following relations:
$$
\langle \delta v \rangle\sim \varepsilon^{1/3}\ell^{1/3},\,\ \langle\delta\omega\rangle\sim \varepsilon^{1/3}\ell^ {-2/3}.
$$
Thus, for $\omega$ fluctuations we have a singularity at $\ell\to 0$. Together with the finite time of energy transfer to small scales $T$, this indicates the possibility of the formation of a collapse, which in the inertial range of scales can be studied using Euler hydrodynamics. It was these considerations that were the main ones when we started our research, in which a direct numerical experiment turned out to be especially useful, which significantly changed our understanding of the mechanisms of transition to developed hydrodynamic turbulence.

As was first noted by V.I. Arnold \cite{arnold1979mathematical}, Euler's hydrodynamics should be considered as a geometric theory. The Euler equations for ideal fluids have a number of common features with the Euler equations for the free rotation of a rigid body. If the motion of a rigid body in three-dimensional space is given by the group $SO (3)$, then the dynamics of the flow of an ideal (incompressible) fluid is determined by an infinite-dimensional group - a group of diffeomorphisms that preserve volume (or area in 2D). In both cases, the equations of motion can be written in the Hamiltonian form by the Poisson brackets. The Poisson brackets for both systems define the corresponding Lie algebras: in the case of a rigid body, this is $so(3)$, and for fluids, we have the algebra of divergence-free vector fields (see, for example, \cite{ZakharovKuznetsov1997}). However, in both cases the Poisson brackets turn out to be degenerate. The degeneracy for the case of a rigid body is well known - it is related to the conservation of the square of the angular momentum (which is the Casimir). For fluids, the degeneration of the (non-canonical) Poisson brackets was first established by Kuznetsov and Mikhailov \cite{KuznetsovMikhailov}: the simplest Casimir found in \cite{KuznetsovMikhailov} turned out to be the helicity $\int({\bf v \cdot \omega}) d{ \bf r}$. This invariant has a topological meaning \cite{fr}: up to a constant factor, the helicity coincides with the Hopf invariant, the number of links of any two vortex lines. 

We would like to emphasize once again that the Euler equations of the hydrodynamics of an ideal incompressible fluid, being Hamiltonian \cite{arnold1969hamiltonian, arnold1979mathematical, KuznetsovMikhailov, ZakharovKuznetsov1997}, are purely nonlinear - they lack any linear part, i.e. the Euler equations themselves refer to systems with an extremely strong nonlinear interaction. Therefore, it should be expected that for hydrodynamic turbulence at high Reynolds numbers, in the inertial range of scales well described by Euler's hydrodynamics, coherent structures play an essential and possibly decisive role. In this case, one of the options for the evolution of coherent structures is collapse - the formation of a singularity in a finite time. 

It should also be noted that the Euler equations for any dimension have an infinitely large number of integrals of motion. These are the so-called Cauchy invariants, which are Lagrangian vector divergence-free invariants advected by the fluid. The existence of these invariants is a local formulation of Kelvin's circulation conservation theorem. This theorem was proved by Kelvin only in 1869. Cauchy found invariants in 1815 as a result of partial integration of the Euler equations written in the Lagrangian form. The theorem named after Kelvin was in fact first proved by Hankel in 1861 on the basis of Cauchy's work. At the end of the twentieth century - the beginning of the twenty-first, this issue was completely clarified (see the works \cite{YakubovichZenkovich, ZakharovKuznetsov1997, Kuznetsov2002}). The history of this issue can be read in the scientific-historical essay \cite{frisch2014cauchy}. 

The presence of the Cauchy invariants significantly complicates the study of developed hydrodynamic turbulence. These invariants under the so-called non-canonical Poisson bracket introduced in \cite{KuznetsovMikhailov}, are, as shown in \cite{KuznetsovRuban} (see also \cite{KuznetsovRuban2000}), Casimirs. That is, the Cauchy invariants as constraints given at each point significantly limit the vortex flows of the fluid. 

In this review, we discuss the role of coherent structures in the onset of developed hydrodynamic turbulence, when the development of these structures can be described in the leading order by the Euler equations for an ideal incompressible fluid. It is shown that for three-dimensional flows, vortex structures of increased vorticity are formed in the form of thinning pancakes, and for two-dimensional flows, narrowing quasi-one-dimensional (filamentous) structures in the form of quasi-shocks of vorticity are formed. The key role in our research is played by the so-called vortex line representation (VRL), which was first introduced in 1998 by Kuznetsov and Ruban~\cite{KuznetsovRuban} and is derived by partial integration of the Euler equations with respect to the conservation of Cauchy invariants. With the help of VLR, we show that the appearance of coherent structures of pancake and filamentary types are due to the compressibility of the so-called frozen-in-fluid fields - the field of continuously distributed vortex lines for three-dimensional hydrodynamics and the field of divorticity lines in two-dimensional geometry. Due to compressibility, large gradients of the corresponding divergence-free fields appear, which in turn has a significant effect on the formation of turbulence spectra.

Recall that in  compressible hydrodynamics - gas dynamics - the formation of a singularity in a finite time (collapse) is due to breaking a phenomenon discovered in gas dynamics by the famous Riemann (see, \cite{landau2013fluid}). In this case, one ``fluid'' particle catches up with another particle, resulting in the formation of infinite gradients for gas characteristics - density and velocity - the so-called gradient catastrophe (see, for example, \cite{arnold}). The main cause of breaking is related to the compressibility of the gas. From a mathematical point of view, this process is the formation of a fold, which can be described in the language of mappings, in this case corresponding to the transition from the Euler description to the Lagrangian one. The singularity appears at the point where the Jacobian of the given mapping vanishes. In incompressible hydrodynamics - Euler hydrodynamics, it would seem that there is no reason for breaking, since the Jacobian of the transformation from the Euler description to the Lagrangian one, due to incompressibility, is identically equal to $1$. Despite this, there are compressible objects in incompressible hydrodynamics - these are continuously distributed vortex lines, which follow from a simple observation.

Consider the equations of motion of the vorticity ${\bf \omega}=[\nabla\times{\bf v}]$ of an ideal fluid, the so-called Helmholtz equations, which are obtained from the Euler equations by applying the rotor operator to them:
\begin{equation} \label{omega}
\frac{\partial{\bf \omega}}{\partial t}=\mbox{rot}[{\bf v}\times{\bf \omega}],\,\,\mbox{div}\, {\bf v}=0.
\end{equation}
As can be seen from this equation, due to the vector product, only the velocity component normal to the vortex line, ${\bf v}_n$, can change ${\bf \omega}$. Moreover, in a generic situation $\mbox{div}\,{\bf v}_n \neq 0$, which is the reason for the compressibility of continuously distributed vortex lines \cite{KuznetsovRuban,Kuznetsov2002}. Thus, despite the incompressibility of the fluid, in Euler hydrodynamics, there are compressible entities - continuously distributed vortex lines. The velocity component parallel to the vorticity, ${\bf v}_{\tau}$, due to (\ref{omega}) does not change the vorticity, providing incompressibility for the full velocity, $\mbox{div}\,{\bf v} = $0. Note that the equation (\ref{omega}) is often called the equation of frozenness. In fact, frozenness is a property of this equation, which says that any fluid particle is pasted to its vortex line, moving along with it. The particle, therefore, has only one ``freedom'' - motion along the vortex line, which, obviously, due to the equation (\ref{omega}) does not change the vorticity. Therefore, ${\bf v}_n$ is the velocity of the vortex line itself. This statement has a simple geometric explanation. For an arbitrary curve, obviously, any deformations along it do not change the curve itself - only transverse deformations lead to its displacement. Therefore, the motion of the vortex line is determined by the velocity ${\bf v}_n$; the position of the vortex line is found from the solution of the system of ordinary differential equations for ``new'' Lagrangian trajectories: 
\begin{equation}  \label{VLR-trajectory}
\frac{d{\bf r}}{dt} = {\bf v}_n({\bf r},t)\,\, \mbox{at}\, {\bf r}|_{t=0} = {\bf a}.  
\end{equation}
The solution of these equations defines a compressible mapping ${\bf r} = {\bf r}({\bf a},t)$. The latter follows directly from the Liouville formula applied to this equation,
\begin{equation} \label{jacobian}
\frac{d{J}}{dt} = \mbox{div}\,{\bf v}_n \,\cdot J,
\end{equation}
where $J=\det{\partial x_i/\partial a_j }$ is the Jacobian of the map. Since $\mbox{div}\,{\bf v}_n \neq 0$, no additional restrictions are imposed on $J$. The Jacobian can take arbitrary values, including zero. It is important that the equation (\ref{omega}) in terms of this mapping admits integration:
\begin{equation} \label{VLR}
{\bf \omega}({\bf r},t) = \frac{({\bf\omega}_0({\bf a}) \nabla_a){\bf r}} { J},
\end{equation}
where $\omega_0({\bf a})$ is the initial vorticity value, which has the meaning of the Cauchy invariants (see, for example, \cite{yakubovich2001matrix, ZakharovKuznetsov1997}). The equations (\ref{VLR},\ref{VLR-trajectory}), together with the incompressibility condition $\mbox{div}\,{\bf v}=0$, form a closed system of vortex line representation (VLR), first introduced by Kuznetsov and Ruban \cite{KuznetsovRuban} (see also \cite{Kuznetsov2002, Kuznetsov2006}).

Later it became clear that the existence of compressible distributions for divergence-free fields is inherent in all frozen-in fields. Moreover, this statement is true for an arbitrary frozen-in-fluid field $\mathbf{B}$, which equations of motion are written in the same form as (\ref{omega}):
\begin{equation} \label{B}
\frac{\partial\mathbf{B}}{\partial t}=\mbox{rot}\lbrack \mathbf{v}\times\mathbf{B}],\,\,\mbox{div}\,\mathbf{v}=0.
\end{equation}

In MHD, $\mathbf{B}$ is a magnetic field (for infinitely large magnetic Reynolds numbers). Less well known is that for two-dimensional flows of an ideal fluid, the vorticity rotor, $B_x=\partial_y \omega,\, B_y=-\partial_x \omega $ ( divorticity), also obeys the equation (\ref{B}) \cite{Sulem} (see also \cite{weiss, kuznetsov2007effects}). 

Since the vorticity in (\ref{VLR}) contains $J$ in the denominator, which can take arbitrary values, a possible collapse scenario can arise due to the vanishing of the Jacobian $J$, which, in a generic situation, should first occur in one separate point. Such a scenario turned out to be possible for three-dimensional integrable hydrodynamics \cite{KuznetsovRuban, KuznetsovRuban2000}. Three-dimensional integrable hydrodynamics can be obtained from Euler's ideal hydrodynamics in the local induction approximation. The equations of three-dimensional integrable hydrodynamics allow the application of the VLR to them. As a result, it turns out that each vortex filament is an autonomous object that does not interact with all others, but with its own nonlinear dynamics, which is described using the integrable Landau-Lifshitz equation, which is gauge-equivalent to the one-dimensional nonlinear Schr{\"o}dinger equation \cite{hasimoto, ZakharovTakhtadzhyan}. Being free, each vortex line can overtake another line, i.e. overturning of the vortex lines occurs. As a result, at some point in a finite time, the VLR Jacobian vanishes, which leads to the appearance of a singularity for the vorticity. The breaking  of vortex lines in three-dimensional integrable hydrodynamics occurs due to the compressibility of vortex lines, despite the incompressibility of the flow itself \cite{KuznetsovRuban2000}.

We note that the question of collapse for Euler hydrodynamics is still debatable, despite numerous numerical and exact analytical results (see the reviews \cite{chae2008incompressible}, \cite{gibbon2008three} and the literature cited there). The solution of this problem, whether there is a collapse or not,  our understanding of the nature of developed hydrodynamic turbulence depends significantly.

The main conclusion that can be drawn based on what has been said is that in Euler hydrodynamics of an incompressible fluid there are compressible entities - this is the vorticity field for three-dimensional flows and the vorticity rotor field - divorticity - in two-dimensional geometry. It is intuitively clear that the compressibility of continuously distributed vortex lines should provide the appearance of structures such as shock waves, which in gas dynamics first arise due to breaking at one separate point, and then the breaking region expands, leading to the formation of a caustic. Precisely such structures, pancake-type structures, are observed in our numerical experiments \cite{agafontsev2015,agafontsev2016development,agafontsev2016asymptotic}. Structures of this type were first found in numerical experiments by M. Brachet et al.~\cite{Brachet} (1992). Subsequently, in \cite{agafontsev2015,agafontsev2016development,agafontsev2016asymptotic} showed that the formation of pancake-shaped structures is similar to the breaking process in gas dynamics, i.e. gradient catastrophe. The appearance of a singularity  does not take place in a finite time, but over an infinite one, with growth exponential in time. 

The vortex line representation  introduced by Kuznetsov and Ruban \cite{KuznetsovRuban}, which takes into account both the presence of an infinite number of Cauchy invariants and the compressibility of continuously distributed vortex lines, is of fundamental importance for understanding the physical nature of vortex line breaking. In this review, the main attention will be paid to the results of numerical integration of the Euler equations, which demonstrate the compressibility of both three-dimensional pancake-type structures \cite{agafontsev2015} and narrowing vorticity quasi-shocks for two-dimensional flows, and their role in the formation of spectra for the developed ($Re\gg 1$) Kolmogorov-type turbulence \cite{AgafontsevKuznetsovMailybaev2019}. We are confident that compressing structures of this type are a property of all vector fields frozen into a fluid, in particular, they are inherent in ideal magnetohydrodynamics. In the works \cite{agafontsev2015}, \cite{agafontsev2016development} on the study of structures of increased vorticity of the pancake type in three-dimensional geometry, it was numerically established that their evolution has a scaling character and is described with high accuracy using the found exact solutions of the three-dimensional Euler equations \cite{agafontsev2016asymptotic}. Scaling between the maximum vorticity in pancake  and its thickness $\ell $,
\begin{equation} \label{scaling}
\omega_{\max}\sim \ell^{-2/3},
\end{equation}
was first found on the basis of direct numerical integration of three-dimensional Euler equations in \cite{agafontsev2015}, and then verified for more than 30 initial conditions \cite{agafontsev2016development}. This gave grounds to assert the universality of this scaling as a Kolmogorov-type relation. In numerical experiments, it was found that the growth of vorticity and narrowing of pancake-type structures depend exponentially on time, without any tendency to explosive behavior. We present both analytical and numerical arguments in favor of the existence of this scaling. Our consideration is based on the vortex line representation \cite{KuznetsovRuban} and its analogs \cite{kuznetsov2007effects, MHD}. In this review, to describe the three-dimensional flows of an ideal incompressible fluid, we will follow the VLR formulation given in \cite{Kuznetsov2002, Kuznetsov2008}. We will discuss the Hamiltonian structure of the vortex line representation, which is based on the existence of an infinite number of local Lagrangian Cauchy invariants and the compressibility of the VLR  map. It is important that the introduction of the Cauchy invariants into the VLR solves the problem of determining all Casimirs for the non-canonical Poisson bracket \cite{KuznetsovMikhailov}. 

The relation (\ref{VLR}) is central to the VLR. It shows that an increase in vorticity is possible due to a decrease in the Jacobian $J$, i.e. is related to the compressibility of the vorticity field. Such a situation, as is known, was first understood for compressible hydrodynamics by Riemann when constructing an exact solution in the form of the so-called simple Riemann wave, which demonstrates the breaking phenomenon when one Lagrangian particle catches up with another. In this case, infinite derivatives appear in the solution profile in a finite time (this is the so-called gradient catastrophe \cite{arnold}). To describe this phenomenon in the three-dimensional case, it is necessary to make a transition from the Euler description to the Lagrangian one in the equations of gas dynamics. The breaking occurs for the first time at the point where the corresponding Jacobian vanishes. Obviously, when approaching the breaking point, it is necessary to take into account dissipation due to viscosity, thermal conductivity, etc. However, away from this region, the overturning process will continue, which leads to the formation of caustics -- pancake-type quasi-two-dimensional structures (see, for example, \cite{shandarin1989large}, \cite{GurbatovSaichevShandarin}). 

In this review, essentially two questions will be considered: first, we will show, using the geometric features of the PVL mapping for three-dimensional Euler equations, that scaling (\ref{scaling}) can be considered as a result of vortex line overturning. In contrast to overturning in compressible gas dynamics, when everything comes down to the appearance of a gradient catastrophe for a scalar quantity - density, in this case, we are talking about overturning of a divergence-free vector field - vorticity. In the first part of the review, we will also discuss how the appearance of pancake-type structures affects the turbulent characteristics during the onset of turbulence, in particular, the turbulence spectrum. It will be shown that, despite the strong anisotropy of turbulence, its spectrum in the inertial interval is close to the Kolmogorov one. Anisotropy has a significant effect on the higher structural velocity functions. In this case, however, the third-order structure functions have the same power dependence on $R=|{\bf r_1}-{\bf r_2}|$ \cite{AgafontsevKuznetsovMailybaev2019} as in isotropic turbulence \cite{kolmogorov1941dissipation} (see also \cite{landau2013fluid}).
 
The second question we will consider concerns the formation of the Kraichnan spectrum for two-dimensional hydrodynamic turbulence - the spectrum for a direct cascade with a constant flux of enstrophy towards the short-wavelength region, and the role of divorticity vector field overturning in this process.

Before proceeding to the presentation of the main provisions of the review, a few words should be said about numerical simulation in the case of three-dimensional geometry both with direct integration of the Euler equations and in the vortex line representation. The entire numerical scheme and all the necessary details of the numerical experiments were presented in \cite{agafontsev2015, agafontsev2016development, agafontsev2016asymptotic}.

The main point in the numerical simulation of the VLR  equations was to find not the direct mapping $\mathbf{r} = \mathbf{r}(\mathbf{a},t)$, but the inverse mapping ${\bf a}= {\bf a} ({\bf r},t)$, which made it possible to represent the VLR equations in the Euler variables $\mathbf{r}$ and $t$. This worked especially effectively when finding an operator inverse to the rotor operator. We emphasize that we everywhere used periodic boundary conditions in all three coordinates.

For two-dimensional simulations, we used approximately the same numerical algorithms and periodic boundary conditions for a square domain. The results of numerical simulation are presented in the papers \cite{kuznetsov2007effects, kudryavtsev2013statistical, kuznetsov2015anisotropic, KuznetsovSereshchenko2017}.

We also note that this review mainly discusses the results obtained by the authors in recent years; the review is written on the basis of lectures read by E.A. Kuznetsov at the Nizhny Novgorod Scientific Schools ``Nonlinear Waves'' in 2016 and 2018~\cite{Kuznetsov-2016,Kuznetsov-2018}. 

\section{Cauchy invariants and the VLR}
As is known (see, for example, the reviews \cite{ZakharovKuznetsov1997, salmon}), the Euler equations for an incompressible fluid
\begin{equation} \label{euler}
\frac{\partial {\bf v}}{\partial t}+({\bf v}\nabla ){\bf v}=-\nabla p, \qquad \mbox {\rm div}~{\bf v}=0
\end{equation}
for both two-dimensional and three-dimensional flows have an infinite number of integrals of motion. These are the Lagrangian Cauchy invariants. The simplest expression for the Cauchy invariants can be obtained from the Kelvin theorem on the conservation of velocity circulation,
\begin{equation} \label{9}
\Gamma = \oint ({\bf v}\cdot d{\bf l}),
\end{equation}
where the integration contour $C[{\bf r}(t)]$ moves along with the fluid. If in this expression we pass from the Euler coordinates ${\bf r}$ to the Lagrangian ones ${\bf a}$, then (\ref{9}) will be rewritten as:
\begin{displaymath} \label{9a}
\Gamma =\oint \dot x_i\cdot \frac{\partial x_i}{\partial a_k}~da_k,
\end{displaymath}
where the contour $C[{\bf a}]$ will already be non-moveable.
 
Because of the arbitrariness of the contour $C[{\bf a}]$ and thanks to the Stokes formula, it immediately follows that the quantity
\begin{equation} \label{10}
{\bf I} = \mbox {\rm rot}_a~\Biggl ( \dot x_i \frac{\partial x_i}{\partial {\bf a}} \Biggr )
\end{equation}
is preserved at every point ${\bf a}$. This is the Lagrangian invariant of Cauchy.
 
The preservation of these invariants, as first shown by Salmon \cite{salmon}, is due to a special symmetry, an infinite symmetry with respect to the relabeling of fluid markers, which is symmetry leaving the action invariant. If the Lagrangian coordinates ${\bf a}$ in (\ref{10}) coincide with the initial positions of the fluid particles, then the invariant ${\bf I}$ coincides with the initial vorticity ${\bf \omega}_0({\bf a})$. The preservation of these invariants is a consequence of the fact that the vortex lines are frozen into the fluid. According to this property,  fluid (Lagrangian) particles cannot leave their own vortex line, where they were at the initial instant of time. For Lagrangian particles, only one unfrozen degree of freedom remains - motion along the vortex line, which, due to (\ref{omega}), does not change the values of ${\bf \omega}$. From this point of view, the vortex line is an invariant object and, therefore, it is natural to pass to a description where this invariance is visible from the very beginning. Such a description is the representation of vortex lines \cite{KuznetsovRuban, KuznetsovRubanPRE}. To obtain it explisetly, we decompose the velocity ${\bf v}$ into two components ${\bf v}_n$ and ${\bf v}_{\tau}$, normal and tangential with respect to the vector ${\bf \omega }$. 

The equation of motion for the transverse velocity ${\bf v}_n$ follows directly from the equation (\ref{euler}). It has the form of the equation of motion of a particle in an electromagnetic field:
\begin{equation} \label{electron}
\frac{\partial {\bf v}_n}{\partial t}+({\bf v}_n\nabla){\bf v}_n={\bf E}+[{\bf v}_n\times {\bf H}],
\end{equation}
where the effective fields - electric and magnetic - are given by the expressions:
\begin{equation} \label{electric}
{\bf E}=-\nabla \left ( p+\frac{v^2_{\tau}}{2} \right )-\frac{\partial {\bf v}_{\tau}}{\partial t},
\end{equation}
\begin{equation} \label{magnetic}
{\bf H}=\mbox {\rm rot}~{\bf v}_{\tau},
\end{equation}
It is interesting to note that the electric and magnetic fields introduced in this way are expressed in terms of the scalar -- $(\varphi)$ and vector -- $({\bf A})$ potentials in the standard way accepted for electrodynamics:
\begin{equation} \label{potentials}
\varphi = p+\frac{{\bf v}^2_{\tau}}{2}, \qquad {\bf A}={\bf v}_{\tau},
\end{equation}
so the two Maxwell equations
\begin{displaymath}
\mbox {\rm div}~{\bf H}=0, \qquad \frac{\partial {\bf H}}{\partial t}= -\mbox {\rm rot}~{\bf E}
\end{displaymath}
are automatically satisfied. In this case, the vector potential ${\bf A}$ is subjected to the gauge
$$
\mbox{div}~ {\bf A}=- \mbox{div}~{\bf v}_n,
$$
which is equivalent to $\mbox{div}~{\bf v}=0$. 

The basic equation here is the equation of motion itself (\ref{electron}) for the normal velocity component, which is the equation of motion for a nonrelativistic particle with charge and mass, equal to unity, the speed of light is also equal to unity.

The equation of motion (\ref{electron}) is written in the Euler representation. To proceed to its Lagrangian formulation, we need to consider the equations for ``trajectories'', which are determined by the velocity ${\bf v}_n$:
\begin{equation} \label{mapping1}
\frac{d {\bf r}}{d t}={\bf v}_n ({\bf r},t)
\end{equation}
with initial conditions
$$
{\bf r}|_{t=0}={\bf a}.
$$
The solution of the equation (\ref{mapping1}) defines the mapping
\begin{equation} \label{19}
{\bf r}={\bf r}({\bf a},t),
\end{equation}
defining the transition from the Euler description to the new Lagrange one. The equations of motion in new variables are the Hamilton equations:
\begin{equation} \label{ham}
\dot{\bf P}=-\frac{\partial h}{\partial {\bf r}}, \qquad \dot{\bf r}=\frac{\partial h}{\partial {\bf P}},
\end{equation}
where the dot means differentiation with respect to time at a fixed value of ${\bf a}$, ${\bf P}={\bf v}_n+{\bf A}\equiv {\bf v}$ is the generalized momentum, and the Hamiltonian of the particle $h$, being a function of momentum ${\bf P}$ and coordinate ${\bf R}$, is given by the standard expression:
\begin{displaymath}
h= \frac 12 ({\bf P}- {\bf A})^2 + \varphi \equiv p+\frac{{\bf v}^2}{2}.
\end{displaymath}
The first equation of the system (\ref{ham}) is the equation of motion (\ref{electron}) written in the variables ${\bf a}$, $t$, and the second equation is the same as (\ref{mapping1}). For the ``new'' hydrodynamics (\ref{electron}) or its Hamiltonian formulation (\ref{ham}), Kelvin's theorem (aka Liouville's theorem) will be valid:
\begin{equation} \label{kelvin}
\Gamma = \oint ({\bf P }\cdot d{\bf r }),
\end{equation}
where the integration is carried out along a closed contour moving together with the ``fluid''. From here, just as it was done above when deriving (\ref{10}), the expression for the ``new'' Cauchy invariant follows:
\begin{equation} \label{18}
{\bf I} = \mbox {\rm rot}_a~\Biggl ( P_i \frac{\partial x_i}{\partial {\bf a}} \Biggr ).
\end{equation}
Its difference from the original Cauchy invariant (\ref{10}) is that in the equation of motion (\ref{mapping1}) instead of the velocity ${\bf v}$ its normal component ${\bf v}_n$ is used. As a consequence, the ``new'' hydrodynamics is compressible: $\mbox {\rm div}~{\bf v}_n \neq 0$. Therefore, no restrictions are imposed on the Jacobian $J$ of the transformation (\ref{19}). 

From the formula (\ref{18}), an expression for the ${\bf \omega}$ vorticity can be easily obtained:
\begin{equation} \label{cauchi}
{\bf \omega}({\bf r},t)=\frac{({\bf \omega_0}({\bf a})\cdot\nabla_a) {\bf r}(a,t)}{J},
\end{equation}
where $J$ is the transformation jacobian (\ref{19}) equal to
\begin{displaymath}
J=\frac{\partial(x_1,x_2,x_3)}{\partial(a_1,a_2,a_3)}.
\end{displaymath}
Here we took into account that the generalized momentum ${\bf P}$ coincides with the velocity ${\bf v}$, including the time $t=0$: ${\bf P}_0({\bf a})\equiv {\bf v}_0({\bf a})$. ${\bf \omega}_0({\bf a})$ in this relation is a new Cauchy invariant (coinciding with the initial vorticity) with zero divergence: $\mbox~{\rm div}_a {\bf \omega } _0(a)=0$.
As noted earlier, the relation (\ref{cauchi}) can be obtained directly by integrating the equation (\ref{omega}), it is the same as (\ref{VLR}). From the above derivation, the meaning of the vector ${\bf \omega}_0({\bf a})$ in (\ref{VLR}) as a Cauchy invariant becomes clear.

The introduction of the VLR (\ref{VLR}) also solves another important problem - the determination of all Casimirs for the Hamiltonian description of the vorticity equation of motion (\ref{omega}) by introducing a Poisson structure. As shown in \cite{KuznetsovMikhailov}, the equations (\ref{omega}) can be represented in the Hamiltonian form using the non-canonical Poisson bracket:
\begin{equation} \label{Omega}
\frac{\partial {\bf \omega}}{\partial t}=\mbox{rot} \left[\mbox{rot}\frac{\delta {\cal H}}{\delta {\bf \omega}}\times {\bf \omega}\right] =\{{\bf \omega },{\cal {H}}\},
\end{equation}
where the Poisson bracket is given by:
\begin{equation} \label{bracket}
\{F,G\}=\int \left( {\bf \omega }\left[ \mbox{rot}\,\frac{\delta F}{\delta	{\bf \omega }}\times \mbox{rot}\,\frac{\delta G}{\delta {\bf \omega }}\right] \right) {d{\bf r}}.
\end{equation}
Here the vector $\mbox{rot}\frac{\delta {\cal H}}{\delta {\bf \omega}}$ has the meaning of generalized velocity. In the case of the Euler equation, the Hamiltonian ${\cal H}$ coincides with the total kinetic energy
\[
{\cal H}=\frac 12\int {\bf v}^2 d{\bf r}.
\] 
For three-dimensional integrable hydrodynamics ${\cal H}=\int |{\bf \omega}| d{\bf r}$ \cite{KuznetsovRuban, KuznetsovRuban2000}.

As was first shown in \cite{KuznetsovMikhailov}, the Poisson bracket (\ref{bracket}) turned out to be degenerate. Its simplest Casimir turned out to be the helicity $\int({\bf v \cdot \omega}) d{\bf r}$ . This invariant has a topological origin \cite{fr}: up to a constant factor, the helicity coincides with the Hopf invariant - the number of links of any two vortex lines.

Casimirs are constraints that are defined in the configuration space, in this case, in the space of divergence-free vector fields ${\bf \omega}$. The presence of the Casimirs does not allow inverting the symplectic operator in the equation (\ref{Omega}), which defines the Poisson bracket. As is known, fixing all Casimirs defines a symplectic leaf. According to the general theory (see, for example, the review \cite{ZakharovKuznetsov1997}), the introduction of coordinates on this sheet makes it possible to establish full-fledged Hamiltonian dynamics, in particular, to write down the variational principle. As shown in \cite{KuznetsovRuban, KuznetsovRubanPRE}, the equations of motion of vortex lines, i.e. in the representation of vortex lines, can be obtained based on the variational principle. At the same time, it is possible to show also that all Casimirs for the bracket (\ref{bracket}) are the Cauchy invariants. This fact was established by computing the Poisson bracket (\ref{bracket}) expressed in terms of ${\bf r}({\bf a})$ and the Cauchy invariants ${\bf \omega}_0({\bf a} )$ using VLR (\ref{VLR}) as the appropriate replacement. Calculations have shown that the bracket does not contain the variational derivatives relative to ${\bf \omega}_0({\bf a})$, i.e. ${\bf \omega}_0({\bf a})$ serve as Casimirs for the bracket (\ref{bracket}).

It should also be noted that the equations of motion (\ref{mapping1}) together with the relation (\ref{cauchi}) are the result of a partial integration of the Euler equation (\ref{euler}). These equations are solved with respect to the Cauchy invariants - an infinite number of integrals of motion, which is fundamentally important for numerical integration (see, for example, \cite{agafontsev2016asymptotic}). For this system, the Cauchy invariants are automatically conserved, while in the case of direct integration of the Euler equations one has to follow the extent to which the Cauchy invariants are conserved quantities. Apparently, this fact is one of the main limitations that determine the accuracy of discrete numerical schemes for direct integration of the Euler equations. 

Another important property of the representation of vortex lines is the absence of any restrictions on the value of the Jacobian $J$, which, for example, takes place when transferring from the Euler description to the Lagrangian one, when the Jacobian is equal to one. In our case, $1/J$ has the meaning of the density of $n$ vortex lines. This quantity, by virtue of the equation (\ref{mapping1}), as a function of ${\bf r}$ and $t$, obeys the continuity equation:
\begin{equation} \label{den}
\frac{\partial n}{\partial t}+\mbox {\rm div}_r(n{\bf v}_n) =0.
\end{equation}
In this equation, $\mbox {\rm div}_r{\bf v}_n\neq 0$, because only the total velocity ${\bf v}$ has zero divergence. 

\section{VLR compressibility and self-similar law of $2/3$}

In this section, we will consider the features of the VLR and its geometric characteristics, based on the exact solution of the Euler equations \cite{agafontsev2016asymptotic}. This solution, as noted earlier, is in good agreement with the results of numerical simulation of pancake-type vortex structures.

Let ${\omega_{\max}}$ be the spatially maximum value of the vorticity norm, which is a function of time $t$. Obviously, at the maximum point $\mathbf{x}=\mathbf{x}_{\max}(t)$ we have $\nabla \omega=0 $. Representing $\boldsymbol{\omega}=\omega\boldsymbol{\tau}$, where $\boldsymbol{\tau}$ is the unit vector ($\boldsymbol{\tau}^2=1$), from (\ref{omega}) it is easy to get the equation for ${\omega_{\max}}$:
\begin{equation}  \label{omega-max}
\omega_{\max}^{-1}\frac{d {\omega_{\max}}}{dt}=\tau_i \frac{\partial 
v_j}{\partial x_i}\tau_j \dot .
\end{equation}
Here the derivative $\frac{\partial v_j}{\partial x_j}$ is taken at the point $\mathbf{x}=\mathbf{x}_{\max}(t)$. In the case when the vorticity field is symmetrical about this point, the expression on the right side of this equality can be written as\begin{equation}  \label{omega-max1}
\tau_i \frac{\partial v_j}{\partial x_i}\tau_j =\mbox{div}\,\mathbf{v_{\tau}}.
\end{equation}

\begin{figure}[t!] 
\centering
\includegraphics[width=0.6\linewidth]{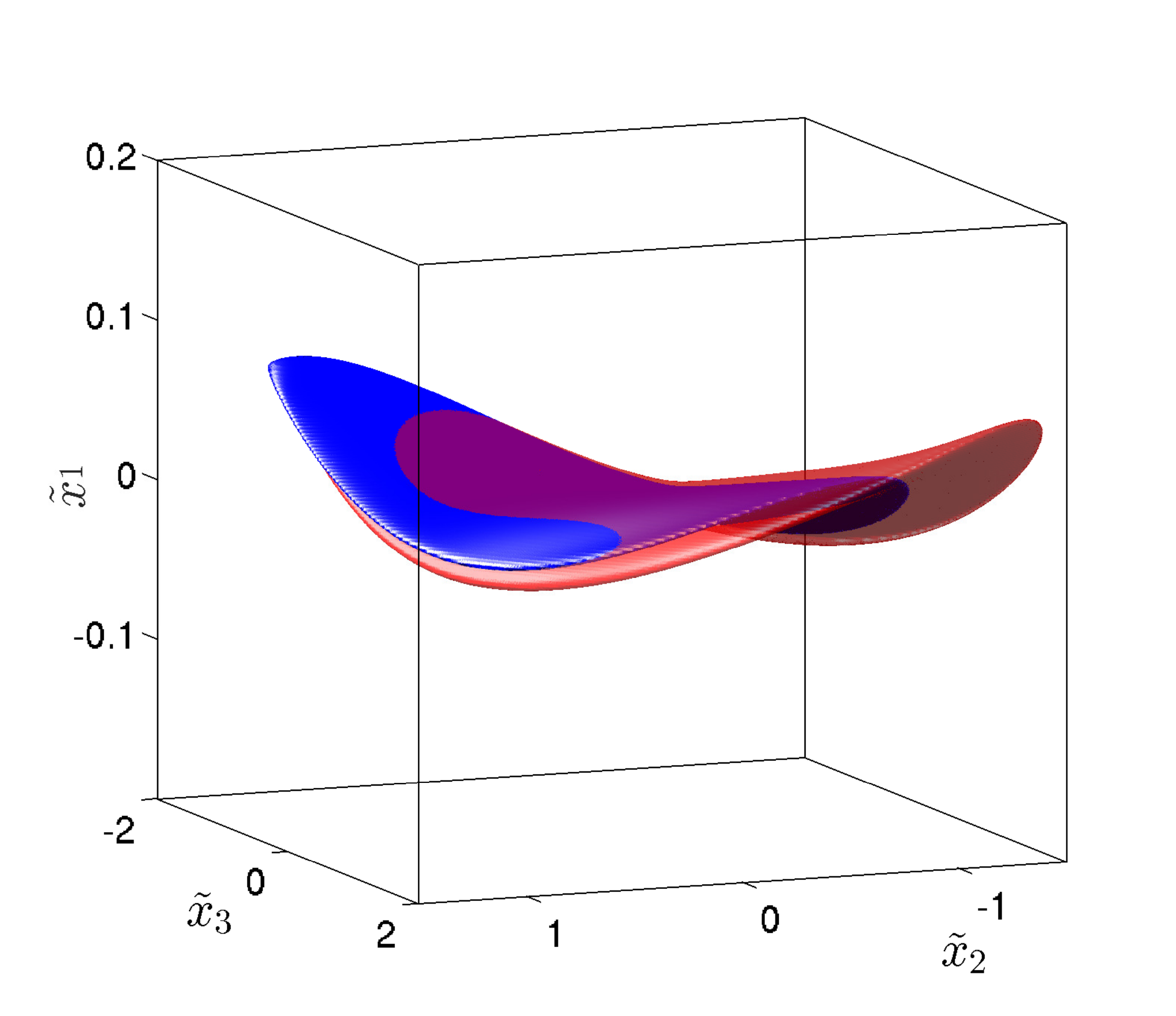}
\caption{Isosurfaces of vorticity $|\omega|=0.8\,\omega_{\max}$ (red) and Jacobian $J=1.25\,J_{\min}$ (blue) at $t=7.5$; VLR simulation \cite{agafontsev2017universal}.} 
\end{figure}

Consider now the equation (\ref{jacobian}). In this equation, the derivative $d/dt$ is taken at a constant value of $\mathbf{a}$. Therefore, in the variables $\mathbf{r}$ and $t$ (i.e., the Eulerian variables) this equation will be written as
\[
\frac{\partial J}{\partial t} + 
({\mathbf{v}_n}\nabla)J=\mbox{div}\,{\mathbf{v}}_n \,\cdot J= -\mbox{div}\,{\mathbf{v}}_{\tau} \,\cdot J, 
\]
This shows that at the minimum point of the Jacobian $\mathbf{x}=\mathbf{x}_{\min}(t)$ we have
\begin{equation}  \label{jacobian1}
\frac{dJ_{\min}}{d t} = \mbox{div}\,{\mathbf{v}}_n \,\cdot J_{\min}.
\end{equation}
If we assume that the maximum point of vorticity coincides with the minimum point of the Jacobian, ${\mathbf{r}_{\max}}(t)={\mathbf{r}_{\min}}(t)$, then in according to (\ref{omega-max})--(\ref{jacobian1}) we arrive at the relation
\begin{equation}  \label{jacobian2}
\omega_{\max}J_{\min} = \mbox{const}.
\end{equation}

Figure 1 shows the results of integrating the VLR equations \cite{agafontsev2017universal} for the vorticity isosurfaces $|\omega|=0.8\,\omega_{\max}$ and the Jacobian $J=1.25\,J_{\min}$ at $t =7.5$, which are practically the same. This coincidence indicates that $|\omega|$ and $J$ near their extreme points (maximum and minimum) have the same spatial dependence: they are determined only by one function of coordinates. In this case, the distance between the maximum vorticity point and the minimum Jacobian in the numerical experiment was of the order of the pancake thickness.

\begin{figure}[t!]
\centering
\includegraphics[width=0.45\linewidth]{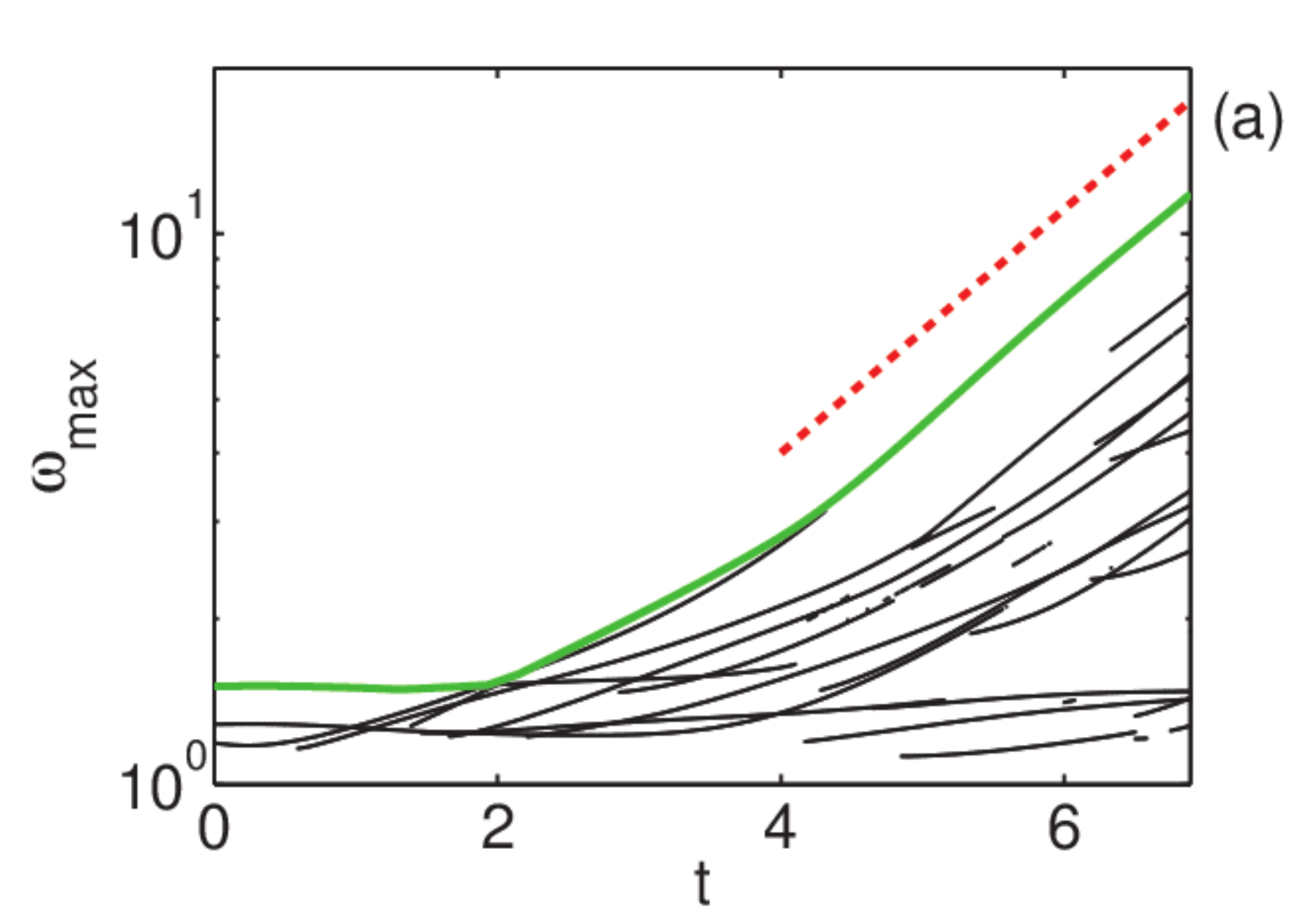}
\includegraphics[width=0.45\linewidth]{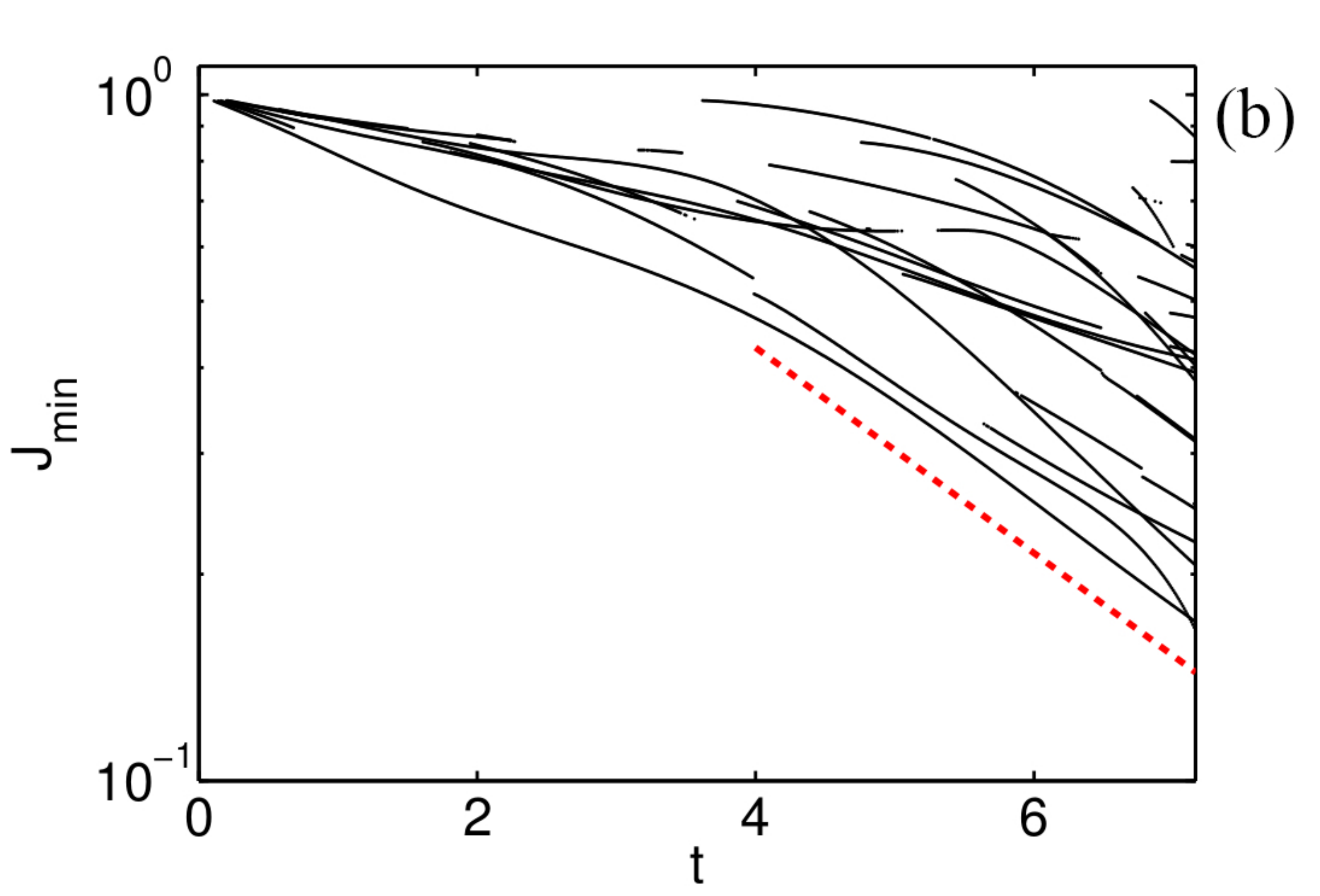}
\caption{(a) Evolution of local vorticity maxima (logarithmic scale). The green line corresponds to the global maximum, the dotted red line - envelope $\propto e^{t/T_{\protect\omega}}$ with $T_{\protect\omega}=2$.
(b) Exponential decrease in time of $J_{\min}$ as a function of time (logarithmic scale) for various pancakes, the dotted red line is the envelope.}
\end{figure}

Figures 2(a) and 2(b) show the time variation of $\omega_{\max}$ and $J_{\min}$ for all observed pancakes. The numerical experiment demonstrates an exponential increase in the maximum vorticity for each of the pancakes and, accordingly, an exponential decrease in the minimum Jacobian, so that their product is $\omega_{\max}J_{\min} \approx \mbox{const}$.

The relation (\ref{jacobian2}) thus shows that the increase in vorticity in the pancake is due to a decrease in the Jacobian. In this case, the numerator changes slightly, especially at long times when exponential growth is clearly observed. The explanation of the latter is related, as will be shown below, to the structure of the exact solution \cite{agafontsev2016asymptotic} of the three-dimensional Euler equation, which models the evolution of a pancake at the exponential stage of its growth with good accuracy.

Exact Cartesian solution $\mathbf{x} = x_{1}\mathbf{n}_{1} + x_{2}\mathbf{n}_{2} + x_{3}\mathbf{n}_ {3}$ for a vorticity that depends only on $x_{1}$ and has only one component parallel to the $\mathbf{n}_{2}$ axis, has the form:
\begin{eqnarray}
\mathbf{v}(\mathbf{x},t) &=& 
-\omega_{\max}(t)\,\ell_{1}(t)\,f\left(\frac{x_1}{\ell_{1}(t)}\right) \mathbf{n}_{3} + 
\left(\begin{array}{c}
-\beta_{1}x_{1} \\ \beta_{2}x_{2} \\ 
\beta_{3}x_{3}\end{array}\right),  \label{solution-1} \\
\boldsymbol{\omega}(\mathbf{x},t) &=& 
\omega_{\max}(t)f^{\prime}\left(\frac{x_1}{\ell_{1}(t)}\right) \mathbf{n}_{2},  \label{solution-2}
\end{eqnarray}
where $\beta_1$, $\beta_2$ and $\beta_3$ are arbitrary constants  satisfying the relation $-\beta_1 + \beta_2 + \beta_3 = 0$. Here $\omega_{\max}(t) = w_0 e^{\beta_{2}t}$ and $\ell_{1}(t) = h_0 e^{-\beta_{1}t}$ are time dependencies of the vorticity maximum and pancake thickness, $w_0$ and $h_0$ are positive (initial) values, $f(\xi)$ is an arbitrary function with $|\max f^{\prime }(\xi) | = $1. The velocity in this solution is a superposition of shear flow and asymmetric potential stretching flow $(-\beta_1 x_1,\,\beta_2 x_2,\,\beta_3 x_3)$.

For this exact solution, the representation of vortex lines is easily constructed:

\begin{equation}  \label{solution-mapping}
x_{1}=a_{1}\,e^{-\beta _{1}t},\quad x_{2}=a_{2}, \quad  x_{3}=a_{3}\,e^{\beta_{3}t}-w_0h_0f\left(\frac{a_{1}}{h_0}\right)\,\frac{\sinh (\beta_{3}t)}{\beta _{3}},
\end{equation}
with a Jacobi matrix of the form:
\begin{equation}  \label{solution-UV}
\hat{J}(\mathbf{a},t)=\left( \begin{array}{ccc} e^{-\beta _{1}t} & 0 & 0 \\ 0 & 1 & 0 \\ 
-w_0f^{\prime }\left(\frac{a_{1}}{h_0}\right)\,\frac{\sinh (\beta _{3}t)}{\beta _{3}} & 0 & e^{\beta _{3}t}\end{array}\right) ,\quad J(\mathbf{a},t)=\mathrm{det}\,\hat{J}=e^{(\beta_3-\beta_1)t}=e^{-\beta _{2}t}.
\end{equation}
It follows that the maximum vorticity is inversely proportional to the Jacobian: $\omega_{\max}\sim J^{-1}$, in full agreement with our previous conclusion. Indeed, for this solution the Jacobian $J$ does not depend on the coordinates. The coordinate dependence arising in the numerical experiment is due to the three-dimensionality of the structure (see Fig. 1). When moving along the pancake surface, the numerical solution of the equations locally (up to distances of the order of ten thicknesses) agrees well with the exact solution (\ref{solution-1}), (\ref{solution-2}). Another important circumstance of the VLR for (\ref{solution-1}), (\ref{solution-2}) is that one (the first) eigenvalue of the Jacobi matrix grows exponentially with time, the second is equal to one, and the third decreases exponentially. Then, the Jacobian decreases exponentially, inversely proportional to $\omega_{\max}$. In this case, the vorticity is directed along the second axis corresponding to the second eigenvalue. This property, as we see in the numerical experiment, is fully preserved and it corresponds to the fact that the numerator in the expression (\ref{VLR}), as noted earlier, due to (\ref{jacobian}) practically does not change in value at the maximum vorticity point, but can change in direction.

\subsection{Singular values}
The Jacobi matrix for the exact solution contains one off-diagonal element $J_{31}$, which grows exponentially $\sim e^{\beta _{3}t}$ with time. The presence of this element significantly affects, as we will see below, the direction of vorticity, as well as the dependence of the Jacobian on the thickness of the pancake, which is observed in the numerical experiment. The second very important circumstance that follows from this is that the eigenvalues of the Jacobi matrix do not represent relative expansions due to the off-diagonal nature of the matrix itself. In order to introduce relative expansions correctly, we need to turn to the singular value problem for the matrix $\widehat{\mathbf{J}} = [\partial x_i/ \partial a_j]$ at the point $J_{\min}$. This problem reduces to finding two rotation matrices $\mathbf{U}$ and $\mathbf{V}$ and a diagonal one $\mathbf{\Sigma}=\mathrm{diag}\{\sigma_{1},\sigma_{2 },\sigma_{3}\}$ containing non-negative elements $0<\sigma_1<\sigma_2<\sigma_3$, called singular values. Moreover, the Jacobi matrix can be represented as $\widehat{\mathbf{J}} = \mathbf{U}\mathbf{\Sigma}\mathbf{V}^T$, where $T$ means the transposition. The rotation matrices $\mathbf{U}$ and $\mathbf{V}$ are constructed from the eigenvectors of the eigenvalue problem for two symmetric matrices $\widehat{\mathbf{J}}\widehat{\mathbf{J}}^T $ and $\widehat{\mathbf{J}}^T\widehat{\mathbf{J}}$, respectively, while the (coinciding) eigenvalues are the squares of the singular values $\sigma_i$. Note that $\mathbf{G}^{(a)}=\widehat{\mathbf{J}}^T\widehat{\mathbf{J}}$ is a metric tensor in the $\mathbf{a} $-space,
\begin{equation}
G_{\alpha\beta}^{(a)} = \bigg[\frac{\partial x_{i}}{\partial a_{\alpha}}\frac{\partial x_{i}}{\partial a_{\beta}}\bigg],\quad d\mathbf{x}^{2} = G_{\alpha\beta}^{(a)}\,da_{\alpha}da_{\beta},  \label{metric-tensor-1}
\end{equation}
while $\mathbf{G}^{(x)}=[\widehat{\mathbf{J}}\widehat{\mathbf{J}}^T]^{-1}$ is the metric tensor in $ \mathbf{x}$-space,
\begin{equation} \label{metric-tensor-2}
G_{ij}^{(x)} = \bigg[\frac{\partial a_{\alpha}}{\partial x_{i}}\frac{\partial a_{\alpha}}{\partial x_{j}}\bigg],\quad d\mathbf{a}^{2} = G_{ij}^{(x)}\,dx_{i}dx_{j}.  
\end{equation}

For the exact solution, the Jacobian matrix has the following singular values:
\begin{equation}
\sigma_1^2 = g-\sqrt{g^2-e^{-2\beta_2t}},\quad \sigma_2^2 = 1, \quad \sigma_3^2 = g+\sqrt{g^2-e^{-2\beta_2t}},  \label{solution-S0}
\end{equation}
where 
\begin{equation} \label{solution-S0b}
g = \frac{1}{2}\left(e^{-2\beta_{1}t}+e^{2\beta_{3}t}+ \left[w_0 f^{\prime }\left(\frac{a_{1}}{h_0}\right)\,\frac{\sinh(\beta_{3}t)}{\beta_{3}}\right]^2 \right).  
\end{equation}
As $t\to \infty$ 
\begin{equation} \label{solution-S}
\sigma_1 \propto e^{-\beta_1t},\quad \sigma_2 = 1, \quad \sigma_3 \propto e^{\beta_3t},  
\end{equation}
which agrees with the numerical results.

Thus, near the minimum of the Jacobian along the first direction, there is a strong contraction: $\sigma_1\propto e^{-\beta_1t}\propto \ell_{1}$, as a result of which all Lagrangian markers must collapse as $t\to\infty$ exactly; in the third direction, a strong expansion of $\sigma_3\propto e^{\beta_3t}\propto \omega_{\max}^{-1}\ell_{1}^{-1}$ occurs; in the intermediate direction $\sigma_2$ is close to unity and changes insignificantly with time.

In this limit for the exact solution, the rotation matrices
$\mathbf{U}=\{\mathbf{n}_{1}^{(x)},\mathbf{n}_{2}^{(x)},\mathbf{n}_{3} ^{(x)}\}$ and
$\mathbf{V}=\{\mathbf{n}_{1}^{(a)},\mathbf{n}_{2}^{(a)},\mathbf{n}_{3} ^{(a)}\}$ in
$\mathbf{x}$- and $\mathbf{a}$-spaces have the form:
\[
\mathbf{U}\simeq \mathbf{1},\quad \mathbf{V}\simeq \left( \begin{array}{ccc}
\frac{1}{\sqrt{1+q^{2}}} & 0 & \frac{q}{\sqrt{1+q^{2}}} \\ 0 & 1 & 0 \\ 
\frac{-q}{\sqrt{1+q^{2}}} & 0 & 
\frac{1}{\sqrt{1+q^{2}}}\end{array}\right) , \]
where 
\begin{equation} \label{solution-UVg}
q=-\frac{w_{0}}{2\beta _{3}}f^{\prime }\left( \frac{a_{1}}{h_{0}}\right) .  
\end{equation}
At large times, as the numerical experiment shows, the matrix $\mathbf{U}$ is close to the identity, and the matrix $\mathbf{V}$ is close to the antidiagonal matrix with elements $V_{13}\approx V_{22}\approx -V_ {31}\approx 1$. Thus, $q$ in (\ref{solution-UV}) can be considered a large value.

\subsection{Scaling law of $2/3$}

Let us now turn to the question of the origin of the scaling (\ref{scaling}), which relates the maximum vorticity and the thickness of the pancake. Let's make one remark. As already noted, the numerical solution is well approximated by the exact solution. Therefore, to find the scaling (\ref{scaling}), the exact solution will be considered as a zero approximation. First of all, this concerns the transition from Lagrangian variables to Eulerian variables in (\ref{VLR}), namely for the Jacobian.

Recall that we determine the thickness of the structure based on the expansion of $\omega$ in the vicinity of the maximum point $\omega_{\max}$:
\begin{equation}
|\omega |=\omega _{\max}-\frac{1}{2}\Gamma _{ij}^{(\omega 
)}{\tilde{x}}_{i}{\tilde{x}}_{j},  \label{expansion}
\end{equation}
where ${\tilde{\mathbf{x}}}=\mathbf{x}-\mathbf{x}_{\max}$. In this case, the eigenvalues $\lambda_{n}^{(\omega)}$ of the matrix of second derivatives $\Gamma_{ij}^{(\omega)}=-\partial_{i}\partial_{j}|\omega |$ calculated in the local vorticity maximum will determine the scales of the structure in three orthogonal directions:
$\ell_{n}=\sqrt{2\omega_{\max}/\lambda_{n}^{(\omega)}}$. The maximum eigenvalue $\lambda_{1}^{(\omega)}$ specifies the thickness of the pancake, and the eigenvector corresponding to this eigenvalue is the direction of the normal to the pancake-type structure. In the $\Gamma _{ij}^{(\omega )}$ eigenaxes, the expansion (\ref{expansion}) is written as
\[
|\omega |=\omega _{\max}\left( 
1-\sum_{n=1,2,3}\frac{{\tilde{x}}_{n}^{2}}{\ell_{n}^{2}}\right).
\]
At the same time, according to our numerical simulation~\cite{agafontsev2016development,agafontsev2016asymptotic}, only the first characteristic size of a pancake (its thickness) decreases with time, while the other two sizes change slightly and remain on the order of unity,
\begin{equation}
\ell_{1}\propto e^{-\beta_{1}t},\quad \ell_{2}\propto 1,\quad \ell_{3}\propto 1. \label{omega-dimensions}
\end{equation}

The geometry of the low Jacobian region near the minimum $J_{\min}$ can be described in the same way:
\begin{equation}
J = J_{\min} - \frac{1}{2}\Gamma_{ij}^{(J)}{\tilde{x}}_{i}{\tilde{x}}_{j},  \label{expansion-J}
\end{equation}
where ${\tilde{\mathbf{x}}}=\mathbf{x}-\mathbf{x}_{\min}$ and the eigenvalues $\lambda_{n}^{(J)}$ of the matrix of second derivatives $\Gamma_{ij}^{J} = \partial_{i}\partial_{j}J$ calculated in the local minimum of the Jacobian determine the size of the structure, $l_{n}=\sqrt{2J_{\min}/\lambda_{n}^{(J)}}$. The regions of high vorticity and low Jacobian largely intersect with each other (see Fig. 1), and the characteristic scales of the second region $l_{n}$ behave in time in the same way as the characteristic dimensions of the first~\cite{agafontsev2017universal} :
\begin{equation}
l_{1}\propto e^{-\beta_{1}t},\quad l_{2}\propto 1,\quad l_{3}\propto 1.  \label{solution-exp-l}
\end{equation}
In this case, the vorticity maximum and the Jacobian minimum behave as $\omega_{\max}\propto J_{\min}^{-1}\propto e^{\beta_{2}t}$, and $\beta_{2} /\beta _{1}\approx 2/3$, i.e. the vortex structure evolves according to the $2/3$ law; see (\ref{scaling}).

\begin{figure}[t]
\centering
\includegraphics[width=0.6\linewidth]{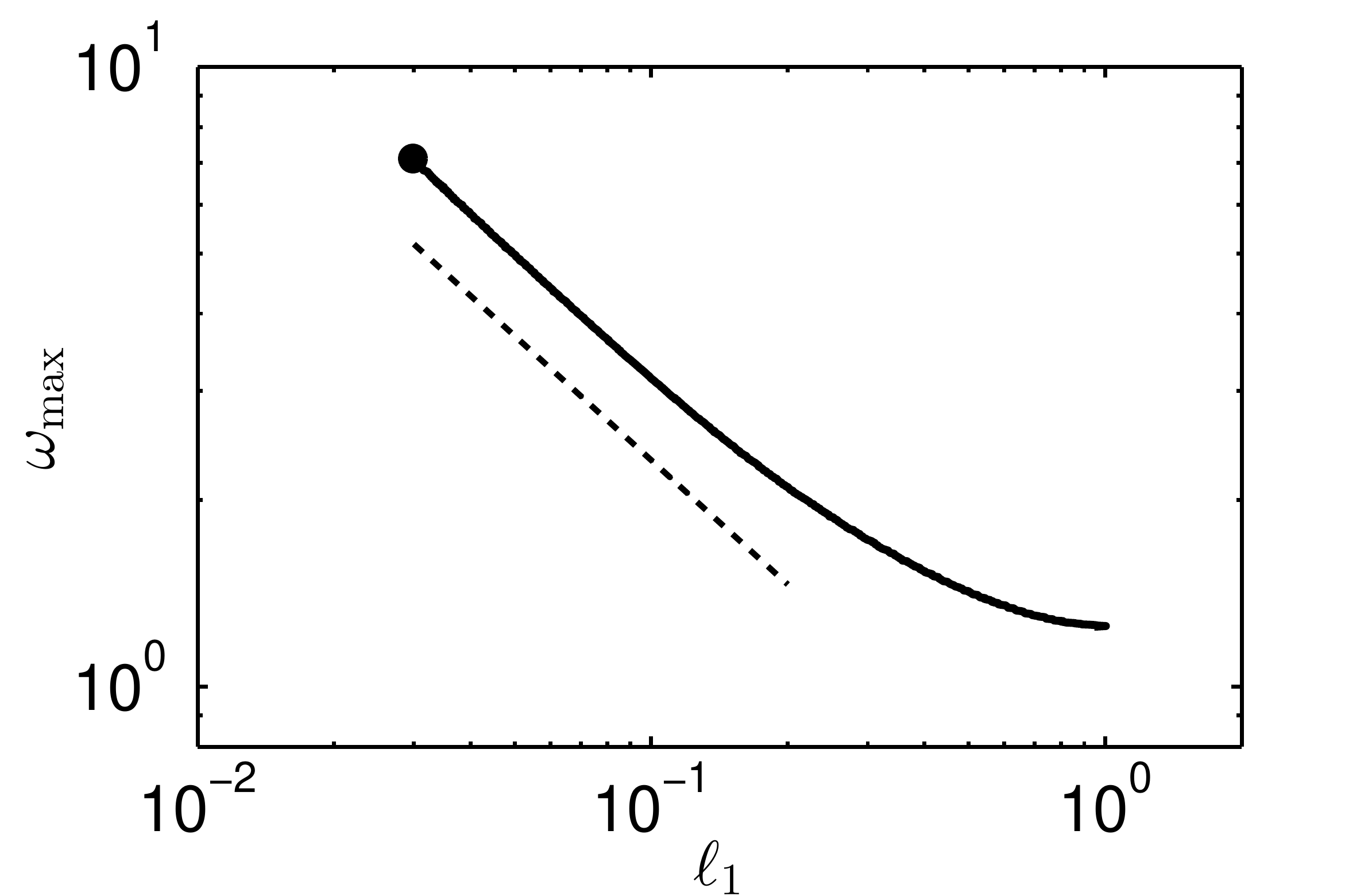}
\caption{Maximum vorticity $\omega_{\max}$ as a function of $\ell_{1}$ on a logarithmic scale. The thick dot on this dependence corresponds to the maximum vorticity at the final simulation time $t=7.5$, the dotted line corresponds to the power law $\omega_{\max}\propto \ell_{1}^{-2/3}$.}
\label{fig:fig3-w-ell}
\end{figure}

In the Lagrangian variables ${\tilde{\mathbf{a}}}$ (everywhere below we omit the tilde sign), the expansion of the Jacobian near the minimum is written as
\begin{equation}  \label{exp-J-a}
J = J_{\min} + \frac 12 \Gamma_{ij}^{(a)}{a}_i{a}_j,
\end{equation}
where $\Gamma_{ij}^{(a)}=\partial^{2}J/\partial a_{i}\partial a_{j}$ is a positive definite matrix.
The matrices $\Gamma_{ij}^{(a)}$ and $\Gamma_{ij}^{(J)}$ (hereinafter, for convenience, the latter will be denoted as $\Gamma_{ij}^{(x)}$) are related according to the chain rule,
\begin{equation}
\Gamma^{(a)}=\hat{\mathbf{J}}^{T}\Gamma^{(x)}\hat{\mathbf{J}}=\mathbf{V}\gamma \mathbf{V}^{T},  \label{HessianJxa}
\end{equation}
where we denoted the matrix
\[
\gamma =\mathbf{\Sigma}\mathbf{U}^{T}\Gamma^{(x)}\mathbf{U}\mathbf{\Sigma }.
\]
In the $\Gamma^{(x)}$ eigenaxes, the $\gamma$ matrix is close to diagonal, because $\mathbf{U}$ tends to the identity, $\Gamma^{(x)}=\mathrm{diag}\{\lambda_{1}^{(J)}, \lambda_{2}^{(J) }, \lambda_{3}^{(J)}\}$ and $\gamma$ turns out to be the product of three diagonal and two almost diagonal matrices. Hence, because $\lambda_{n}^{(J)} = 2J_{\min}/l_{n}^{2}$, for the diagonal elements $\gamma$ we approximately have:
\[
\gamma _{ii}\approx \sigma _{i}^{2}\lambda _{i}^{(J)}=2J_{\min}\sigma _{i}^{2}/l _{i}^{2}.
\]
Remembering that $\sigma_{1}\propto l_{1}\propto e^{-\beta_{1}t}$, $\sigma_{2}\propto l_{2}\propto 1$, $\sigma_ {3}\propto e^{\beta_{3}t}$ and $l_{3}\propto 1$, we get
\[
\gamma_{11} \propto J_{\min},\quad \gamma_{22} \propto J_{\min},\quad \gamma_{33} \propto J_{\min}\sigma_{3}^{2}.
\]
According to this estimate, the first two diagonal elements must decrease with time as $J_{\min}$. The numerical experiment indeed demonstrates a decrease in $\gamma_{11}$ and $\gamma_{22}$, although they do not follow the exponential dependence exactly, which may be due to the small difference between the $\mathbf{U}$ matrix and the diagonal one, which we observed in experiments. The off-diagonal elements of $\gamma$ also turn out to be small, and only the $\gamma_{33}$ component remains on the order of unity and varies little with time. Recalling that the third singular value is expressed in terms of the vorticity maximum and the vorticity pancake thickness as $\sigma_{3}\propto \omega_{\max}^{-1}\ell_{1}^{-1}$, and $J_{ \min}\propto\omega_{\max}^{-1}$, we get
\[
\gamma_{33}\propto \omega_{\max}^{-3} \ell_{1}^{-2},
\]
which leads to the relation~(\ref{scaling}) observed in the numerical simulations, see Fig.~\ref{fig:fig3-w-ell}:
\begin{equation}  \label{23extended}
\omega_{\max}\propto \gamma_{33}^{-1/3}\ell_{1}^{-2/3}.
\end{equation}

In conclusion, it should be said that if the rotation $\mathbf{V}$ is applied to the matrix $\gamma$, then as a result the largest element is the component $(1,1)$, which practically coincides with $\gamma_{33}$ . Thus, the Jacobian and, accordingly, the vorticity mainly depend on the $x_1$ coordinate, while the influence of other coordinates turns out to be exponentially weak. This once again emphasizes that this structure is quasi two-dimensional, but the occurrence of scaling~(\ref{scaling}) is a purely three-dimensional phenomenon, which owes its existence to the compressibility of continuously distributed vortex lines.

\section{Statistics of the 3D turbulence onset}
In this Section, we discuss the statistical properties of the turbulence onset, confining ourselves to considering the regime when the number of emerging pancake-like structures is sufficiently large and statistical analysis can be applied to them. It is clear from the outset that in this case the turbulence is highly anisotropic. Each pancake-like structure generates strongly anisotropic jet-type distribution in the $k$-space, elongated in the direction perpendicular to the pancake plane with characteristic thickness $\sim \ell_{\perp}^{-1}\ll \ell_{1}^{-1}$. It is the interaction between jets (in fact, between pancake-like structures), as will be shown below, that determines the behavior of the structure functions of the velocity field.

In our numerical experiments, we observe formation of the pancake-like structures for all considered initial conditions (more than 30)~\cite{agafontsev2015, agafontsev2016development, agafontsev2016asymptotic}. During the evolution of all such structures of high vorticity, the Kolmogorov-type relation $\omega_{max}(t)\sim \ell (t)^{-2/3}$ is satisfied to some extent. In the first paper on this topic~\cite{agafontsev2015} it was shown that pancakes in the $x$-space generate highly anisotropic distributions in the $k$-space in the form of jets elongated in the directions perpendicular to pancakes (see Fig.4).

\begin{figure}[t]
\centering
\includegraphics[width=0.6\linewidth]{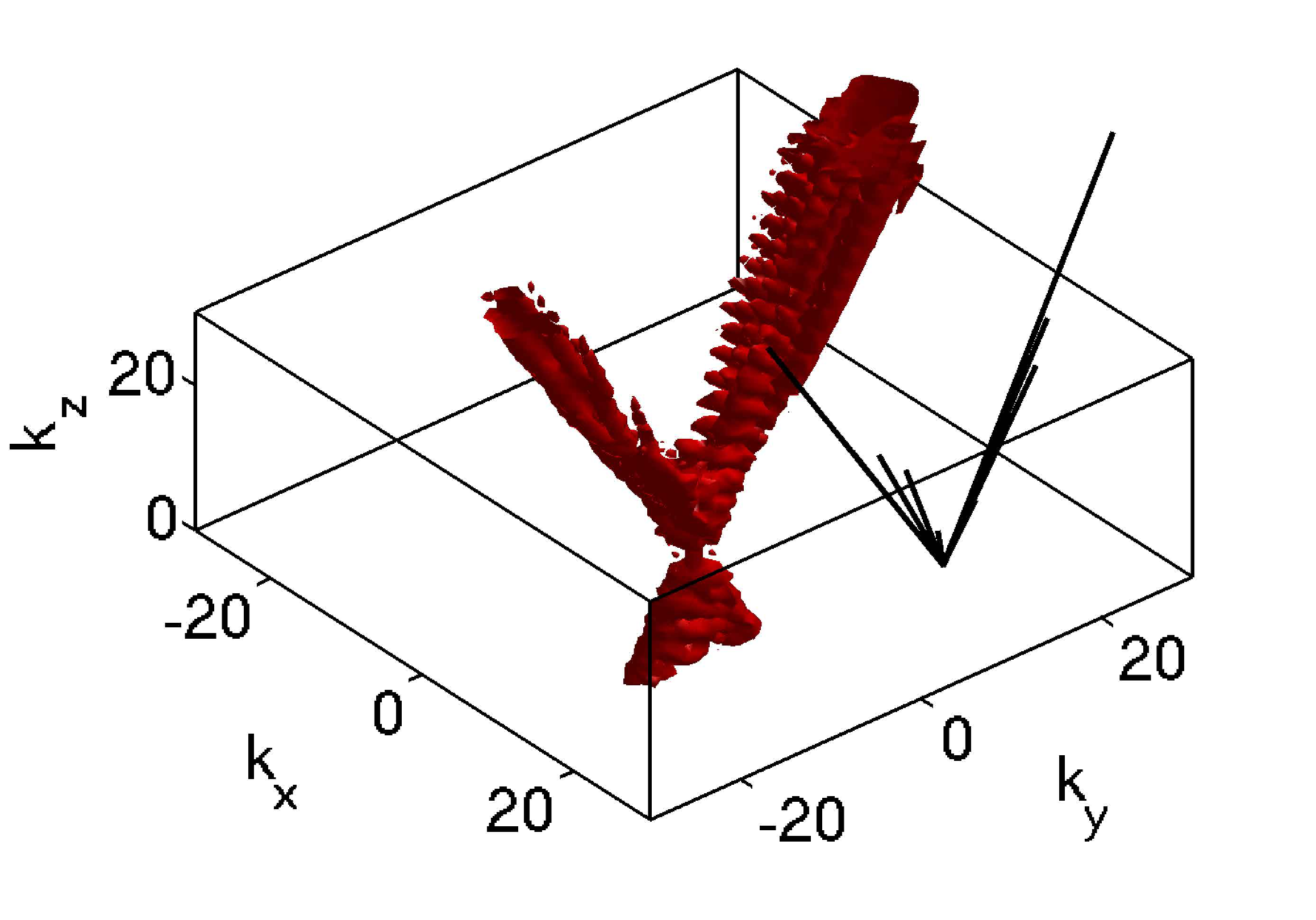}
\caption{Isosurface of the absolute value of the Fourier-transformed vorticity (normalized to the maximum value within the shell) $|\tilde{\bf\omega}(\mathbf{k})| = 0.2$ in the $k$-space for $t_{m1}$. The solid lines show the characteristic $k$-vectors of the pancakes normalized to $\ell_{1}^{-1}$ in length. \cite{agafontsev2015}}
\label{fig5}
\end{figure}

Since the ratio $\ell_1/\ell_{\perp}$ decreases with the pancake evolution, where $\ell_{\perp}$ is the characteristic longitudinal size of the pancake, the jet anisotropy increases with time, respectively. Due to this, the instantaneous turbulence spectrum turns out to be strongly jagged and anisotropic. Meanwhile the number of pancake-like structures increases and, accordingly, the number of jets increases too. Each jet has its own angular distribution of order $(k\ell_{\perp})^{-1}$. As the number of jets increases, jets begin to overlap in the $k$-space. When there are enough such overlaps, the Kolmogorov spectrum $E(k) \sim k^{-5/3}$ is formed in these regions.

\begin{figure}[t]
\centering
\includegraphics[width=0.3\linewidth]{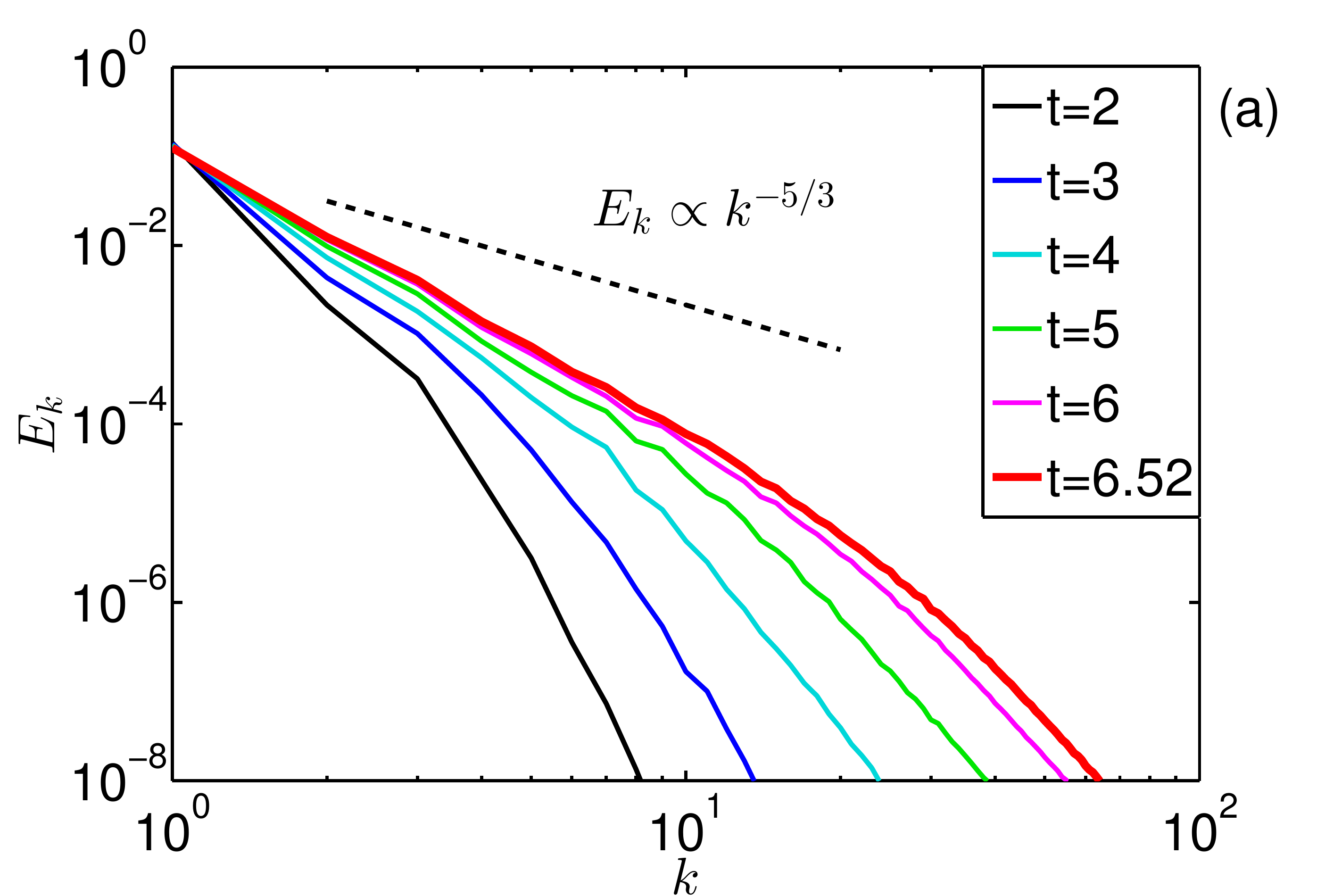}
\includegraphics[width=0.3\linewidth]{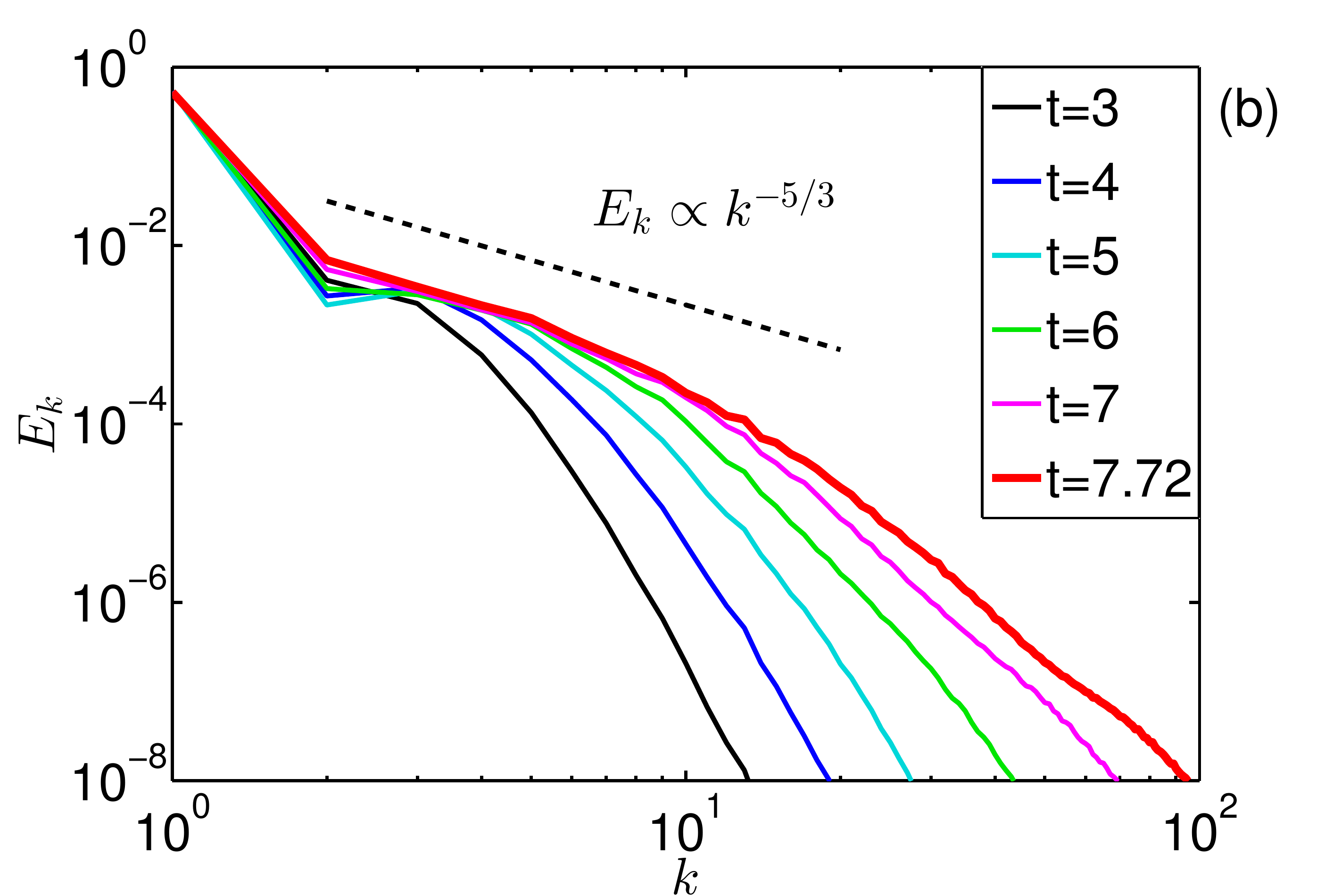}
\includegraphics[width=0.3\linewidth]{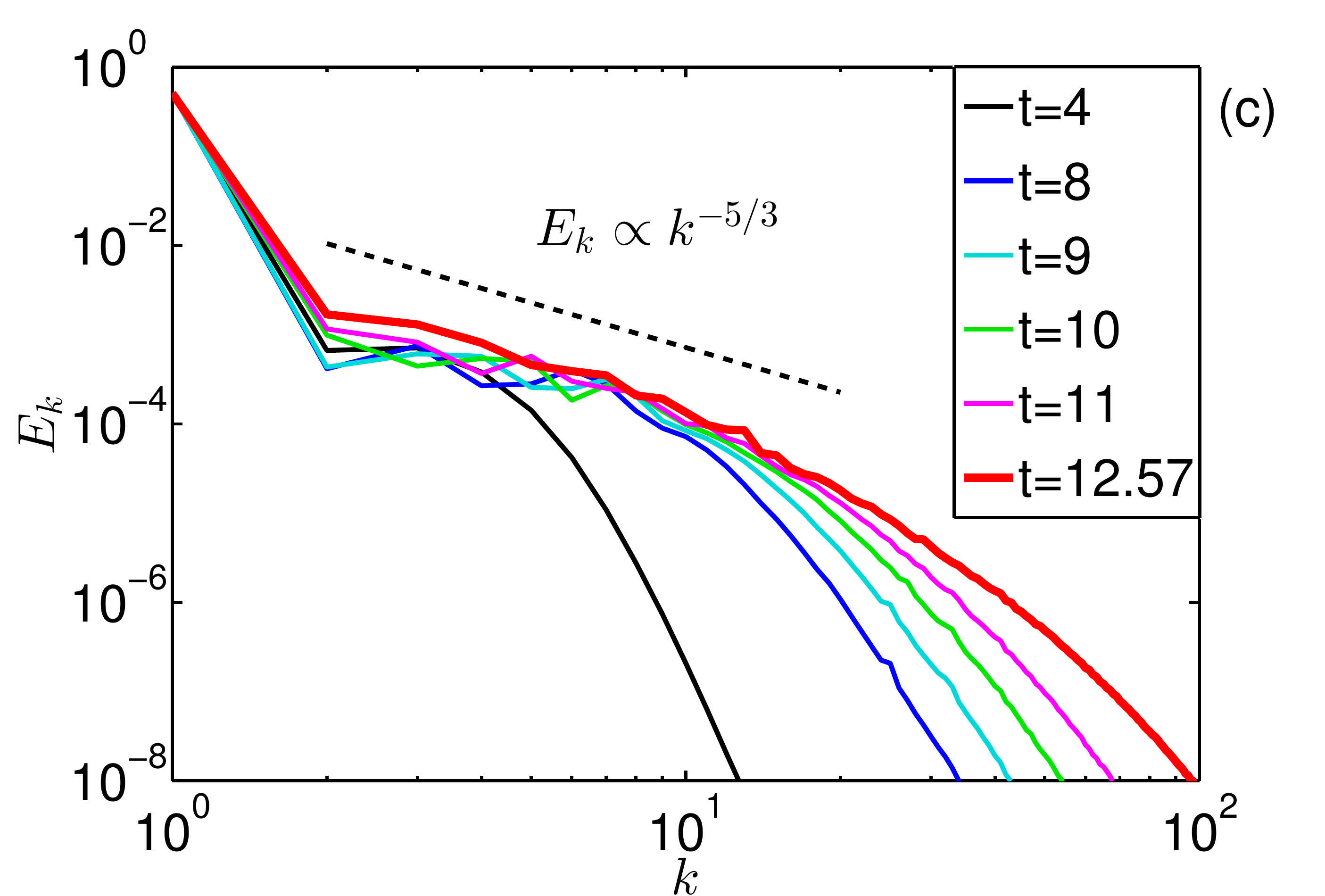}
\caption{Energy spectrum $E_{k}(t)$ at different times for initial conditions from~\cite{agafontsev2016development}: (a) $IC_{1}$ (type I), (b) $IC_{2}$ (type II), and (c) $IC_{3}$ (type III). \cite{agafontsev2016development}}
\label{fig6}
\end{figure}

Note that for the initial conditions $IC_{1}$ from~\cite{agafontsev2016development}, which do not contain anisotropy, the Kolmogorov spectrum is not observed after averaging over the angle, see Fig~\ref{fig6}(a). The initial conditions $IC_{2}$ and $IC_{3}$ are a superposition of the shear flow (the degenerate ABC flow) and isotropic Gaussian noise; here the ABC flow is the Arnold - Beltrami - Childress flow. The anisotropy generated by the shear flow causes the appearance of jets with a strong angular overlap, due to which the Kolmogorov spectrum is formed, see Fig~\ref{fig6}(b,c). It should be emphasized that the spectrum not averaged over the angle turns out to be strongly anisotropic. Thus, despite the presence of the Kolmogorov behavior of the spectrum, turbulence in the inertial interval is far from isotropic and homogeneous, at least for the times calculated in the numerical experiment. Recall that in the case of isotropic turbulence, one of the exact results of Kolmogorov's theory is the so-called $4/5$ law ~\cite{kolmogorov1941dissipation,landau2013fluid,frisch1999turbulence}. In the inertial interval of scales $r$ this law is written as
\begin{equation} \label{K45}
\langle\delta v_{\parallel}^{3}\rangle = -(4/5)\,\varepsilon\,r,
\end{equation}
where $\delta v_{\parallel}$ is the longitudinal velocity increment (projected onto the vector direction ${\bf r}={\bf r}_1-{\bf r}_2$) and $\langle...\rangle $ means averaging over the statistical ensemble. From here, based on dimensional considerations, we obtain relations for the second-order structure functions, $\langle\delta v^{2}\rangle\propto \varepsilon^{2/3}r^{2/3}$, and the Kolmogorov spectrum itself: $E_{k}\propto\varepsilon^{2/3} k^{-5/3}$. We emphasize that the key assumption in this case is the one that the nonlinear interaction is local on the scales of the inertial interval.  In this case, at high Reynolds numbers $Re\gg 1$, the dynamics on these scales can be described by the Euler equations, and the appearance of Kolmogorov relations can be expected before the excitation of viscous scales, as evidenced by numerous numerical experiments~\cite{orlandipirozzoli2010,holm2002transient,cichowlas2005effective,holm2007}. We emphasize that in our numerical experiments~\cite{agafontsev2015,agafontsev2016development} a power-law energy spectrum with scaling, close to Kolmogorov's, was observed in a completely inviscid flow, the dynamics of which is determined mainly by pancake-like structures of high vorticity~\cite{brachet1992numerical,agafontsev2016asymptotic,agafontsev2017universal}. The main contribution to the spectrum of emerging turbulence is made by jets, despite the fact that they occupy a small fraction of the entire spectral space, which leads to the formation of a power-law interval $E_{k}\propto k^{-\alpha}$ with the exponent $\alpha$ close to $5/3$, and the expansion of the Kolmogorov region over time to smaller scales. Moreover, power-law scaling occupies a much larger interval of scales if the jets have close orientations, thereby increasing the anisotropy of the flow.

In numerical experiments~\cite{AgafontsevKuznetsovMailybaev2019}, the behavior of two-point structure functions (moments) of velocity in the turbulence onset regime has been studied. In particular, how the power-law scaling forms for the longitudinal and transverse moments $[M_{\parallel}^{(n)}(r)]^{1/n}\propto r^{\xi_{n}}$ and $[M_{\perp}^{(n)}(r)]^{1/n}\propto r^{\zeta_{n}}$ in the same interval of scales, as for the energy spectrum $E_{k}$. As shown by the experiments, the exponents $\xi_{n}$ and $\zeta_{n}$ exhibit the same key properties as in the case of developed (stationary) turbulence: they decrease with decreasing order $n$ of the moment, indicating intermittency and anomalous scaling. The longitudinal exponents are somewhat larger than the transverse ones, and the approximate relation $\xi_{3}\simeq\alpha/5$ holds for the third-order structure functions of velocity.

Numerical results have shown that, despite the strong anisotropy of the longitudinal and transverse third-order moments, a power-law dependence on $r$ is observed for almost all directions ${\bf r}/r$ with an exponent close to $1$, as in the Kolmogorov law $4/5$~(\ref{K45}). Thus, when the power-law of the energy spectrum is close to Kolmogorov's, the longitudinal moment of the third order as a function of distance shows a scaling close to linear, compatible with the law~(\ref{K45}). In this case, the vorticity distribution is characterized by a strongly non-Rayleigh shape due to intermittency, and the power-law tail of this distribution indicates the nontrivial geometry of the pancake-like structures of high vorticity.

\subsection{Numerical scheme for studying the statistical characteristics of the 3D turbulence onset} 
Before presenting the results of numerical experiments for incompressible three-dimensional Euler equations~(\ref{omega}) (in the vorticity terms), a few words should be said about the numerical scheme used in the studies~\cite{agafontsev2015, agafontsev2016development, agafontsev2016asymptotic}.

Numerical simulation of the Euler equations~(\ref{omega}) was carried out in the periodic box $\mathbf{r} = (x,y,z)\in [-\pi ,\pi ]^{3}$ using the pseudo-spectral Runge-Kutta method of the fourth order. The initial conditions were chosen as a superposition of the shear flow
\begin{equation}\label{IC}
\bf\omega_{sh}(\mathbf{r}) = (\sin z, \cos z, 0),\quad |\bf\omega_{sh}(\mathbf{r})|=1,
\end{equation}
representing a stationary solution of the Euler equations, and a random periodic perturbation. The inverse rotor operator and all spatial derivatives were calculated in the Fourier space. An adaptive anisotropic rectangular grid was used, uniform for each direction. The grid adaptation was carried out independently for each of the three coordinates based on the analysis of the Fourier spectrum of the vorticity. The time step was chosen using the Courant - Friedrichs - Lewy (CFL) stability criterion with the Courant number $0.5$. The initial size of the cubic grid was $128^3$, and then it increased due to adaptation until the total number of nodes reached $2048^{3}$ ($1024^3$ for some numerical experiments). The grid adaptation was carried out as follows. Using the three functions
\[
S_{j}(k)=\int|\boldsymbol\omega(\mathbf{p})|^{2} \delta(|p_{j}|-k)d^{3}\mathbf{p},\quad j=x,y,z,
\]
representing the vorticity spectrum integrated in a plane perpendicular to each of the three axes, the breaking point between the energy-containing region and the numerical short-wavelength noise in each direction was tracked. As soon as the breaking point approached $2K_{\max }^{(j)}/3$ along any of the three directions, the grid was made denser in that direction. Here $K_{\max}^{(j)}=N_{j}/2$ are the maximum wavenumbers and $N_{j}$ are the dimensions of the grid in the directions $j = x,y,z$, and the factor $2/3$ takes into account the aliasing effect. The transition from one grid to another was performed using the Fourier interpolation. After reaching the maximum allowed number of nodes, numerical simulation continued on a fixed grid, i.e. without further modification. After that, the calculation was completely stopped if the Fourier spectrum of vorticity at $2K_{\max }^{(j)}/3$ in any direction exceeded $10^{-13}$ of its maximum value, $S_{j }(2K_{\max}^{(j)}/3) \geq 10^{-13} \max_{k}(S_j(k))$ (for more details see ~\cite{agafontsev2015,agafontsev2016development}).

More detailed information about the simulation of the Euler equations in the vortex line representation is given in the papers~\cite{agafontsev2015,agafontsev2016development,agafontsev2017universal}, where it has been shown that the accuracy on the simulation time interval is very high, allowing to get exactly the same vorticity field as in direct simulation. 

For some numerical experiments, a gradual formation of power-law scaling was observed in the energy spectrum $E_{k}\propto k^{-\alpha}$ at small and medium wavenumbers starting from $k\ge 2$. The first harmonic $k=1$, in which the initial energy is concentrated, contains most of the total energy (up to 97\% at the final time) and does not belong to this interval. To exclude its influence on the structure functions of the velocity, the moments were calculated for the modified velocity $\tilde{\mathbf{v}}$ obtained from the original one by removing nine harmonics $\mathbf{k}=(k_{x},k_{y} ,k_{z})$ with $k_{x,y,z}= -1,0,1$.

Calculation of the moments for (non-stationary) turbulence onset requires much more computational resources than a similar problem for developed (stationary) turbulence using time averaging (see~\cite{ishihara2009study} for instance). Such a simulation was carried out as follows. First, for a given radius $r$, a sufficient number of points $\mathbf{r}$ are set uniformly distributed on the sphere $|\mathbf{r}|=r$. Then for each $\mathbf{r}$ the velocity increment $\mathbf{\delta \tilde{v}} = \tilde{\mathbf{v}}(\mathbf{x}+\mathbf{r},t)-\tilde{\mathbf{v}}(\mathbf{x},t)$ is calculated at each node of the grid $\mathbf{x}$ using nearest neighbor interpolation for the ``shifted'' velocity $\tilde{\mathbf{v}}(\mathbf{x}+\mathbf{r},t)$. Finally, the longitudinal and transverse moments of order $n$ are calculated as the corresponding integral sums over all points on the sphere $\mathbf{r}$ and all nodes $\mathbf{x}$, 
\begin{eqnarray}
M_{\parallel}^{(n)}(r) &=& \frac{1}{4\pi r^{2}} \int_{|\mathbf{r}|=r}d^{3}\mathbf{r}\int\frac{d^{3}\mathbf{x}}{(2\pi)^{3}} \,(\mathbf{\delta \tilde{v}}\cdot \mathbf{m}_{r})^{n},  \label{moments-parallel}\\
M_{\perp}^{(n)}(r) &=& \frac{1}{4\pi r^{2}} \int_{|\mathbf{r}|=r}d^{3}\mathbf{r}\int\frac{d^{3}\mathbf{x}}{(2\pi)^{3}} \,\bigg|\mathbf{\delta \tilde{v}}\times \mathbf{m}_{r}\bigg|^{n},  \label{moments-perpendicular}
\end{eqnarray}
where $\mathbf{m}_{r}=\mathbf{r}/r$ is the unit vector.\\

\subsection{Numerical results} 
Let us present the results of numerical simulations for the initial conditions $I_{1}$ from~\cite{agafontsev2015} on a grid with total number of nodes $2048^{3}$. We emphasize that for these initial conditions the grid at the final time $t_f=7.75$ was anisotropic rectangular with dimensions $972\times 2048\times 4096$. The vorticity maximum $\omega_{\max}$, equal to $1.5$ at the initial moment of time, reached $18.4$ at $t=t_f$. At the final time, the thinnest region of high vorticity was resolved with $10$ grid points at the level of vorticity half maximum.

\begin{figure}[t]
\centering
\includegraphics[width=0.6\linewidth]{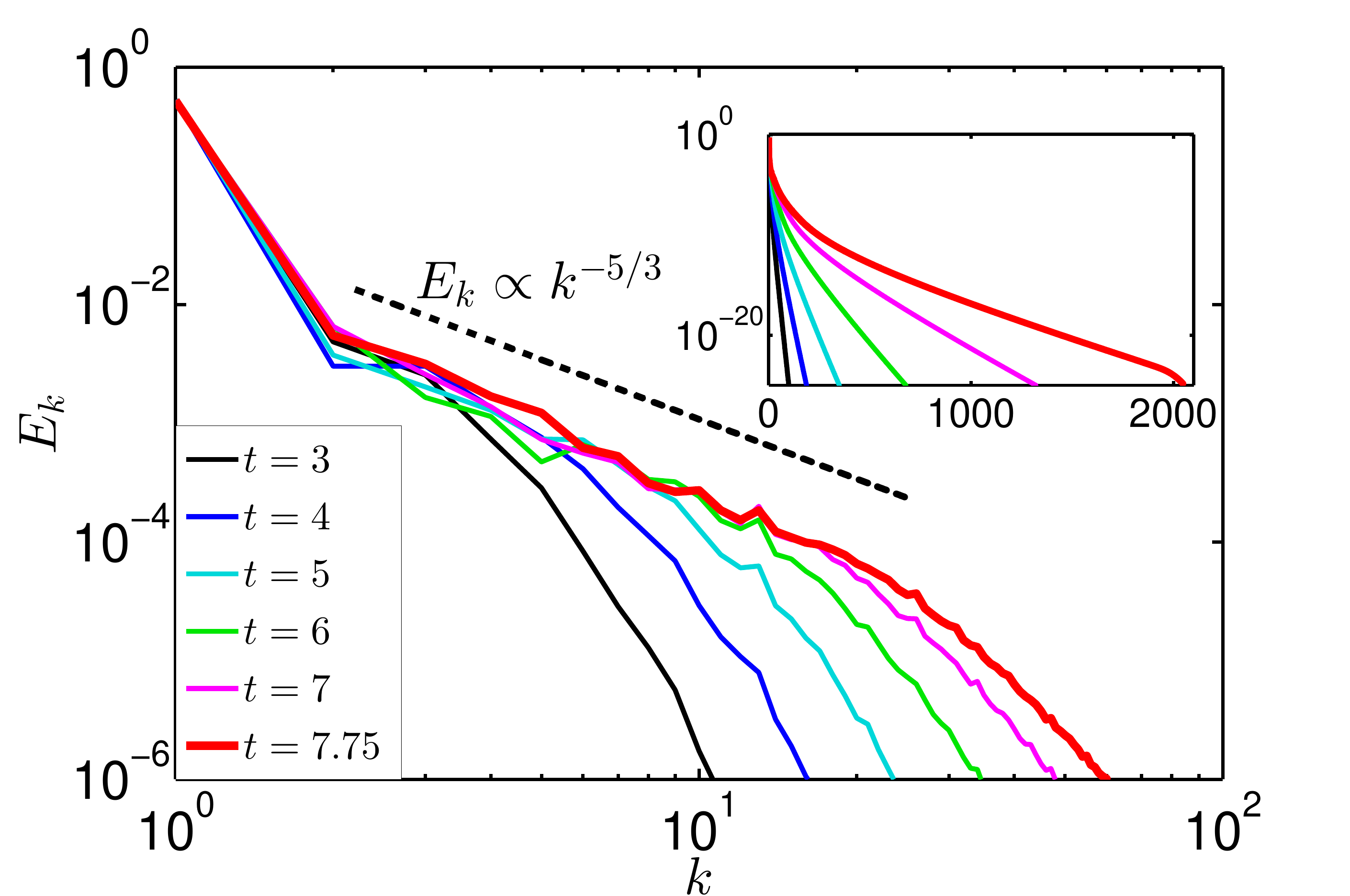}
\caption{Energy spectrum $E_{k}$ in double logarithmic scales. The inset shows the spectrum in semi-logarithmic scales. \cite{AgafontsevKuznetsovMailybaev2019}}
\label{fig:fig1}
\end{figure}

Evolution of the energy spectrum for this simulation is shown in Fig.~\ref{fig:fig1}. At large wavenumbers $k$, the spectrum decays exponentially, as shown in the inset to the figure. At small and medium $k$, a gradual formation of a power-law interval with a scaling close to the Kolmogorov's one $E_{k}\propto k^{-5/3}$ is clearly observed. The power-law interval is characterized by a frozen spectrum, in contrast to strong changes with time at large wavenumbers. By the end of the simulation, this interval expanded to slightly more than one decade, $2 \lesssim k \lesssim 30$. It should be noted that this interval contains only a small fraction of the energy: even at the final time, 97.2\% of energy is still contained in the first harmonic $k=1$, while the wave numbers $2\le k\le 30$ and $k>30$ receive only 2.8\% and less than 0.1\% of energy, respectively.

\begin{figure}[t]
\centering
\includegraphics[width=0.3\linewidth]{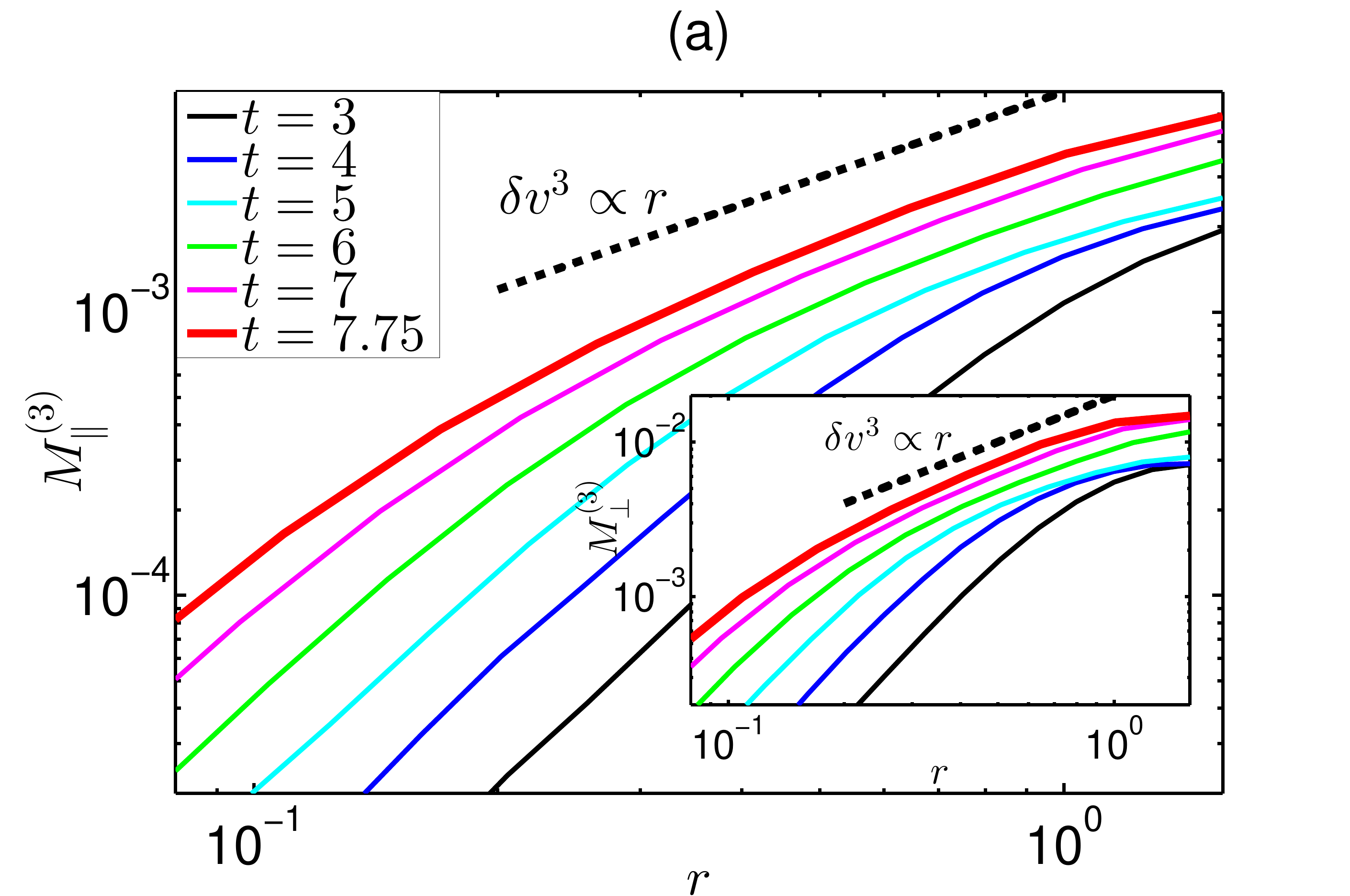}
\includegraphics[width=0.3\linewidth]{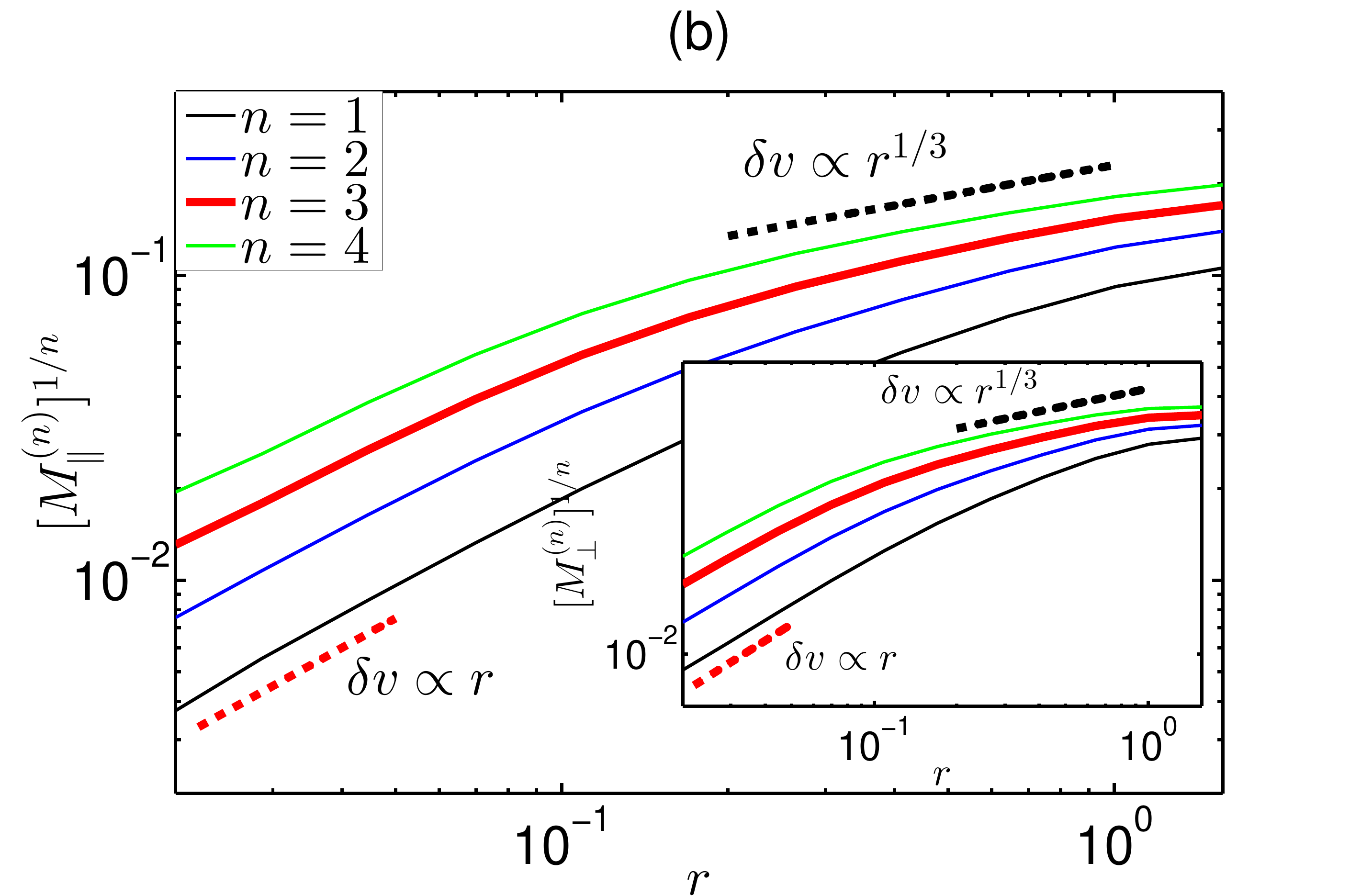}
\includegraphics[width=0.3\linewidth]{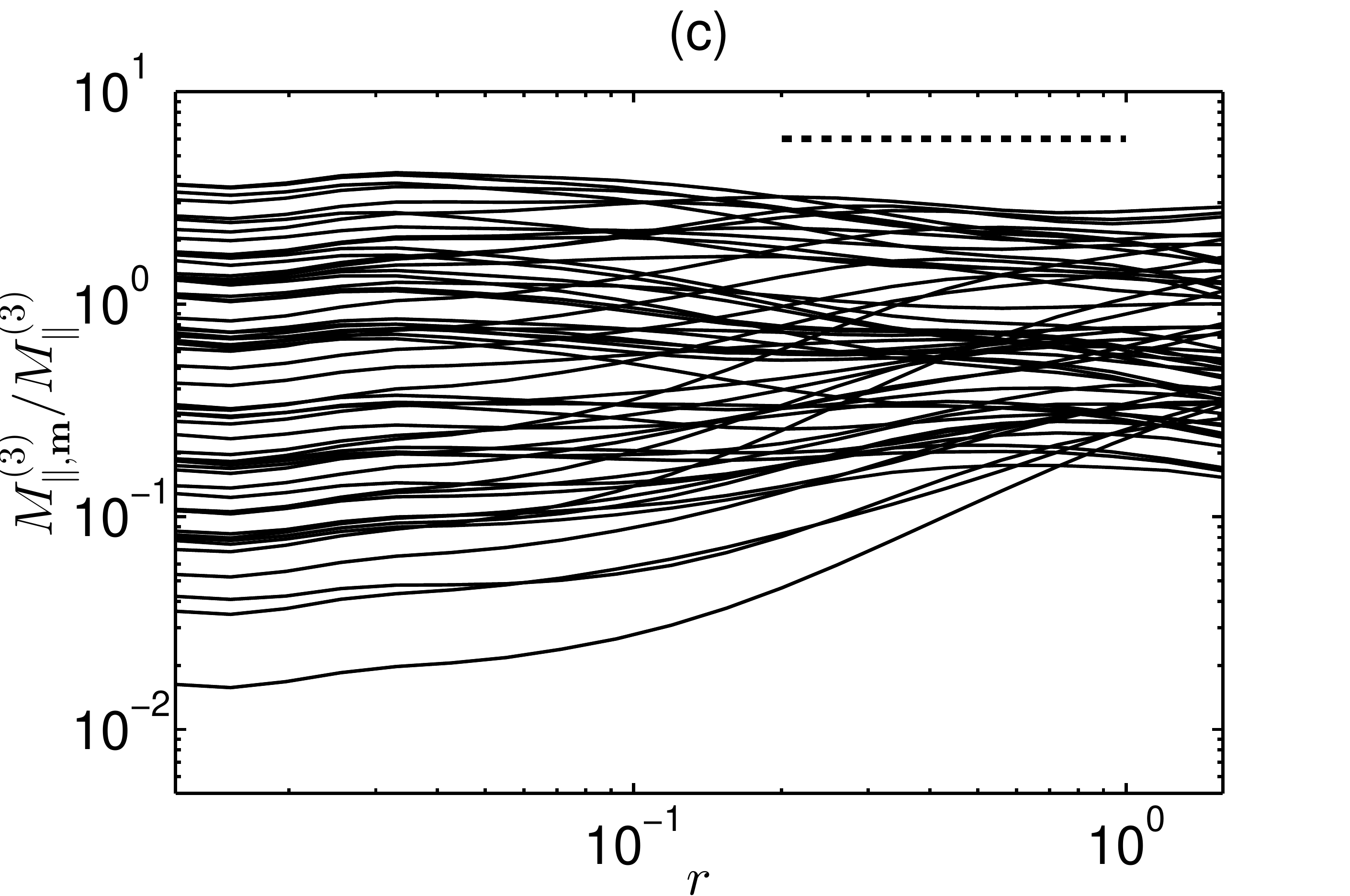}
\caption{(a) Longitudinal moments of the third order $M_{\parallel}^{(3)}$. The black dotted line shows the scaling of $M^{(3)}\propto r$.
(b) Longitudinal moments $[M_{\parallel}^{(n)}]^{1/n}$ of orders $n=1,2,3,4$ at the final time $t=7.75$. The black dashed line shows the scaling $[M^{(n)}]^{1/n}\propto r^{1/3}$ for the power-law interval, and the red dashed line shows the scaling $[M^{(n)}]^{1/n}\propto r$ at smaller scales. The insets in panels (a) and (b) show transverse moments. 
(c) Third-order compensated longitudinal moments $M_{\parallel,\mathbf{m}}^{(3)}/M_{\parallel}^{(3)}$ in the $114$ directions $\mathbf{m}$ uniformly distributed over spherical coordinates, at the final simulation time. The dotted horizontal line shows the power-law interval for $M_{\parallel}^{(3)}$ in the panel (b).
\cite{AgafontsevKuznetsovMailybaev2019}}
\label{fig:fig2}
\end{figure}

Evolution of the third-order moments is shown in Fig.~\ref{fig:fig2}(a); see also Fig.~\ref{fig:fig2}(b) where the final moments of time are shown on a larger scale. The power-law interval with a scaling close to the linear $M^{(3)}\propto r$ is gradually formed over time for both longitudinal and transverse moments at sufficiently large scales, expanding up to $0.2\lesssim r\lesssim 1$ at the final time. These scales correspond to the wavenumbers $6\lesssim k\lesssim 30$, which belong to the power-law interval in the energy spectrum in Fig.~\ref{fig:fig1}.

The exponents $\xi_{n}$ and $\zeta_{n}$ for longitudinal and transverse moments $[M_{\parallel}^{(n)}]^{1/n}\propto r^{\xi_{n}}$ and $[M_{\perp}^{(n)}]^{1/n}\propto r^{\zeta_{n}}$ decrease with the order of moment $n$, indicating intermittency and anomalous scaling; see Fig.~\ref{fig:fig2}(b). The first four longitudinal exponents have the following values: $\xi_{1}=0.60 \pm 0.06$, $\xi_{2} = 0.48 \pm 0.04$, $\xi_{3} = 0.39 \pm 0.03$ and $\xi_{4} = 0.32 \pm 0.03$. The corresponding transverse exponents $\zeta_{1} = 0.55 \pm 0.07$, $\zeta_{2} = 0.42 \pm 0.06$, $\zeta_{3} = 0.33 \pm 0.05$ and $\zeta_{4} = 0.26 \pm 0.04$ are slightly less than the longitudinal ones, $\xi_{n}\gtrsim\zeta_{n}$, but remain within the standard deviations from each other. It should be noted that, for the developed turbulence, the transverse exponents also turn out to be somewhat smaller than the longitudinal ones, see e.g.~\cite{gotoh2002velocity,zybin2015stretching}.

Using the moments of velocity along different directions (directional moments), one can examine the anisotropy of the velocity distribution, for example, for the longitudinal moment of the third order,
\begin{eqnarray} \label{moments-parallel-d}
M_{\parallel,\mathbf{m}}^{(n)}(r) &=& \int\frac{d^{3}\mathbf{x}}{(2\pi)^{3}} \,(\mathbf{\delta \tilde{v}}\cdot \mathbf{m})^{n},  
\end{eqnarray}
where $\mathbf{r} = \mathbf{m}\,r$ and $\mathbf{m}$ is the unit vector of the direction.  Fig.~\ref{fig:fig2}(c) shows behavior of the directional moment $M_{\parallel,\mathbf{m}}^{(3)}$ relative to the angular-averaged moment $M_{\parallel}^{(3)}$ for $114$ directions uniformly distributed over spherical coordinates. At the scales of the power-law interval, the change in the moments $M_{\parallel,\mathbf{m}}^{(3)}$ with the change in direction reaches one order of magnitude. Moreover, for some directions, the moments $M_{\parallel,\mathbf{m}}^{(3)}$ increase much faster (slower) with increasing distance $r$ compared to the angular-averaged moment $M_{\parallel}^{(3)}$. Note, however, that for most directions the moments of $M_{\parallel,\mathbf{m}}^{(3)}$ change with distance almost in the same way as $M_{\parallel}^{(3)}$. Such behavior, as will be shown in the next section, was first discovered for two-dimensional hydrodynamic turbulence in the direct cascade mode~\cite{kuznetsov2015anisotropic}, when the Kraichnan spectrum appears due to the formation of vorticity quasi-shocks~\cite{kuznetsov2007effects,kudryavtsev2013statistical}, similar to pancake-like structures of high vorticity in the three-dimensional case.

In order to study the relationship between the energy spectrum and the moments of the velocity field in more detail, an additional $30$ numerical experiments were performed on grids with a total number of $1024^{3}$ nodes for $30$ different initial flows taken as a superposition of the shear flow~(\ref{IC}) and a random periodic perturbation
\begin{equation} \label{IC1}
\bf\omega_{p}(\mathbf{r}) = \sum_{\mathbf{h}} \left[\mathbf{A}_\mathbf{h}\cos(\mathbf{h}\cdot\mathbf{r}) +\mathbf{B}_\mathbf{h}\sin(\mathbf{h}\cdot\mathbf{r})\right]. 
\end{equation}
Here $\mathbf{h} = (h_x,h_y,h_z)$ is a vector with integer components $|h_{j}|\le 2$, $j=x,y,z$, while $\mathbf{A}_\mathbf{h}$ and $\mathbf{B}_\mathbf{h}$ are real random coefficients with zero mean and standard deviation $\sigma_{\mathbf{h}}^2 \sim \exp(-|\mathbf{h}|^2)$ which satisfy the orthogonality conditions, $\mathbf{h}\cdot\mathbf{A}_\mathbf{h} = \mathbf{h}\cdot\mathbf{B}_\mathbf{h} = 0$ necessary for self-consistency. The initial conditions are chosen as a mix of flows~(\ref{IC}) and~(\ref{IC1}),
\begin{equation} \label{IC2}
{\bf\omega_{0}}(\mathbf{r}) = (1-p)\,{\bf\omega_{sh}}(\mathbf{r}) + p\, R\, {\bf\omega_{p}}(\mathbf{r}),
\end{equation}
where $p$ is the mixing coefficient and $R=\sqrt{4\pi^{3}/E_{p}}$ is the renormalization coefficient. Here $4\pi^{3}$ and $E_{p}$ are the energies of the shear flow~(\ref{IC}) and perturbation~(\ref{IC1}) in the simulation box $[-\pi ,\pi]^{3}$, so that the coefficient $R$ renormalizes the perturbation to the same energy as the shear flow. Three groups of experiments were performed with $p=1$ (random periodic flows), $p=0.1$ and $p=0.02$, for $10$ random realizations of initial flows for each group.

\begin{figure}[t]
\centering
\includegraphics[width=0.6\linewidth]{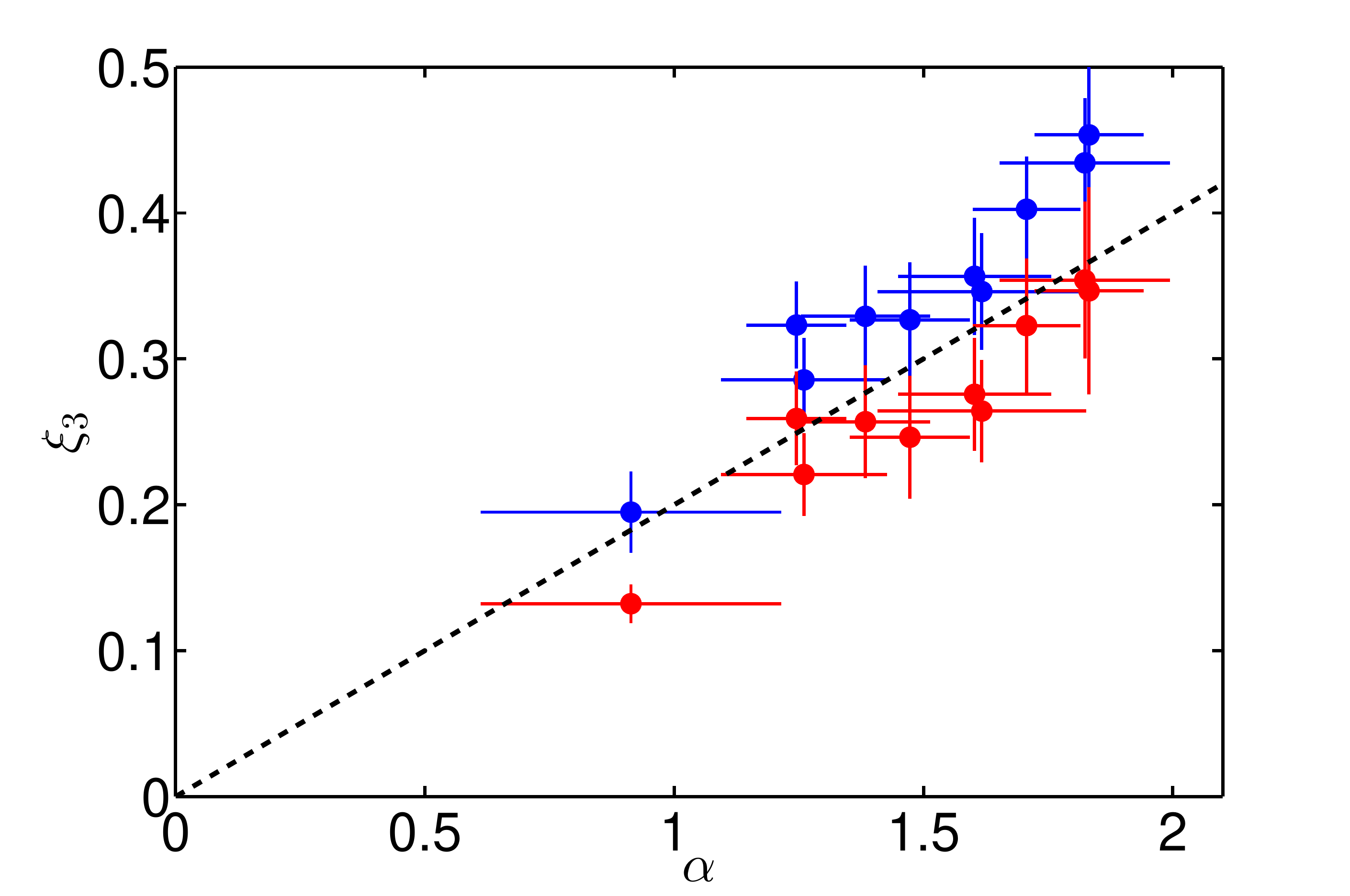}
\caption{Exponents $\xi_{3}$ (blue) and $\zeta_{3}$ (red) for power-law scaling of the longitudinal and transverse moments of the third order depending on the exponent $\alpha$ of the energy spectrum scaling; $10$ simulations from the third group of experiments with $p=0.02$. The horizontal and vertical lines show the standard deviations, the dotted black line shows the relation $\xi_{3}=\zeta_{3}=\alpha/5$. \cite{AgafontsevKuznetsovMailybaev2019}}
\label{fig:fig3}
\end{figure}

For the first group of experiments with random periodic flows, none of the ten simulations showed a power-law interval for the energy spectrum or for the velocity moments. For the second group, $p=0.1$, all ten simulations show a power-law interval for the energy spectrum, and six out of ten simulations demonstrate a power-law interval for the velocity moments; the intervals reach $2\lesssim k\lesssim 20$ for the spectrum and $0.3\lesssim r\lesssim 0.8$ for the moments. The third group with $p=0.02$ shows the power-law intervals for both the spectrum and the moments for all ten simulations; the intervals reach $2\lesssim k\lesssim 40$ and $0.15\lesssim r\lesssim 0.8$ respectively. For all simulations, the lower bound $r_{l}$ of the power-law interval $r_{l}\lesssim r\lesssim r_{h}$ for the moments (if this interval exists) is related to the upper bound $k_{h}$ of the power-law interval $k_{l}\lesssim k\lesssim k_{k}$ for the spectrum as $r_{l}\approx 2\pi/k_{h}$. The upper bound $r_{h}$ roughly corresponds to the wave number $2\pi/r_{h} \simeq 6$.

For the third group of experiments, a power-law scaling $E_{k}\propto k^{-\alpha}$ is observed for the energy spectrum with the exponent $\alpha$ between $0.9$ and $1.8$; for most simulations $\alpha$ is close to $1.6$. The exponents $\xi_{3}$ and $\zeta_{3}$ describing the power-law scaling of the velocity moments $[M_{\parallel}^{(3)}(r)]^{1/3}\propto r^{\xi_{3}}$ and $[M_{\perp}^{(3)}(r)]^{1/3}\propto r^{\zeta_{3}}$ take the values $0.2 \le\xi_{3}\le 0.45$ and $0.13 \le\zeta_{3}\le 0.35$. Longitudinal exponents turn out to be somewhat larger than the transverse ones, $\xi_{3}\gtrsim\zeta_{3}$, and most of the ten simulations show $\xi_{3}$ near $0.35$ and $\zeta_{3}$ near $0.25$. As shown in Fig.~\ref{fig:fig3}, simulations that have a larger exponent $\alpha$ also demonstrate larger exponents $\xi_{3}$ and $\zeta_{3}$, and vice versa, with an approximate relation for the longitudinal exponent
\begin{equation}
\xi_{3}\simeq \alpha/5. \label{zeta3-alpha}
\end{equation}
Note that such a relation cannot be obtained from Fourier analysis. Indeed, a velocity increment satisfying $\delta v\propto r^{\zeta}$ in the physical space has a scaling of $\delta v_{k}\propto k^{-\zeta-1}$ in the Fourier space, which leads to the energy spectrum $E_{k}\propto k^{-2\zeta-1}$. The relations $\zeta = \alpha/5$ and $\zeta = (\alpha-1)/2$ intersect only at one point: $\alpha = 5/3$, $\zeta = 1/3$.

\begin{figure}[t]
\centering
\includegraphics[width=0.6\linewidth]{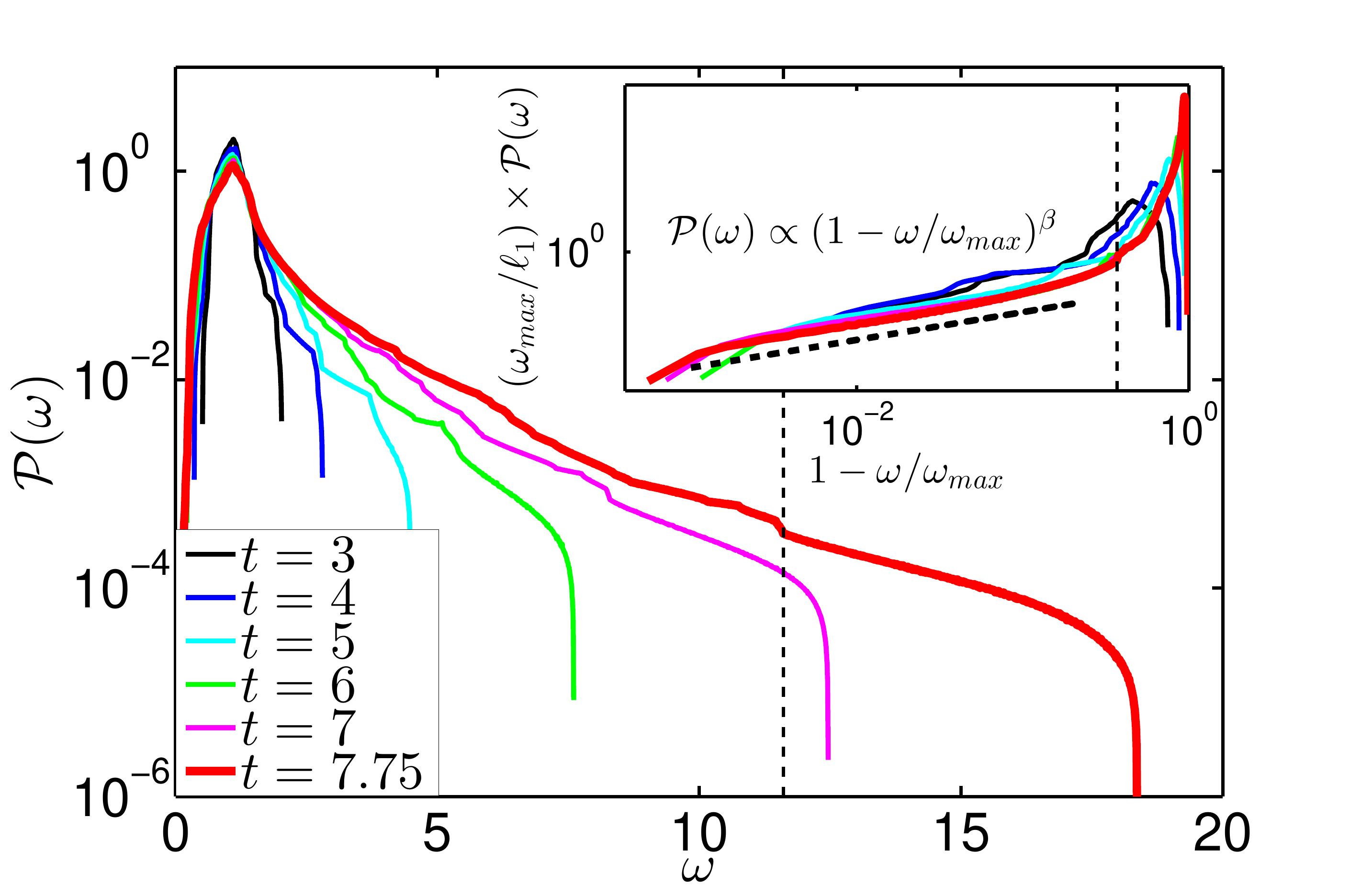}
\caption{Vorticity distribution for $I_{1}$ simulation. The inset shows the normalized distribution as a function of $(1-\omega/\omega_{\max})$. The vertical dashed line shows the second local vorticity maximum, and the thick dashed line in the inset shows the scaling $\mathcal{P}(\omega)\propto (1-\omega/\omega_{\max})^{\beta}$ with $\beta=0.6$. \cite{AgafontsevKuznetsovMailybaev2019}}
\label{fig:fig4-sf}
\end{figure}

One of the functions that may indicate intermittency is the distribution $\mathcal{P}(\omega)$ of the absolute value of the vorticity. The evolution of this function for the initial flow $I_{1}$ is shown in Fig.~\ref{fig:fig4-sf}. As the maximum vorticity increases with time, the distribution acquires a strongly non-Rayleigh shape with a so-called ``heavy tail'' expanding to larger vorticity values. The value of the second local vorticity maximum (shown in Fig.~\ref{fig:fig4-sf} with a dotted vertical line) turns out to be much smaller than that of the first one, which makes it possible to study the vorticity distribution inside an isolated pancake-like region corresponding to the global vorticity maximum. In a local orthonormal basis $\mathbf{x} = \mathbf{x}_{m} + a_{1}\mathbf{w}_{1} + a_{2}\mathbf{w}_{2} + a_{3}\mathbf{w}_{3}$ of the pancake, the vorticity modulus can be described using the quadratic approximation~\cite{agafontsev2015},
\begin{equation}\label{pancake-quadratic}
\frac{|\bf\omega(\mathbf{x})|}{\omega_{\max}} = 1 - \sum_{j=1}^{3}\bigg(\frac{a_{j}}{\ell_{j}}\bigg)^{2} + o(|\mathbf{x}-\mathbf{x}_{m}|^{2}),
\end{equation}
where $\mathbf{x}_{m}$ is the position of the local maximum, $\ell_{j}=\sqrt{2\omega_{\max}/|\lambda_{j}|}$ are the characteristic scales of the pancake, $\ell_{1}\ll\ell_{2}\lesssim\ell_{3}$, while $\lambda_{1}<\lambda_{2}<\lambda_{3}<0$ and $\mathbf{w}_{j}$ are the eigenvalues and eigenvectors for the (symmetric) matrix $\partial^{2}|\bf\omega|/\partial x_{i}\partial x_{j}$ computed at $\mathbf{x}_{m}$. Using this approximation, we get
$$
\mathcal{P}(f)\propto |dV/df| \propto (\ell_{1}\ell_{2}\ell_{3})(1-f)^{1/2},\quad f=\omega/\omega_{\max},
$$
where $V=(4\pi/3)\ell_{1}\ell_{2}\ell_{3}(1-f)^{3/2}$ is the volume of the ellipsoid~(\ref{pancake-quadratic}). It was shown in~\cite{agafontsev2015} that only the thickness of the pancake $\ell_{1}$ changes significantly with time, while the other two scales $\ell_{2,3}$ remain of unity order. This allows $\ell_{2,3}$ to be eliminated from the relation above, giving
\begin{equation}\label{PDF-scaling}
\mathcal{P}(\omega)\propto (\ell_{1}/\omega_{\max})(1-\omega/\omega_{\max})^{\beta},\quad \beta=1/2.
\end{equation}
Note that the results of the simulations show slightly larger exponent $\beta\gtrsim 1/2$, see Fig.~\ref{fig:fig4-sf}, that hints to a non-trivial geometry of the pancake structure.

\section{2D turbulence: from breaking to the Kraichnan spectrum}
In 1967 Kraichnan \cite{kraichnan} demonstrated that for the developed two-dimensional hydrodynamic turbulence there exist two Kolmogorov spectra, generated by two integrals of motion -- energy $E=1/2\int (\bf v)^2 d{\bf r}$ and enstrophy $1/2\int \omega^{2}d\mathbf{r}$. The first spectrum corresponds to a constant energy flux $\epsilon$ directed toward the region of small wave numbers (inverse cascade). This spectrum has the same dependence on $k$, as the famous Kolmogorov \cite{kolmogorov1941dissipation} spectrum for three-dimensional hydrodynamic turbulence. The second spectrum - the Kraichnan spectrum \cite{kraichnan}, 
\begin{equation} 
E(k)~\sim ~\eta^{2/3}k^{-3}, \label{Kraichnan}
\end{equation}
corresponds to a constant enstrophy flux $\eta$ towards the small-scale region (direct cascade). The existence of these two spectra has been confirmed in many numerical experiments simulating two-dimensional turbulence at high Reynolds numbers (see, e.g., \cite{boffetta} and references therein). However, just after the Kraichnan paper \cite{kraichnan}, in the first numerical experiments \cite{lilly} there was observed the emergence of sharp vorticity gradients corresponding to the formation of jumps (quasi-shocks) with thickness small compared to their length. Based on these numerical observations, Saffman \cite{saffman} proposed another spectrum $E(k)~\sim ~k^{-4}$, the main contribution in which comes  from isotropically distributed quasi-shocks (in this meaning, the Saffman spectrum is analogous to the Kadomtsev-Petviashvili spectrum \cite{KP} for acoustic turbulence). On the other hand, the Fourier amplitude from the vorticity jump $\omega _{k} \propto k^{-1}$, that immediately yields the Kraichnan type spectrum $E(k)~\sim ~k^{-3}$. However, the energy distribution from one such jump is anisotropic and has the form of a jet with an apex angle of the order of $\left( kL\right) ^{-1}$, where $L$ is characteristic length of the jump. It should be emphasized that for isotropically distributed vorticity shocks we should arrive at the Saffman spectrum. In this sense {\it the Kraichnan type spectrum generated by quasi-singularities must be anisotropic}. That was confirmed by both analytical arguments and numerical experiments in the case of a freely two-dimensional turbulence \cite{kuznetsov2007effects,kudryavtsev2013statistical,K-04} when anisotropy in turbulence spectra is due to the presence of jets. In these papers, there was revealed the physical mechanism of quasi-shocks formation due to a tendency to breaking (note that according to strong theorems \cite{wolibner} this process in a finite time is forbidden). This mechanism, as noted in the Introduction, is associated with the property of frozenness into fluid of the divorticity field ${\bf B}=\mbox{rot}\,\omega$, which allows us to express $\mathbf{B}$ in a representation similar to the VLR (\ref{VLR}):
\begin{figure}[t]
	\centerline{
		\includegraphics[width=0.6\linewidth]{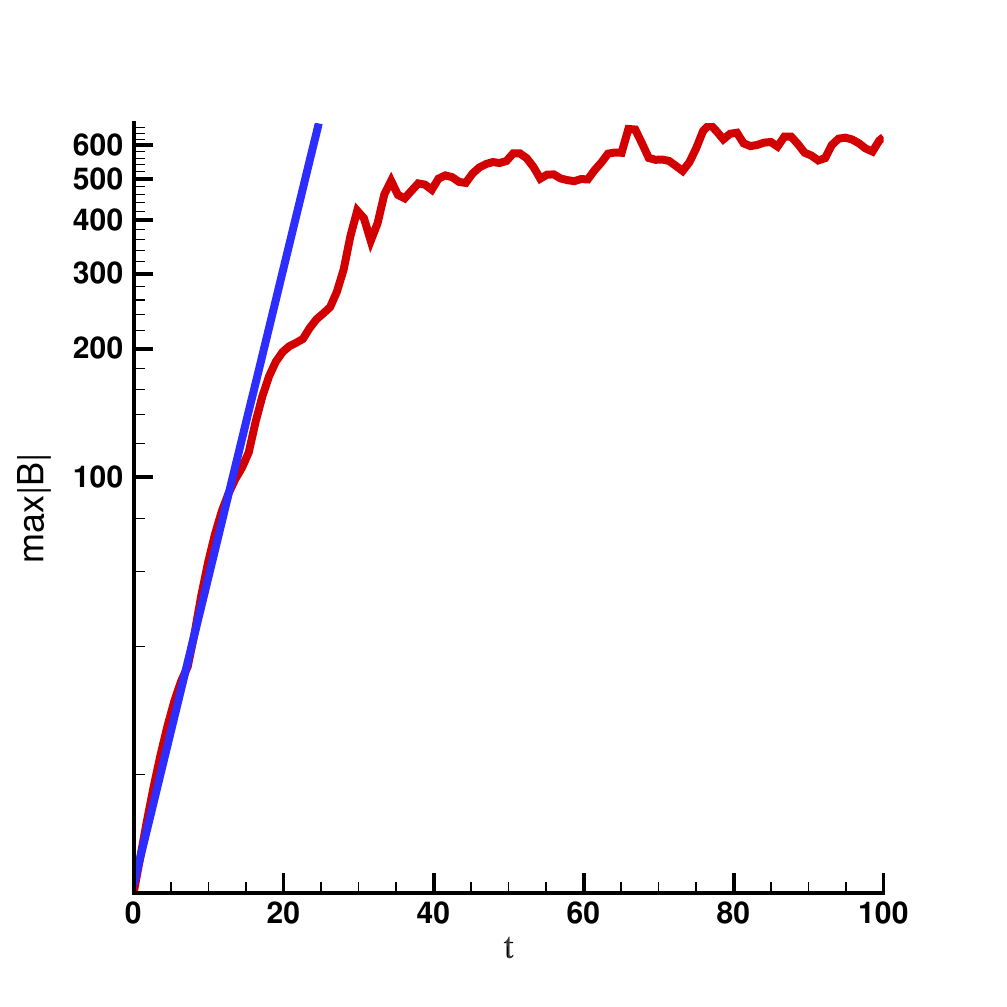}}
	\caption{Maximum value of $|\mathbf{B}|$ versus time (logarithmic scale, the straight line corresponds to exponential growth). \cite{kudryavtsev2013statistical}}
	\label{fig:fig-gr}
\end{figure}
\begin{equation} \nonumber
\mathbf{B}(\mathbf{x},t) = \frac{(\mathbf{B}_0(\mathbf{a}) \cdot\nabla_a)\mathbf{r}(\mathbf{a},t)}{J} ,
\end{equation}
where $\mathbf{B_0}(\mathbf{a})$  is the initial  field $\mathbf{B}$, and $J$ is the Jacobian of the mapping ${\bf x}={\bf x} ({\bf a}, t)$, equation for which has the same form as (\ref{VLR-trajectory}), and ${\bf v_n}$ in this case is the normal component of velocity with respect to the field $\mathbf{B}$. As in the previous case of the 3D Euler equations in the VLR (\ref{VLR}), the Jacobian $J$ can take arbitrary values, including zero. However, in the two-dimensional Euler hydrodynamics there is only a tendency to form sharp vorticity gradients in the form of quasi-shocks, that was confirmed in numerical experiments on freely decaying turbulence  \cite{kuznetsov2007effects, KNNR-10, kudryavtsev2013statistical}. In particular, the increase of maximum value of $B$ in these experiments was 2 - 2.5 orders of magnitude (see for example Fig. \ref{fig:fig-gr}), and the spatial distribution of $|B|$ was concentrated around the lines (positions of quasi-shocks) with significantly less values of $|B|$ between these lines (see for example Fig. \ref{fig:B-distr}). 

As was found in the numerical experiments \cite{KuznetsovSereshchenko2019} with high spatial resolution, the maximum value of the divorticity $B_{max}$ at the stage of quasi-shocks formation increases exponentially in time, while the thickness $\ell(t)$ of the maximum area in the transverse direction to the vector ${\bf B}$ decreases in time (see Fig. \ref{fig:fig2-f} left) also exponentially. It is important to note that this process is similar to the formation of three-dimensional vortex structures of pancake type \cite{agafontsev2017universal}.  

\begin{figure}[t]
	\centerline{
		\includegraphics[width=0.6\linewidth]{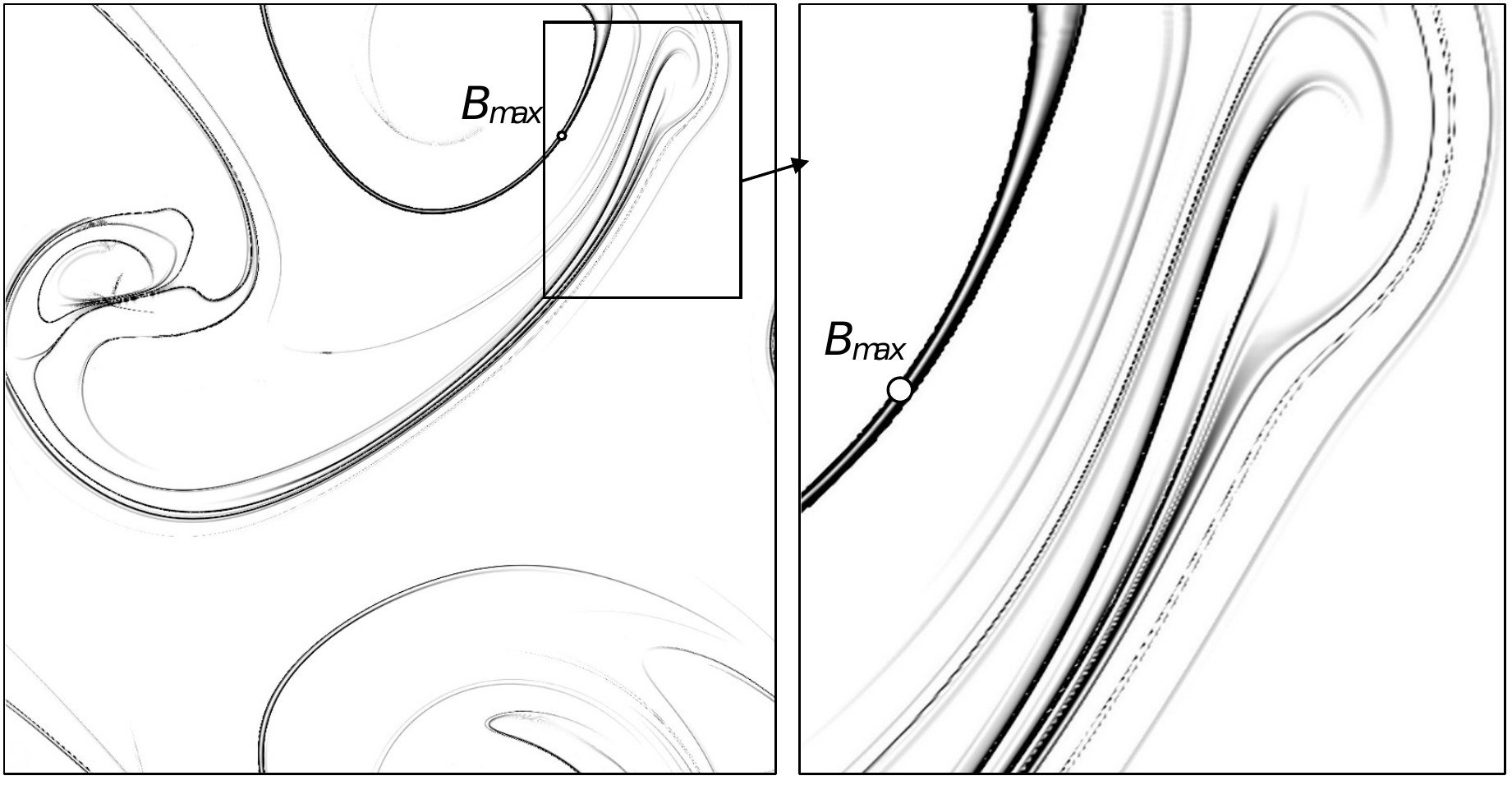}}
	\caption{Distribution $|B|$ at $t=12$. \cite{KuznetsovSereshchenko2019}}
	\label{fig:B-distr}
\end{figure}

\begin{figure}[b]
\centerline{ 
\includegraphics[width=0.45\textwidth]{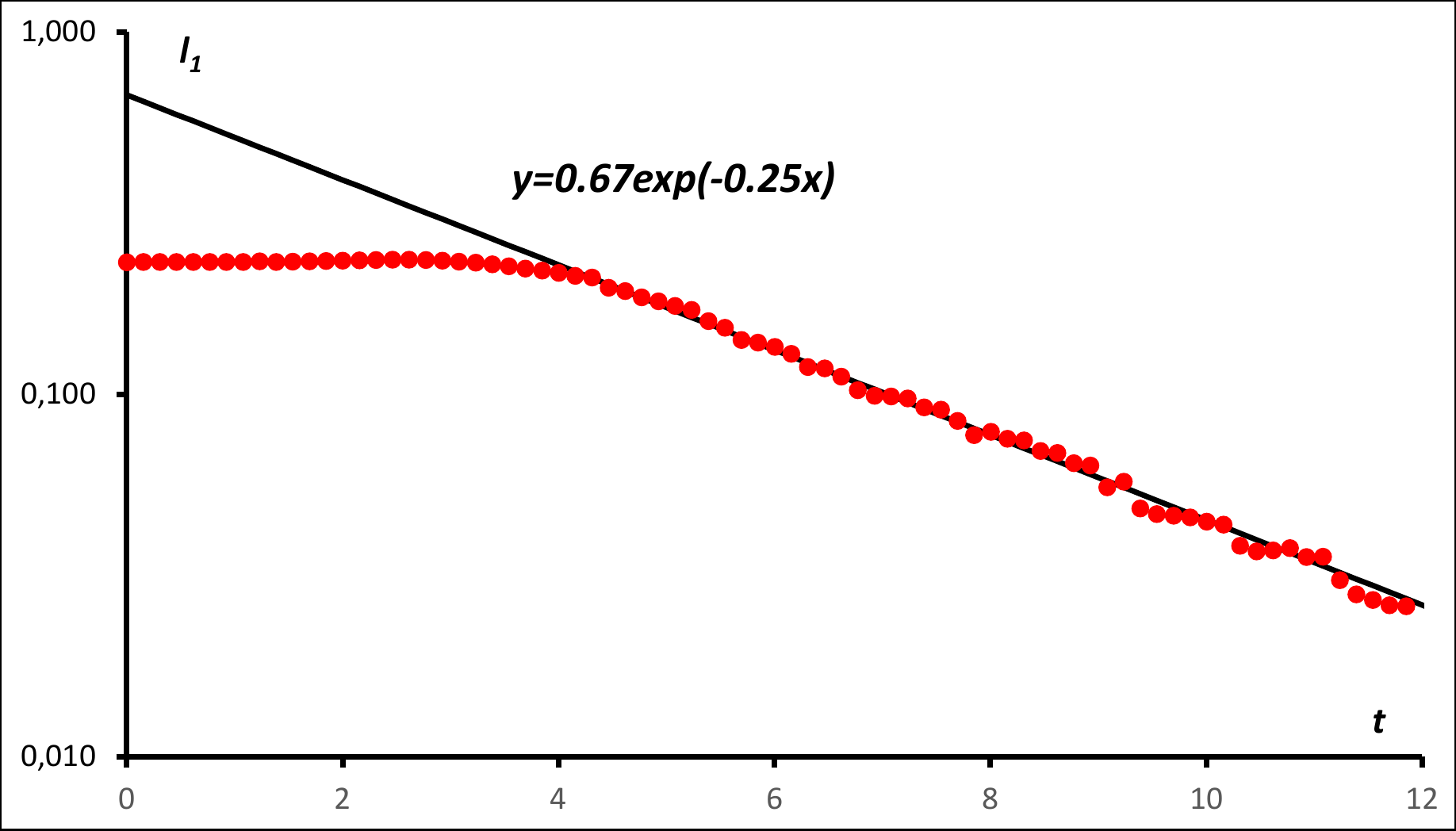}
\includegraphics[width=0.45\textwidth]{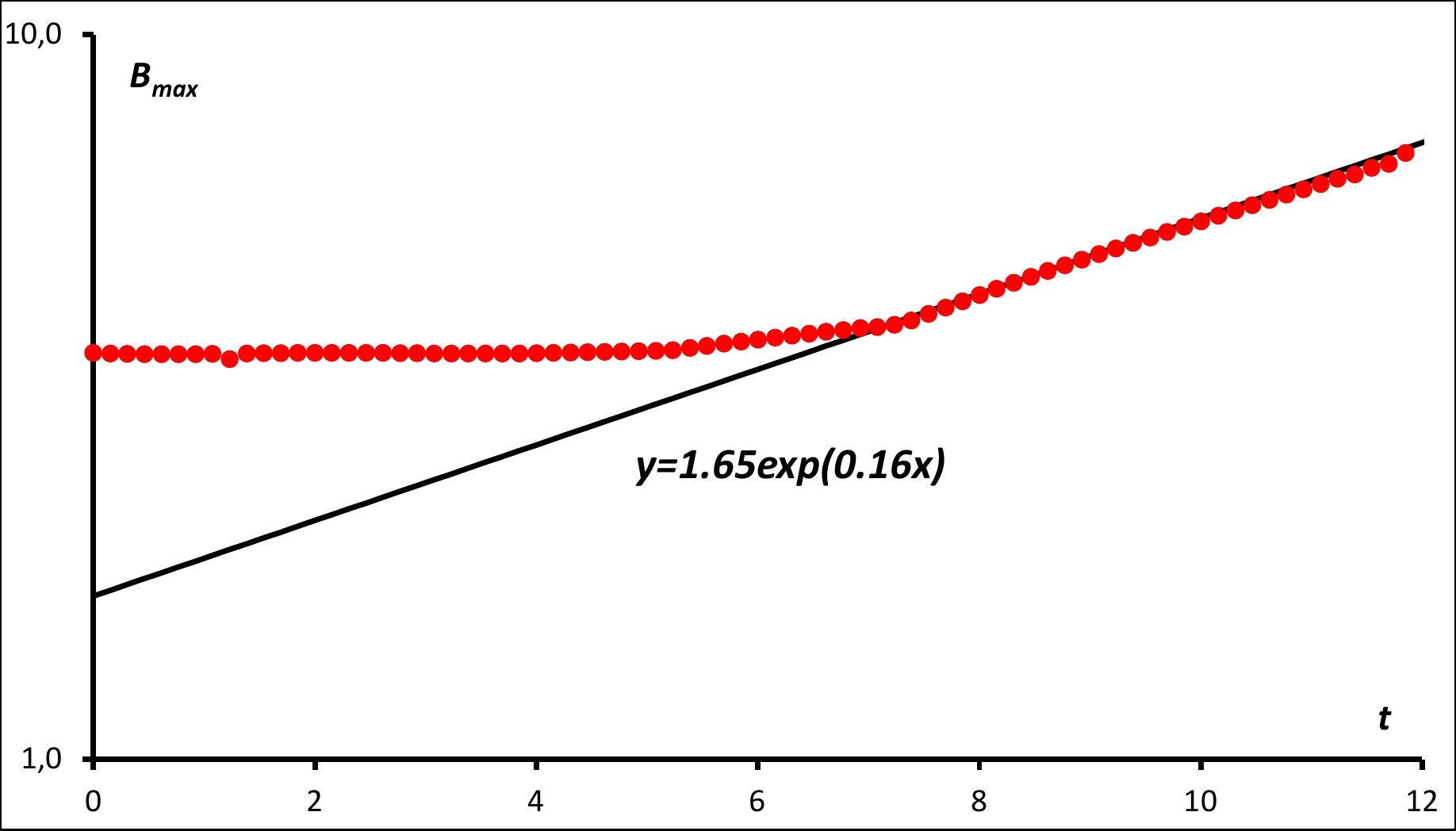}}
\caption{Time dependencies of maximum divorticity (left) and the thickness $\ell_1$ (right); scale is logarithmic. The dots correspond to the numerical results and the lines to the exponents. \cite{KuznetsovSereshchenko2019}}
\label{fig:fig2-f}
\end{figure}

In the energy spectrum, each such quasi-shock corresponded to its own jet \cite{kuznetsov2007effects,kudryavtsev2013statistical,K-04}. Along of each such jet the energy distribution decreases in accordance with the Kraichnan type power law $E\sim k^{-3}$. Our first results of the numerical experiments of the direct cascade for two-dimensional turbulence, i.e., in the presence of  both pumping and damping were presented in \cite{kuznetsov2015anisotropic}. The pumping given by growth rate $\Gamma(k)$ was concentrated in small $k$ value with a strong (singular at $k=0$) dissipation, providing suppression of inverse cascade. At large wave numbers, at $k=k_0 \sim 2/3 k_{max}$, we introduced a viscous type dissipation that allowed us to simultaneously solve the aliasing problem. At short times, in the inertial interval, turbulence was developed by the same scenario as in the case of a freely decaying turbulence with the formation of both quasi-shocks and jets in the turbulence spectrum. In these experiments, at the initial stage we observed the formation of Kraichnan-type dependence of spectrum on the module $k$ ($E\sim k^{-3}$) at all angles, and the dependence of third-order velocity structure function $S_3=\langle\delta v_{\|}^3\rangle$ on the separation length $r$ with strong anisotropy characteristic of the freely decaying turbulence. However, the spectrum averaging over angles $E(k) = C_K\eta^{2/3}k^{-3}$, where  $C_K\simeq 1.3$ is the Kraichnan constant, coincided with the spectrum previously obtained numerically, see e.g.~\cite{boffetta}). It is important to note that the structure function  $S_3$ averaging over angles gave the answer very different from its isotropic value. Analysis of these results testified in favor of the fact that the reason for this lies in the lack of spatial and temporal resolutions (our first experiments were performed on a $4096 \times 4096$ point grid). In this regard, we have been increased spatial resolution up to $16384 \times 16384$ and doubled the calculation time in comparison with the best experiments  \cite{kuznetsov2015anisotropic}. The main difference of the obtained results from the previous ones is that, at times of the order of $10\,\Gamma_{max}^{-1}$  ($\Gamma_{max}^{-1}$ is a characteristic pumping time inversely proportional to the growth rate maximum), the jet structure of the spectrum in the direct cascade is destroyed and turbulence tends to be isotropic. In particular, on these times any significant anisotropy in angular fluctuations of energy spectrum (for a fixed value of  $k$) is not observed. In the regime of an isotropic distribution, we found probability distribution functions $P$ for both vorticity and divorticity module $B$. The $P(\omega)$ structure of the corresponds to the predictions of the isotropic theory  \cite{FalkovichLebedev2011}.

\subsection{Main equations and numerical scheme}
Let us briefly considered the equations of motion and the numerical scheme, which are completely coincide with those in \cite{kuznetsov2015anisotropic}. The equation of motion (\ref{omega}) for two-dimensional flows, depending only on the coordinates $x,\,y$ in the flow plane, are written only for one $z$-component $\omega$:
\begin{equation} \label{NS}
\frac{\partial \omega }{\partial t}+(\mathbf{v}\nabla )\omega =\hat{\Gamma}\omega +\hat{\gamma}\omega.
\end{equation}
In this equation, two terms responsible for pumping and damping are introduced into the right-hand side to simulate turbulence, in particular the direct cascade, which is initially formed due to the appearance of quasi-shocks of the vorticity. In the absence of the right-hand side responsible for pumping and damping, the  vorticity $\omega$ is a Lagrangian invariant advected by the fluid with velocity $\mathbf{v}$. This is the situation realized in the freely decaying two-dimensional turbulence regime. 

To model the direct cascade of two-dimensional hydrodynamic turbulence, the right-hand side (\ref{NS}) contains two operators: the $\hat{\Gamma}$ operator is responsible for both injection of the energy and its dissipation on large scales to exclude inverse cascade, and the $\hat{\gamma}$ operator is responsible for the entropy dissipation at large $k$. Both of these operators were set by their Fourier transforms (see \cite{kuznetsov2015anisotropic}): 
\begin{eqnarray}
\Gamma _{k} &=&A\frac{(b^{2}-k^{2})(k^{2}-a^{2})}{k^{2}}\quad \mbox{for}\quad 0\leq k\leq b,\quad \nonumber \\ \quad \Gamma _{k} &=&0\quad \mbox{for}\quad k>b, \nonumber
\end{eqnarray}
and 
\begin{eqnarray}
\gamma _{k} &=&0\quad \mbox{\rm for}\quad k\leq k_{c},\quad \nonumber \\ \quad \gamma _{k} &=&-\nu (k-k_{c})^{2}\quad \mbox{\rm for}\quad k>k_{c}. \nonumber
\end{eqnarray}
In the numerical integration of the equation (\ref{NS}), the parameters $a$ and $b$ were chosen from the conditions of the most rapid transition of the  system to the steady-state regime at small $k$ value. In subsections 5.2 and 5.3 are the results with $A=0.004$, $a=3$ and $b=6$. For dissipation in the viscous-type form, providing enstrophy absorption, coefficient of viscosity was $\nu=1.5$ and $k_{c}$ was $0.6k_{\max }$ where $k_{\max}=8192$, that simultaneously solves the problem of aliasing. The initial conditions were the same as in our previous papers \cite{kudryavtsev2013statistical,kuznetsov2015anisotropic}. The maximum grid size was $16384 \times 16384$.

Numerical simulations of equation (\ref{NS}) for both freely decaying turbulence and direct cascade were performed in a square box $L=1$ with periodic boundary conditions.

\subsection{Folding in 2D turbulence}
Let us now present the results of numerical integration of the two-dimensional Euler equation (with zero right-hand side (\ref{NS})) for the vorticity rotor or divorticity:
\[
B_x=\frac{\partial\omega}{\partial y}, \,\, B_y=-\frac{\partial\omega}{\partial x}. 
\]
As can be seen from the definition of this vector, ${\bf B}$ is directed tangentially to the isoline $\omega(x,y)=\mbox{const}$. It follows that the growth of ${\bf B}$ leads to the appearance of a jump $\omega(x,y)$ in the direction perpendicular to the vector ${\bf B}$. Thus, the formation of vorticity jumps (quasi-shocks) corresponds to the growth of divorticity ${\bf B}$. As mentioned above, the growth of ${\bf B}$ in the experiments \cite{kuznetsov2007effects, KNNR-10, kudryavtsev2013statistical} was 2 - 2.5 orders of magnitude. At the same time, ${\bf B}$ concentrated in the vicinity of the lines, i.e., the formation of quasi-one-dimensional structures took place. The growth of ${\bf B}$ is due to a decrease of the Jacobian $J$, which causes the compressibility of continuously distributed divorticity lines and, accordingly, a tendency to breaking, resulting in the formation of vorticity quasi-shocks.

Figure \ref{fig:B-distr}, taken from \cite{KuznetsovSereshchenko2019}, shows the structure of $|B|$ at $t=12$. Two sets of Gaussian vortices with positive and negative vorticity with zero (the maximum value of $|\omega|$ equal to unity) total vorticity were used as initial conditions. The size of each pair was random in the range of 0.2 - 0.6, the location of vortices was also random.  Unlike previous works \cite{kuznetsov2007effects, KNNR-10, kudryavtsev2013statistical}, we limited the number of vortices to 8 (4 positive and 4 negative) to more accurately determine the required dependencies of the ${\bf B}$ field and its geometric characteristics:  positions of maximums, longitudinal and transverse quasi-shock sizes, etc dependencies - the ${\bf B}$ field and its geometric characteristics: maximum positions, longitudinal and transverse quasi-shock sizes, etc \cite{KuznetsovSereshchenko2019}. Zoom shows that between the maximum lines values of $|B|$ are significantly less than its maximums. For vorticity this corresponds to a system of terraces with steps of variable height. Each of these steps is a vorticity quasi-shock. 

Figure \ref{fig:fig2-f} on the left shows the dependence of $B_{max}$ on time at the initial stage. As can be seen from these figure, $B_{max}$ grows exponentially. The thickness of the maximum area in the transverse direction to the vector ${\bf B}$ decreases in time also exponentially (Fig. \ref{fig:fig2-f} on the right).

\begin{figure}[t]
\centerline{
\includegraphics[width=0.45\textwidth]{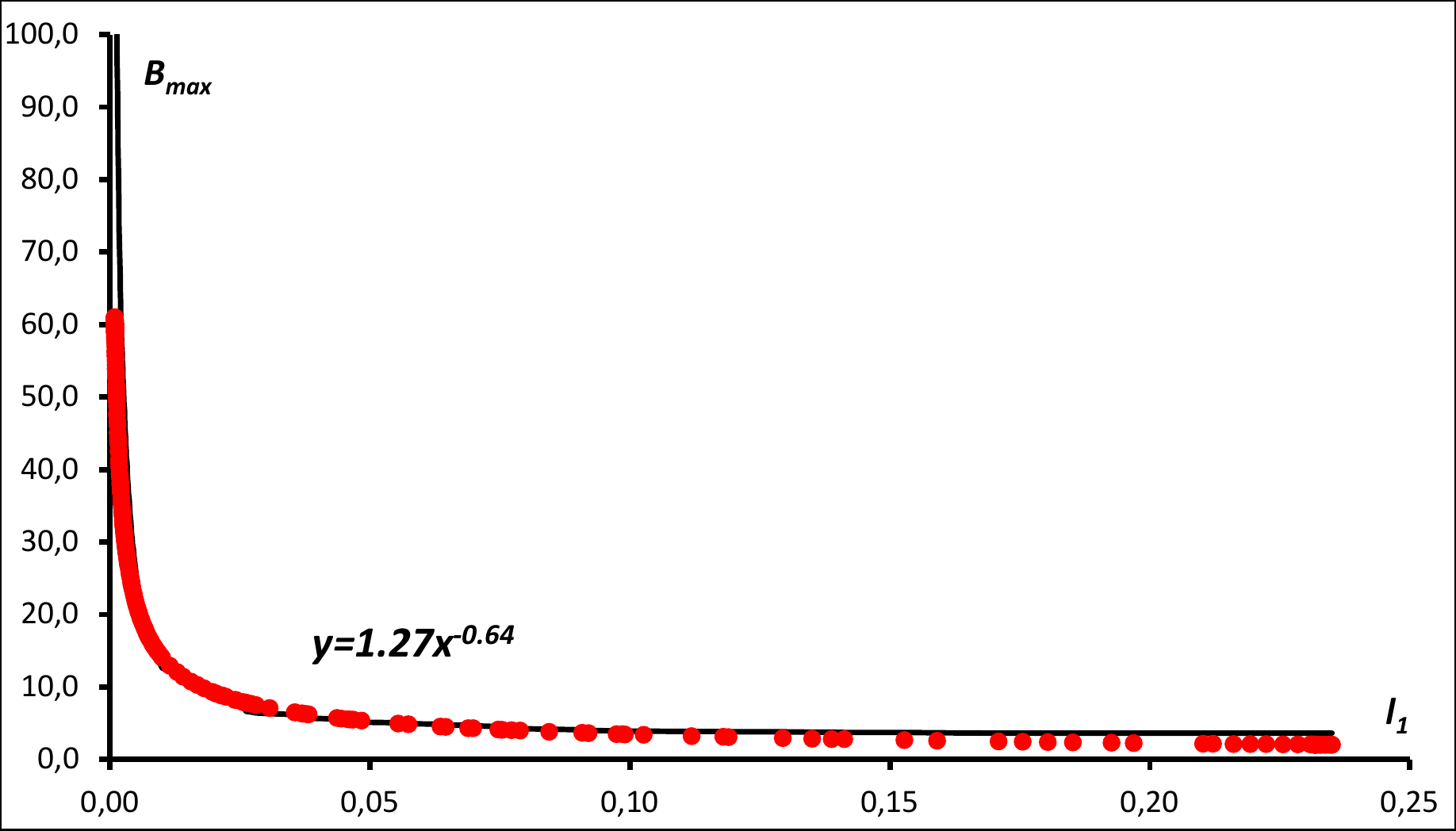}}
\caption{Maximum divorticity versus the thickness $\ell_1$. The points correspond to numerical results, and the line is the power dependence $B_{max}\sim \ell_{1}^{-2/3}$. \cite{KuznetsovSereshchenko2019}}
\label{fig:fig4-f}
\end{figure}

The obtained dependencies for $B_{max}$ and thickness $\ell$ show that at the exponential stage between these values there is a power dependence $B_{max}=C \ell^{\alpha}$ with $\alpha= 0.16/(-0.25) = -0.64 \approx -2/3 $, $C$ is a constant  (see Fig. \ref{fig:fig4-f}).

It is worth noting that this dependence of $B_{max}$ on $\ell$ in the form of the $2/3$ law was also verified  for another initial conditions (recall that the positions of vortices and their sizes were random). This allows ones to believe that this relation can be considered as universal. 

\subsection{Statistical Properties of 2D Turbulence}
In the case of freely decaying turbulence, the process of breaking is dominant, leading to a strong anisotropy of the turbulence spectrum due to the presence of jets generated by quasi-shocks \cite{kuznetsov2007effects, KNNR-10, kudryavtsev2013statistical}. This process turns out to be the fastest, as a result the turbulence spectrum of the direct cascade at the initial stage forms a power dependence on the wave number $k$ with the Kraichnan type power law: $E_k\sim k^{-3}$ (see the original paper of Kraichnan \cite{kraichnan}), even in the presence of pumping, as shown by numerical experiments \cite{KuznetsovSereshchenko2017}. At the same time, the formation of the vorticity quasi-shocks  is exponential; in accordance with this, the regions of the maximum of divorticity are decreased in the direction perpendicular to the lines of the constant vorticity. As shown by the numerical experiments \cite{kuznetsov2007effects, KNNR-10, kudryavtsev2013statistical, kuznetsov2015anisotropic}, for typical initial conditions the growth of the divorticity is 2 -- 2.5 orders of magnitude, and the transverse size of the maximal area ${\bf B}$ decreases significantly. The explanation of this growth is related to the possibility of partial integration of the equation (\ref{B}) in terms of mapping ${\bf r}={\bf r} ({\bf a},t)$:
\begin{equation} \nonumber
\mathbf{B}(\mathbf{r},t) = \frac{(\mathbf{B}_0(\mathbf{a}) \cdot\nabla_a)\mathbf{r}(\mathbf{a},t)}{J} ,
\end{equation}
where $\mathbf{B_0} (\mathbf{a})$ is the initial $\mathbf{B}$, which is an analogue of the Cauchy invariant. A similar formula for the three-dimensional Euler equations is basic in the so-called vortex line representation  \cite{KuznetsovRuban, KuznetsovRubanPRE}. The key point here for understanding is the compressibility of the divorticity field and the possibility of $J$ to vanish. As is known, breaking in the gas dynamics occurs due to the compressibility of the gas. The formation of quasi-two-dimensional caustics occurs when approaching the breaking point (see, e.g., \cite{shandarin1989large}). Similarly, the formation of the vorticity quasi-shocks happens.

In all numerical experiments simulating both freely decaying turbulence and direct cascade, the initial vorticity distribution was chosen as 10 positive and 10 negative Gaussian vortices with the same value of the maximum value of $|\omega|$ equal to unity and zero mean vorticity. The location and size of the vortices were random. 

\begin{figure}[t]
	\centerline{ 
		\includegraphics[width=0.3\textwidth]{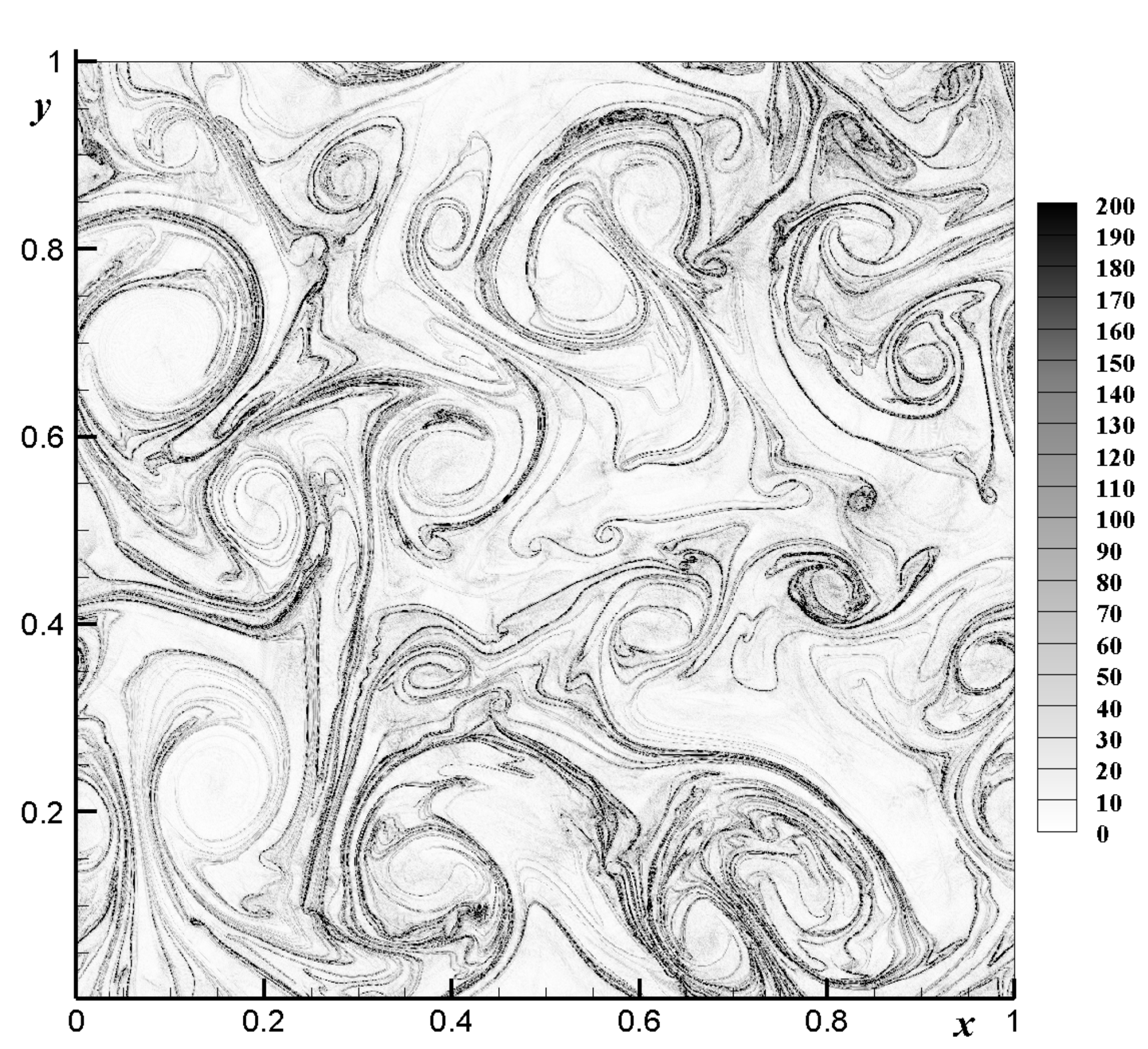}
		\includegraphics[width=0.3\textwidth]{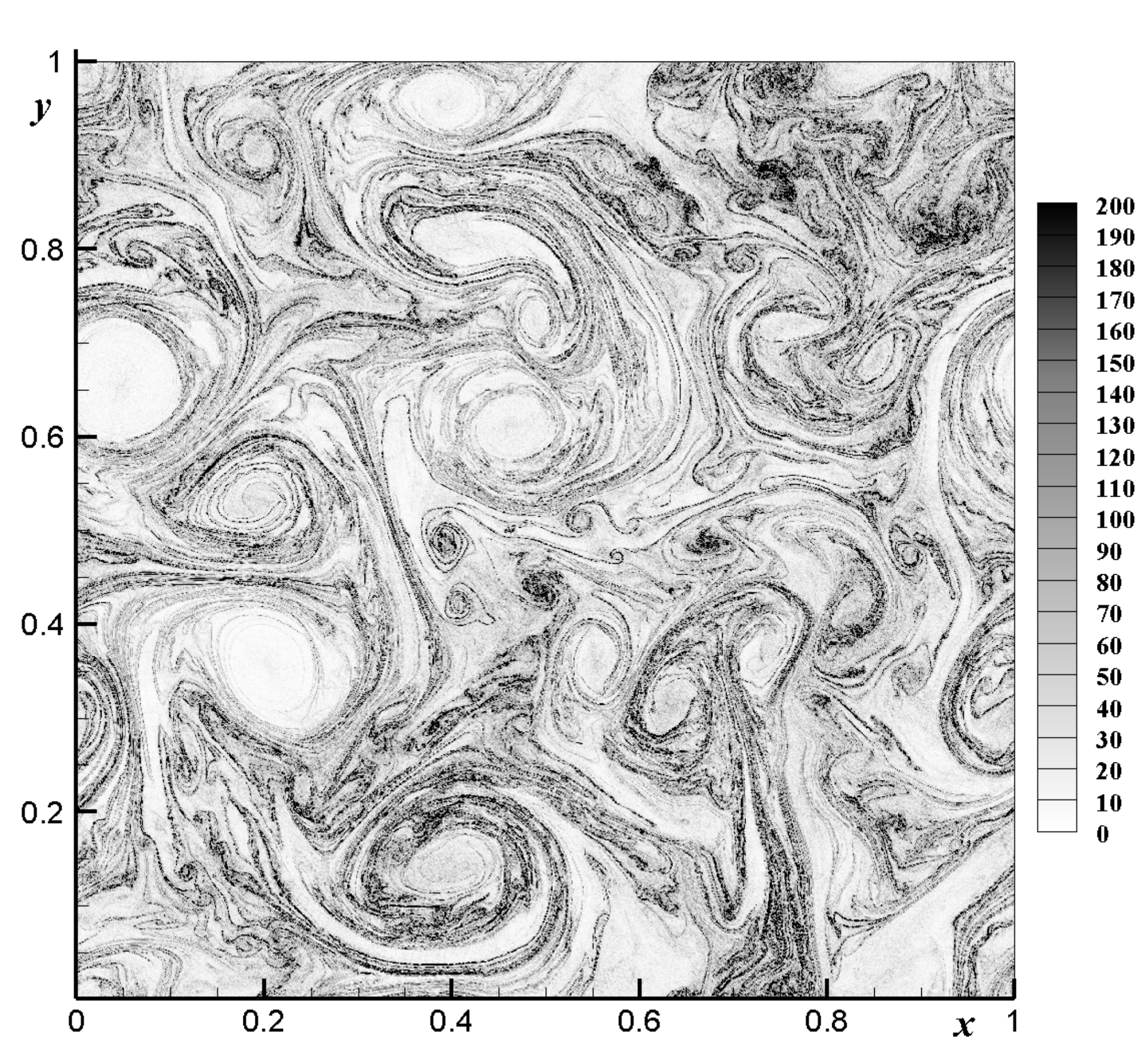}
		\includegraphics[width=0.3\textwidth]{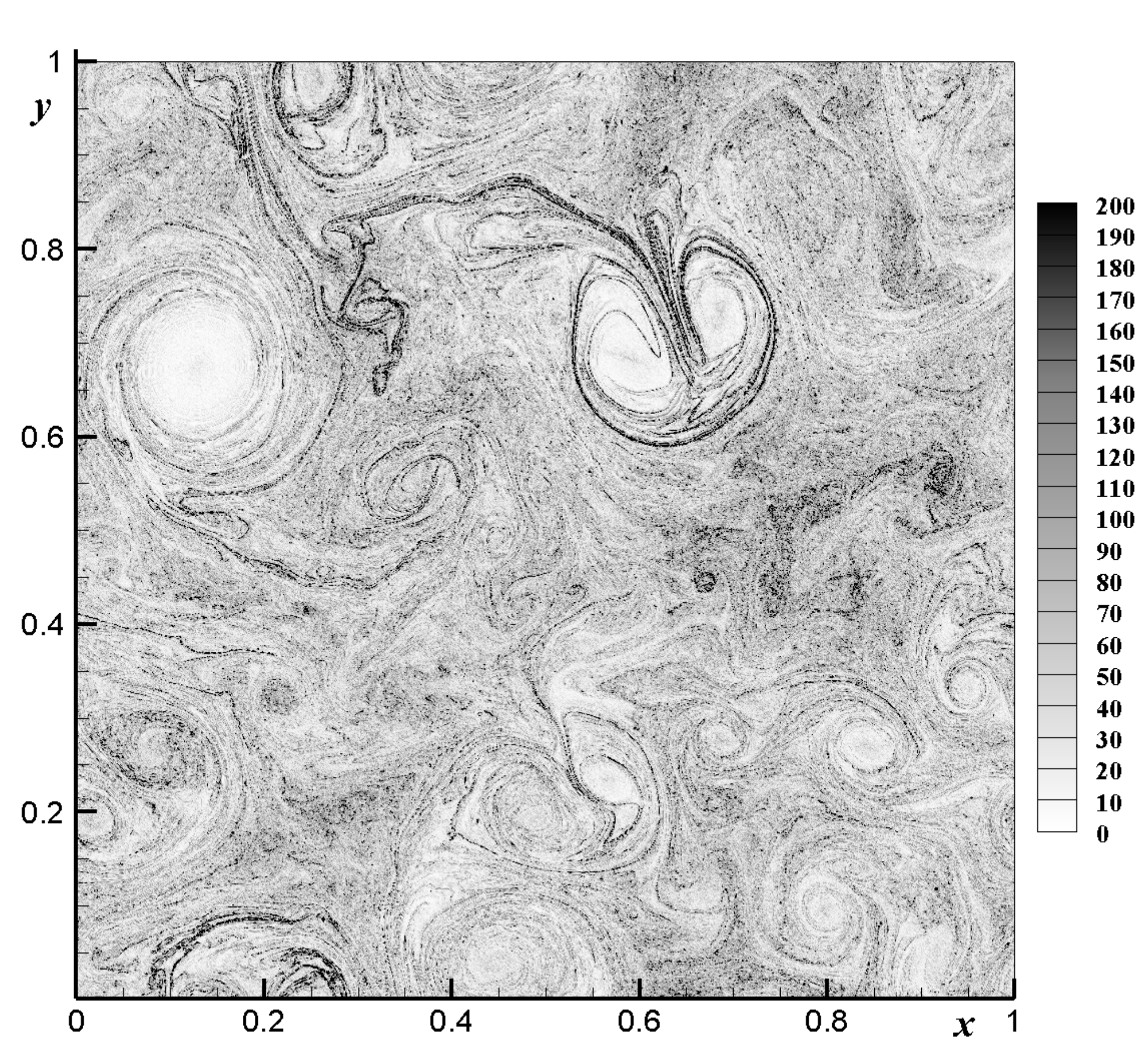}}
	\begin{tabular}{p{0.42\linewidth}p{0.13\linewidth}p{0.42\linewidth}}
		\centering a) & \centering b) & \centering c) \\
	\end{tabular}
	\caption{Distribution of $|B|$ at $t=150, 250, 450$. \cite{KuznetsovSereshchenko2017}}
	\label{fig:fig5-dc}
\end{figure}

\begin{figure}[b]
\centerline{
\includegraphics[width=0.3\textwidth]{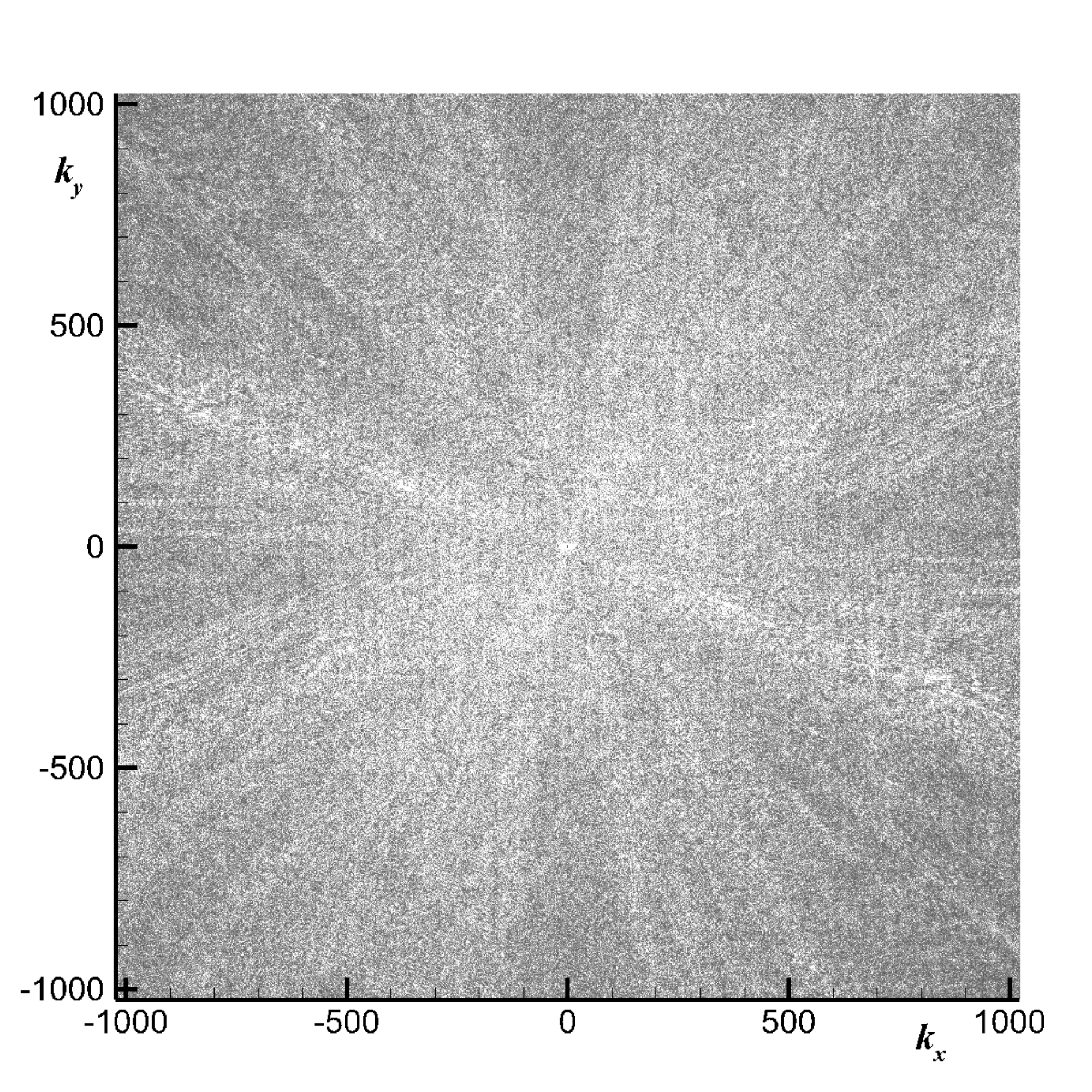}
\includegraphics[width=0.3\textwidth]{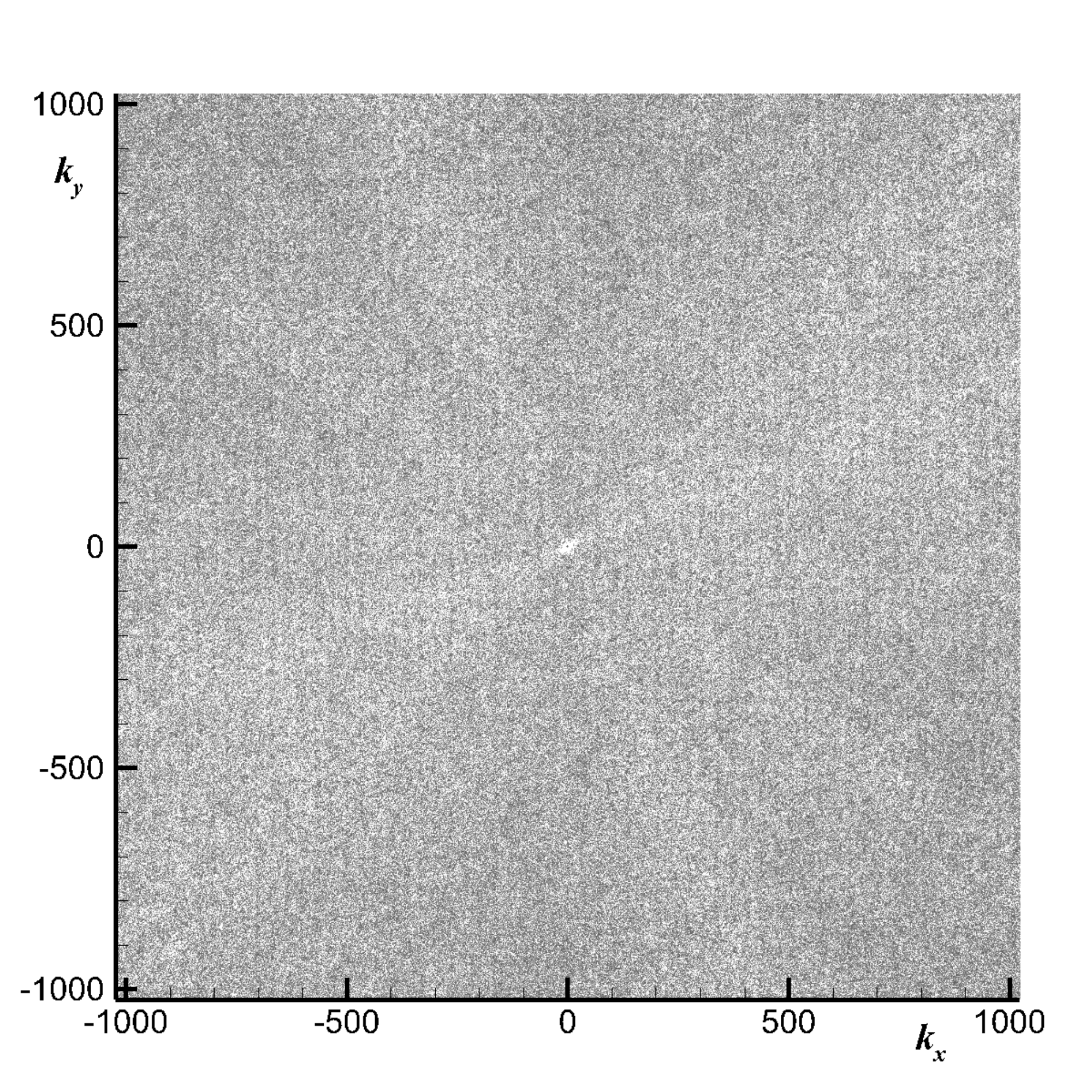}
\includegraphics[width=0.3\textwidth]{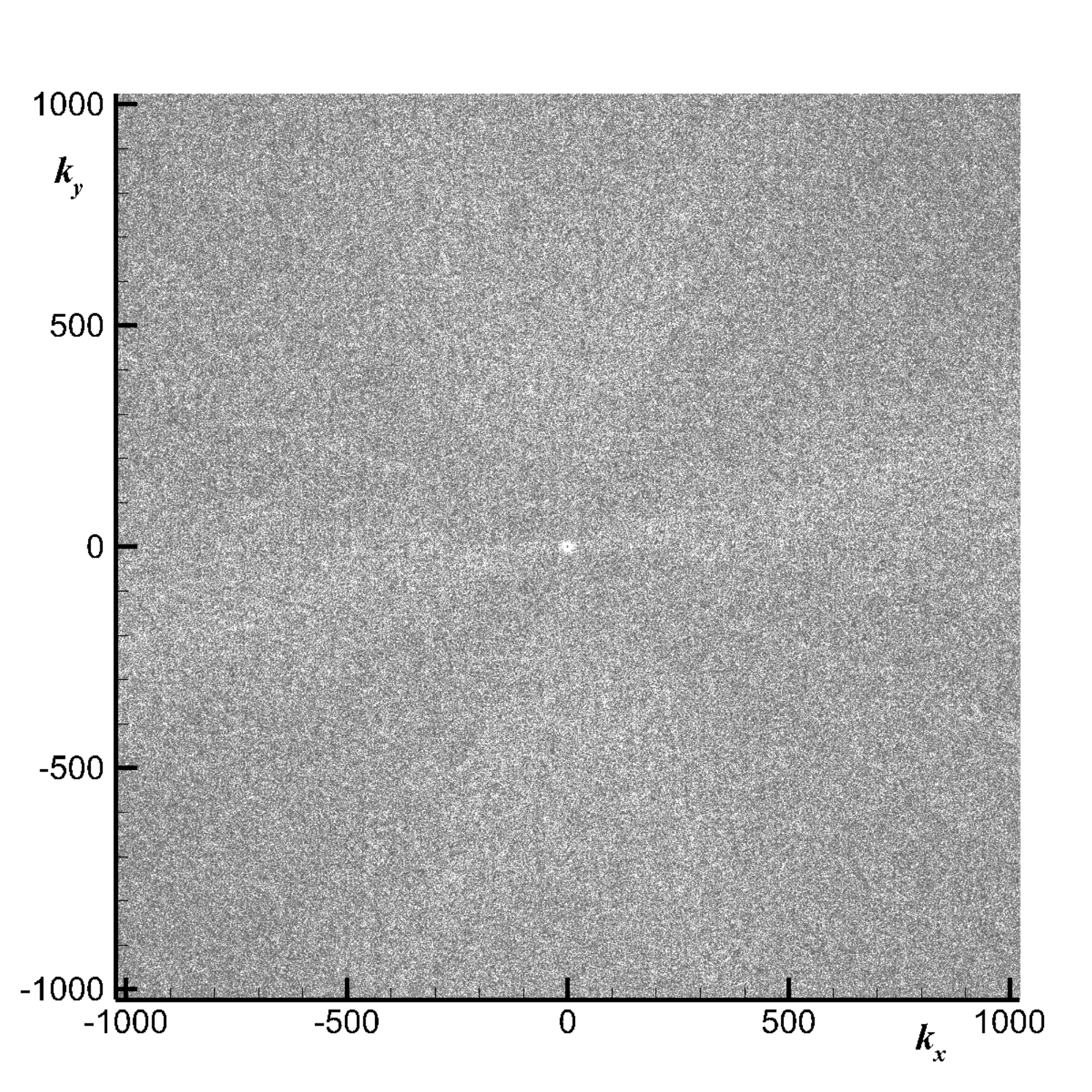}}
\begin{tabular}{p{0.42\linewidth}p{0.13\linewidth}p{0.42\linewidth}}
\centering a) & \centering b) & \centering c) \\
\end{tabular}
\caption{Energy density distribution of fluctuations $\epsilon({\bf k})$ normalized to $k^{-4}$ at $t=150, 250, 450$. \cite{KuznetsovSereshchenko2017}}
\label{fig:fig6-dc}
\end{figure}

In the direct cascade mode, at the initial stage, for the times of order of the inverse  pumping growth rate $\Gamma_{max}^{-1}$, the development of turbulence is about the same scenario as in the case of a freely decaying turbulence \cite{kudryavtsev2013statistical}: quasi-singular distributions of divorticity are formed, which in $k$-space correspond to jets, leading to a strong turbulence anisotropy. Fig. \ref{fig:fig5-dc} shows a typical distribution of the divorticity module $|B|$, which is the most concentrated on the lines (positions of quasi-shocks). Between these lines the value of $|B|$ is significantly lower. Accordingly, jets (with weak and strong overlapping in the k-space) are observed in the spectrum, as a result, the turbulence spectrum has a large anisotropic component. Fig. \ref{fig:fig6-dc} shows in k-space the distributions of the energy density of fluctuations  $\epsilon({\b k})$, normalized by $k^{-4}$.

At each angle in the $k$-space in the inertial interval value of $\epsilon({\bf k})k^{4}$ at a given time fluctuates greatly, and after averaging  in the interval ($k-\Delta k/2, k+\Delta k/2$) is almost constant (see Fig. 3 of \cite{KuznetsovSereshchenko2017}). It is important to note that the formation of the Kraichnan-type dependence on the module $k$ occurs on the first stage of the direct cascade development, when enstrophy transfer reaches the ``viscous'' region. According to our estimates \cite{KuznetsovSereshchenko2017}, the time of such transfer is order of the inverse pumping growth rate $\Gamma_{max}^{-1}$ (for the numerical experiment presented in this section, this time was on the order of $50$). At this stage, the energy spectrum depends strongly on the angle. It is surprising that after averaging over the angles spectrum $E(k)$ having both the Kraichnan-type dependence on $k$ and enstrophy flux $\eta$, defined as $1/2\int \gamma(k) |\omega_k |^2 d{\bf k}$, gives a value for the Kraichnan constant $C_K\simeq 1.3$ which coincides with that previously obtained in numerical experiments (see \cite{boffetta}).

\begin{figure}[t]
	\centerline{
		\includegraphics[width=0.45\textwidth]{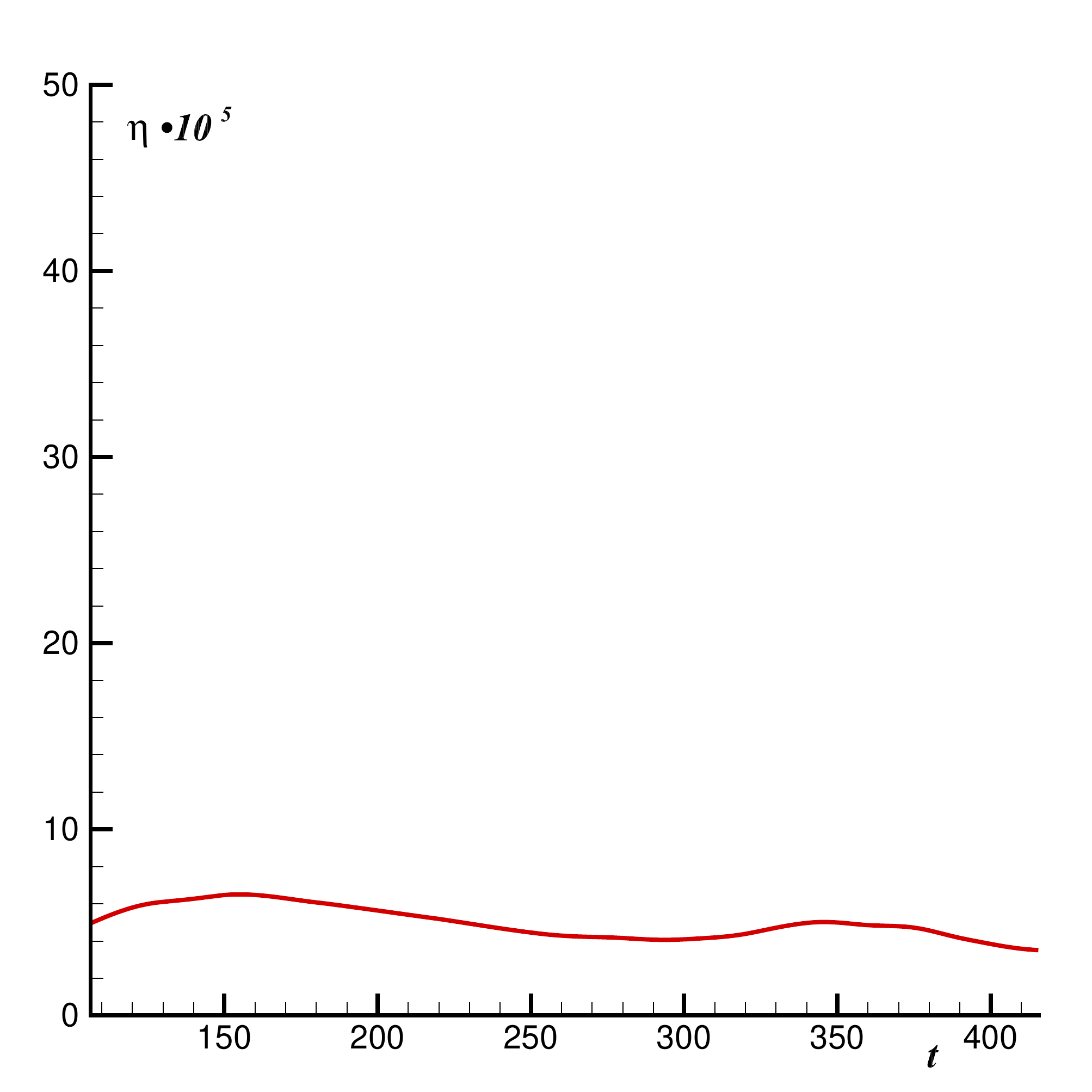}}
	\caption{Time dependence of the enstrophy flux $\eta$. \cite{KuznetsovSereshchenko2017}}
	\label{fig:fig4}
\end{figure}

In the next stage, the net of quasi-shocks lines becomes  more complicated (turbulent) (\ref{fig:fig5-dc}b). The distances between quasi-shocks lines are reduced, and as a result the anisotropy in energy spectrum decreases (Fig. \ref{fig:fig6-dc}b). Finally, for times of the order of $10\, \Gamma_{max}^{-1}$ jets practically disappear (Fig. \ref{fig:fig5-dc}c  and Fig. \ref{fig:fig6-dc}c) and turbulence in the direct cascade becomes almost isotropic. It also appears that at all times, starting from occurrence of jets up to their disappearance, enstrophy flux is almost constant (Fig. \ref{fig:fig4}).  The total energy of sufficiently fast becoming a constant (at the first stage), which is not for the total enstrophy. It is close to a constant value only at the isotropization stage \cite{KuznetsovSereshchenko2017}.

\begin{figure}[t]
	\centerline{
		\includegraphics[width=0.45\textwidth]{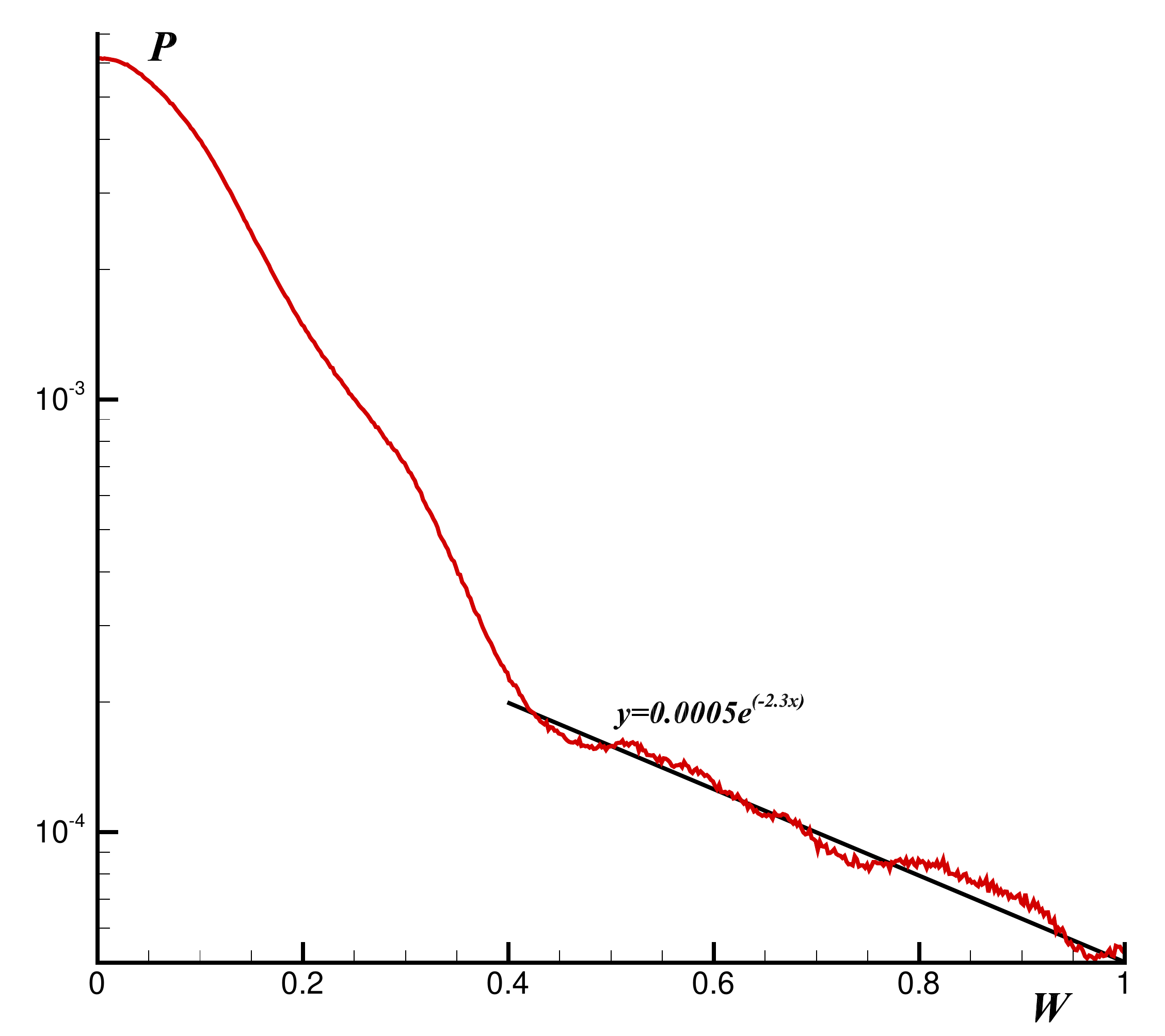}
		\includegraphics[width=0.45\textwidth]{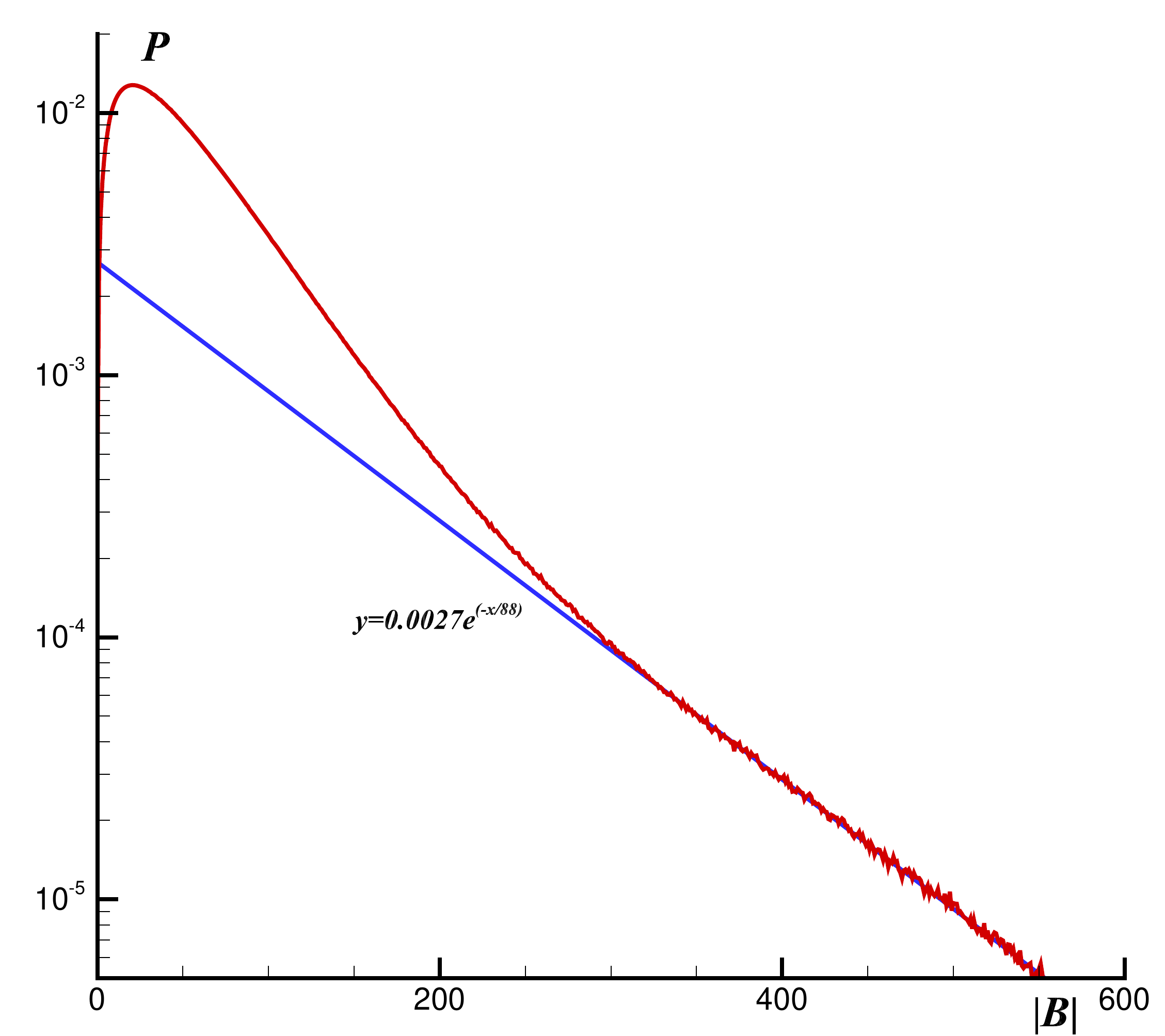}}
	\begin{tabular}{p{0.50\linewidth}p{0.40\linewidth}}
		\centering a) & \centering b)\\
	\end{tabular}
	\caption{(a) Probability distribution function for vorticity and (b) probability distribution function for divorticity at $t=450$. \cite{KuznetsovSereshchenko2017}}
	\label{fig:fig7-dc}
\end{figure}

Another indication of turbulence isotropization at times of the order of $10\, \Gamma_{max}^{-1}$ found in \cite{KuznetsovSereshchenko2017} is a probability distribution function of vorticity $P$ (Fig. \ref{fig:fig7-dc}), which for large arguments has an exponential tail with exponent $\beta$, linearly dependent on vorticity $\bar {\omega}$, in agreement with the theoretical predictions \cite{FalkovichLebedev2011}. According to these predictions, the angle slope of the exponent \ is order of ${\bar\omega}_{rms}^{-1}$, where $\bar{\omega}_{rms}$ is rms vorticity fluctuations. Numerical experiment (Fig. \ref{fig:fig7-dc}a) gives the asymptotic behavior of $P=0.0005exp(-2.3{\bar\omega})$ with ${\bar\omega}_{rms}=0.43$. If calculate the enstrophy flux as the integral $\eta=1/2\int \gamma|\omega_k|^2d{\bf k}$, then $\bar{\omega}_{rms}=0.15$. Calculation of $\bar{\omega}_{rms}$ by a given distribution function gives the value of $0.2566$. Thus, the values $\bar{\omega}_{rms}$ are close to each other with accuracy of the order of unity.

The corresponding distribution function $P$ for divorticity $B$ also has two specific regions (Fig. \ref{fig:fig7-dc}b): in the first one the distribution function is close to the Poisson distribution, $\sim B\exp (-B^2/B_0^2)$, in the second region (large value of divorticity $B$) the distribution function $P$ is of exponential behavior with more pronounce  linear dependence of exponent on $B$ than the similar one for vorticity. For this numerical experiment $B_{rms}$, calculated from the angle slope, is equal to $88$ (see \cite{KuznetsovSereshchenko2017}). If calculate the $B_{rms}$ by means of the distribution function $P(B)$, then this value is equal to $84.6$. 

\section{Conclusion}
The main conclusion of this work is that in both three-dimensional and two-dimensional hydrodynamics of an incompressible fluid, at the stage of turbulence initiation at high Reynolds numbers, the main role is played by coherent vortex structures, whose evolution is due to the compressibility of the corresponding fields, despite their divergence-free nature. For the 3D case, these are pancake-type structures with Kolmogorov-type scaling: the ratio between the maximum vorticity and the pancake thickness is given by $\omega_{\max}\sim \ell^{-2/3}$. The compression of these structures is exponential and can be interpreted as an overturning process similar to the formation of shock waves in gas dynamics due to the compressibility of the gas. 

Using a combined analytical-numerical approach based on the representation of vortex lines, we have shown that scaling for pancake-type structures arises by taking into account the three-dimensionality of these structures. At the same time, the velocity demonstrates the H\"older behavior, which is used by some mathematicians to construct a theory of 3D turbulence. 

For two-dimensional turbulence, it was found that the formation of a direct cascade, the Kraichnan cascade with a constant enstrophy flow, is due to the appearance of vorticity quasi-shocks due to the compressibility of the vorticity rotor field. This process turns out to be the fastest, as a result of which the turbulence spectrum of the direct cascade almost from the very beginning acquires a power law dependence on the wave number $k$ with the Krauchnan exponent. But the turbulence spectrum at this stage turns out to be strongly anisotropic due to jets (Fourier transforms of quasi-shocks). At the next, slower, stage, the complication (turbulization) of the structure of quasi-shock lines occurs. The distances between them are reduced, and the spectrum becomes more isotropic. It is important to note that in the isotropic state, the vorticity probability distribution function for large arguments $\omega$ forms an exponential tail with an exponent that can be extrapolated as a linear dependence on vorticity, in accordance with the theoretical prediction of \cite{FalkovichLebedev2011} for a two-dimensional isotropic turbulence.

The review also presents results on two-point structure functions (moments) of velocity. Despite the strong anisotropy inherent in the (nonstationary) problem at the stage of onset of three-dimensional hydrodynamic turbulence, a power-law scaling is formed for both longitudinal and transverse moments in the same scale interval as for the energy spectrum. The exponents of structure functions have the same key properties as for developed (stationary) turbulence. In particular, the exponents depend nontrivially on the order of the moment, indicating intermittency and anomalous scaling, while the longitudinal exponents turn out to be somewhat larger than the transverse ones. Analyzing simulations for various initial conditions, a rather rough estimate for $\xi_{3}\simeq \alpha/5$ is found, which relates the scaling exponents for the longitudinal moment of the third order and the energy spectrum. Thus, when the energy spectrum has a near-Kolmogorov power-law scaling, the third-order longitudinal moment shows a near-linear scaling with distance consistent with Kolmogorov's $4/5$ law~(\ref{K45}). 

It should be noted that, before angle averaging, the third-order moments show a highly anisotropic behavior, although the linear scaling obtained after angle averaging can be traced for most directions. The vorticity distribution is characterized by a strongly non-Rayleigh shape, which also indicates intermittency. The power scaling~(\ref{PDF-scaling}) for the tail of this distribution has the exponent $\beta\gtrsim 1/2$, which indicates the non-trivial geometry of pancake-shaped vorticity structures. We also note that the third-order structure velocity functions for two-dimensional turbulence, while having a large anisotropy due to vorticity quasi-shocks, nevertheless have the same power-law dependence on the coordinates as in the isotropic case.

In conclusion, we would like to note that the emergence of quasi-one-dimensional structures that grow exponentially at the stage of turbulence onset is due to the frozenness property of the corresponding fields. A classic example of frozen-in fields is the magnetic field in MHD in the absence of ohmic dissipation. A well-known example of the formation of magnetic filaments in the kinematic dynamo approximation with a given velocity field at zero magnetic viscosity, first considered by Parker in 1963 \cite{parker1963kinematical} as applied to the convective zone of the Sun, indicates that the growth of the magnetic field is also exponential in time. In the case of the kinematic approximation for the induction equation, the appearance of filaments occurs in regions with a hyperbolic velocity profile \cite{KuznetsovMikhailov2020}.

The work of D.S.A, E.A.K. and E.V.S. was supported by the Russian Science Foundation (grant 17-01-00622). Numerical experiments were performed in the information and computing center of Novosibirsk State University and the data center of the Institute of Pure and Applied Mathematics (IMPA, Rio de Janeiro). D.S.A. grateful for the support from IMPA during the visits to Brazil. The authors thank S.N. Gurbatov for helpful remarks.


\begin{thebibliography}{72}%
\makeatletter
\providecommand \@ifxundefined [1]{%
 \@ifx{#1\undefined}
}%
\providecommand \@ifnum [1]{%
 \ifnum #1\expandafter \@firstoftwo
 \else \expandafter \@secondoftwo
 \fi
}%
\providecommand \@ifx [1]{%
 \ifx #1\expandafter \@firstoftwo
 \else \expandafter \@secondoftwo
 \fi
}%
\providecommand \natexlab [1]{#1}%
\providecommand \enquote  [1]{``#1''}%
\providecommand \bibnamefont  [1]{#1}%
\providecommand \bibfnamefont [1]{#1}%
\providecommand \citenamefont [1]{#1}%
\providecommand \href@noop [0]{\@secondoftwo}%
\providecommand \href [0]{\begingroup \@sanitize@url \@href}%
\providecommand \@href[1]{\@@startlink{#1}\@@href}%
\providecommand \@@href[1]{\endgroup#1\@@endlink}%
\providecommand \@sanitize@url [0]{\catcode `\\12\catcode `\$12\catcode
  `\&12\catcode `\#12\catcode `\^12\catcode `\_12\catcode `\%12\relax}%
\providecommand \@@startlink[1]{}%
\providecommand \@@endlink[0]{}%
\providecommand \url  [0]{\begingroup\@sanitize@url \@url }%
\providecommand \@url [1]{\endgroup\@href {#1}{\urlprefix }}%
\providecommand \urlprefix  [0]{URL }%
\providecommand \Eprint [0]{\href }%
\providecommand \doibase [0]{https://doi.org/}%
\providecommand \selectlanguage [0]{\@gobble}%
\providecommand \bibinfo  [0]{\@secondoftwo}%
\providecommand \bibfield  [0]{\@secondoftwo}%
\providecommand \translation [1]{[#1]}%
\providecommand \BibitemOpen [0]{}%
\providecommand \bibitemStop [0]{}%
\providecommand \bibitemNoStop [0]{.\EOS\space}%
\providecommand \EOS [0]{\spacefactor3000\relax}%
\providecommand \BibitemShut  [1]{\csname bibitem#1\endcsname}%
\let\auto@bib@innerbib\@empty
%</preamble>
\bibitem [{\citenamefont {Richardson}(1926)}]{Richardson}%
  \BibitemOpen
  \bibfield  {author} {\bibinfo {author} {\bibfnamefont {L.~F.}\ \bibnamefont
  {Richardson}},\ }\href@noop {} {\bibfield  {journal} {\bibinfo  {journal}
  {Proc. R. Soc. Lond. A}\ }\textbf {\bibinfo {volume} {110}},\ \bibinfo
  {pages} {709} (\bibinfo {year} {1926})}\BibitemShut {NoStop}%
\bibitem [{\citenamefont {Kolmogorov}(1941)}]{kolmogorov1941dissipation}%
  \BibitemOpen
  \bibfield  {author} {\bibinfo {author} {\bibfnamefont {A.~N.}\ \bibnamefont
  {Kolmogorov}},\ }\href@noop {} {\bibfield  {journal} {\bibinfo  {journal}
  {Dokl. Akad. Nauk SSSR}\ }\textbf {\bibinfo {volume} {30}},\ \bibinfo {pages}
  {301} (\bibinfo {year} {1941})}\BibitemShut {NoStop}%
\bibitem [{\citenamefont {Obukhov}(1941)}]{Obukhov}%
  \BibitemOpen
  \bibfield  {author} {\bibinfo {author} {\bibfnamefont {A.~M.}\ \bibnamefont
  {Obukhov}},\ }\href@noop {} {\bibfield  {journal} {\bibinfo  {journal} {Dokl.
  Akad. Nauk SSSR}\ }\textbf {\bibinfo {volume} {32}},\ \bibinfo {pages} {22}
  (\bibinfo {year} {1941})}\BibitemShut {NoStop}%
\bibitem [{\citenamefont {Zakharov}\ \emph {et~al.}(2012)\citenamefont
  {Zakharov}, \citenamefont {L'vov},\ and\ \citenamefont {Falkovich}}]{ZLF}%
  \BibitemOpen
  \bibfield  {author} {\bibinfo {author} {\bibfnamefont {V.~E.}\ \bibnamefont
  {Zakharov}}, \bibinfo {author} {\bibfnamefont {V.~S.}\ \bibnamefont
  {L'vov}},\ and\ \bibinfo {author} {\bibfnamefont {G.}~\bibnamefont
  {Falkovich}},\ }\href@noop {} {\emph {\bibinfo {title} {{Kolmogorov spectra
  of turbulence I: Wave turbulence}}}}\ (\bibinfo  {publisher} {Springer
  Science \& Business Media},\ \bibinfo {year} {2012})\BibitemShut {NoStop}%
\bibitem [{\citenamefont {Zakharov}(2009)}]{zakharov2009turbulence}%
  \BibitemOpen
  \bibfield  {author} {\bibinfo {author} {\bibfnamefont {V.~E.}\ \bibnamefont
  {Zakharov}},\ }\href@noop {} {\bibfield  {journal} {\bibinfo  {journal}
  {Stud. Appl. Math.}\ }\textbf {\bibinfo {volume} {122}},\ \bibinfo {pages}
  {219} (\bibinfo {year} {2009})}\BibitemShut {NoStop}%
\bibitem [{\citenamefont {Suret}\ \emph {et~al.}(2011)\citenamefont {Suret},
  \citenamefont {Picozzi},\ and\ \citenamefont {Randoux}}]{suret2011wave}%
  \BibitemOpen
  \bibfield  {author} {\bibinfo {author} {\bibfnamefont {P.}~\bibnamefont
  {Suret}}, \bibinfo {author} {\bibfnamefont {A.}~\bibnamefont {Picozzi}},\
  and\ \bibinfo {author} {\bibfnamefont {S.}~\bibnamefont {Randoux}},\
  }\href@noop {} {\bibfield  {journal} {\bibinfo  {journal} {{Opt. Express}}\
  }\textbf {\bibinfo {volume} {19}},\ \bibinfo {pages} {17852} (\bibinfo {year}
  {2011})}\BibitemShut {NoStop}%
\bibitem [{\citenamefont {Picozzi}\ \emph {et~al.}(2014)\citenamefont
  {Picozzi}, \citenamefont {Garnier}, \citenamefont {Hansson}, \citenamefont
  {Suret}, \citenamefont {Randoux}, \citenamefont {Millot},\ and\ \citenamefont
  {Christodoulides}}]{picozzi2014optical}%
  \BibitemOpen
  \bibfield  {author} {\bibinfo {author} {\bibfnamefont {A.}~\bibnamefont
  {Picozzi}}, \bibinfo {author} {\bibfnamefont {J.}~\bibnamefont {Garnier}},
  \bibinfo {author} {\bibfnamefont {T.}~\bibnamefont {Hansson}}, \bibinfo
  {author} {\bibfnamefont {P.}~\bibnamefont {Suret}}, \bibinfo {author}
  {\bibfnamefont {S.}~\bibnamefont {Randoux}}, \bibinfo {author} {\bibfnamefont
  {G.}~\bibnamefont {Millot}},\ and\ \bibinfo {author} {\bibfnamefont {D.~N.}\
  \bibnamefont {Christodoulides}},\ }\href@noop {} {\bibfield  {journal}
  {\bibinfo  {journal} {{Phys. Rep.}}\ }\textbf {\bibinfo {volume} {542}},\
  \bibinfo {pages} {1} (\bibinfo {year} {2014})}\BibitemShut {NoStop}%
\bibitem [{\citenamefont {Walczak}\ \emph {et~al.}(2015)\citenamefont
  {Walczak}, \citenamefont {Randoux},\ and\ \citenamefont
  {Suret}}]{walczak2015optical}%
  \BibitemOpen
  \bibfield  {author} {\bibinfo {author} {\bibfnamefont {P.}~\bibnamefont
  {Walczak}}, \bibinfo {author} {\bibfnamefont {S.}~\bibnamefont {Randoux}},\
  and\ \bibinfo {author} {\bibfnamefont {P.}~\bibnamefont {Suret}},\
  }\href@noop {} {\bibfield  {journal} {\bibinfo  {journal} {{Phys. Rev.
  Lett.}}\ }\textbf {\bibinfo {volume} {114}},\ \bibinfo {pages} {143903}
  (\bibinfo {year} {2015})}\BibitemShut {NoStop}%
\bibitem [{\citenamefont {Agafontsev}\ and\ \citenamefont
  {Zakharov}(2015)}]{agafontsev2015integrable}%
  \BibitemOpen
  \bibfield  {author} {\bibinfo {author} {\bibfnamefont {D.~S.}\ \bibnamefont
  {Agafontsev}}\ and\ \bibinfo {author} {\bibfnamefont {V.~E.}\ \bibnamefont
  {Zakharov}},\ }\href@noop {} {\bibfield  {journal} {\bibinfo  {journal}
  {Nonlinearity}\ }\textbf {\bibinfo {volume} {28}},\ \bibinfo {pages} {2791}
  (\bibinfo {year} {2015})}\BibitemShut {NoStop}%
\bibitem [{\citenamefont {Agafontsev}\ and\ \citenamefont
  {Zakharov}(2016)}]{agafontsev2016integrable}%
  \BibitemOpen
  \bibfield  {author} {\bibinfo {author} {\bibfnamefont {D.~S.}\ \bibnamefont
  {Agafontsev}}\ and\ \bibinfo {author} {\bibfnamefont {V.~E.}\ \bibnamefont
  {Zakharov}},\ }\href@noop {} {\bibfield  {journal} {\bibinfo  {journal}
  {Nonlinearity}\ }\textbf {\bibinfo {volume} {29}},\ \bibinfo {pages} {3551}
  (\bibinfo {year} {2016})}\BibitemShut {NoStop}%
\bibitem [{\citenamefont {Gelash}\ and\ \citenamefont
  {Agafontsev}(2018)}]{gelash2018strongly}%
  \BibitemOpen
  \bibfield  {author} {\bibinfo {author} {\bibfnamefont {A.~A.}\ \bibnamefont
  {Gelash}}\ and\ \bibinfo {author} {\bibfnamefont {D.~S.}\ \bibnamefont
  {Agafontsev}},\ }\href@noop {} {\bibfield  {journal} {\bibinfo  {journal}
  {Phys. Rev. E}\ }\textbf {\bibinfo {volume} {98}},\ \bibinfo {pages} {042210}
  (\bibinfo {year} {2018})}\BibitemShut {NoStop}%
\bibitem [{\citenamefont {Agafontsev}\ \emph {et~al.}(2021)\citenamefont
  {Agafontsev}, \citenamefont {Randoux},\ and\ \citenamefont
  {Suret}}]{agafontsev2020integrable}%
  \BibitemOpen
  \bibfield  {author} {\bibinfo {author} {\bibfnamefont {D.~S.}\ \bibnamefont
  {Agafontsev}}, \bibinfo {author} {\bibfnamefont {S.}~\bibnamefont
  {Randoux}},\ and\ \bibinfo {author} {\bibfnamefont {P.}~\bibnamefont
  {Suret}},\ }\href@noop {} {\bibfield  {journal} {\bibinfo  {journal} {Phys.
  Rev. E}\ }\textbf {\bibinfo {volume} {103}},\ \bibinfo {pages} {032209}
  (\bibinfo {year} {2021})}\BibitemShut {NoStop}%
\bibitem [{\citenamefont {Zakharov}\ and\ \citenamefont
  {Kuznetsov}(2012)}]{ZakharovKuznetsov2012}%
  \BibitemOpen
  \bibfield  {author} {\bibinfo {author} {\bibfnamefont {V.~E.}\ \bibnamefont
  {Zakharov}}\ and\ \bibinfo {author} {\bibfnamefont {E.~A.}\ \bibnamefont
  {Kuznetsov}},\ }\href@noop {} {\bibfield  {journal} {\bibinfo  {journal}
  {Physics-Uspekhi}\ }\textbf {\bibinfo {volume} {55}},\ \bibinfo {pages} {535}
  (\bibinfo {year} {2012})}\BibitemShut {NoStop}%
\bibitem [{\citenamefont {Zakharov}\ and\ \citenamefont
  {Kuznetsov}(1997)}]{ZakharovKuznetsov1997}%
  \BibitemOpen
  \bibfield  {author} {\bibinfo {author} {\bibfnamefont {V.~E.}\ \bibnamefont
  {Zakharov}}\ and\ \bibinfo {author} {\bibfnamefont {E.~A.}\ \bibnamefont
  {Kuznetsov}},\ }\href@noop {} {\bibfield  {journal} {\bibinfo  {journal}
  {Physics-Uspekhi}\ }\textbf {\bibinfo {volume} {40}},\ \bibinfo {pages}
  {1087} (\bibinfo {year} {1997})}\BibitemShut {NoStop}%
\bibitem [{\citenamefont {Landau}\ and\ \citenamefont
  {Lifshitz}(2013)}]{landau2013fluid}%
  \BibitemOpen
  \bibfield  {author} {\bibinfo {author} {\bibfnamefont {L.~D.}\ \bibnamefont
  {Landau}}\ and\ \bibinfo {author} {\bibfnamefont {E.~M.}\ \bibnamefont
  {Lifshitz}},\ }\href@noop {} {\emph {\bibinfo {title} {{Fluid Mechanics.
  Course of Theoretical Physics}}}},\ Vol.~\bibinfo {volume} {6}\ (\bibinfo
  {publisher} {Elsevier},\ \bibinfo {year} {2013})\BibitemShut {NoStop}%
\bibitem [{\citenamefont {Arnol'd}(2013)}]{arnold1979mathematical}%
  \BibitemOpen
  \bibfield  {author} {\bibinfo {author} {\bibfnamefont {V.~I.}\ \bibnamefont
  {Arnol'd}},\ }\href@noop {} {\emph {\bibinfo {title} {{Mathematical methods
  of classical mechanics}}}},\ Vol.~\bibinfo {volume} {60}\ (\bibinfo
  {publisher} {Springer Science \& Business Media},\ \bibinfo {year}
  {2013})\BibitemShut {NoStop}%
\bibitem [{\citenamefont {Kuznetsov}\ and\ \citenamefont
  {Mikhailov}(1980)}]{KuznetsovMikhailov}%
  \BibitemOpen
  \bibfield  {author} {\bibinfo {author} {\bibfnamefont {E.~A.}\ \bibnamefont
  {Kuznetsov}}\ and\ \bibinfo {author} {\bibfnamefont {A.~V.}\ \bibnamefont
  {Mikhailov}},\ }\href@noop {} {\bibfield  {journal} {\bibinfo  {journal}
  {Phys. Lett. A}\ }\textbf {\bibinfo {volume} {77}},\ \bibinfo {pages} {37}
  (\bibinfo {year} {1980})}\BibitemShut {NoStop}%
\bibitem [{\citenamefont {Moreau}(1961)}]{fr}%
  \BibitemOpen
  \bibfield  {author} {\bibinfo {author} {\bibfnamefont {J.~J.}\ \bibnamefont
  {Moreau}},\ }\href@noop {} {\bibfield  {journal} {\bibinfo  {journal} {C. r.
  hebd. s{\'e}ances Acad. sci.}\ }\textbf {\bibinfo {volume} {252}},\ \bibinfo
  {pages} {2810} (\bibinfo {year} {1961})}\BibitemShut {NoStop}%
\bibitem [{\citenamefont {Arnol'd}(1969)}]{arnold1969hamiltonian}%
  \BibitemOpen
  \bibfield  {author} {\bibinfo {author} {\bibfnamefont {V.~I.}\ \bibnamefont
  {Arnol'd}},\ }\href@noop {} {\bibfield  {journal} {\bibinfo  {journal}
  {Uspekhi Mat. Nauk}\ }\textbf {\bibinfo {volume} {24}},\ \bibinfo {pages}
  {225} (\bibinfo {year} {1969})}\BibitemShut {NoStop}%
\bibitem [{\citenamefont {Yakubovich}\ and\ \citenamefont
  {Zenkovich}(2001{\natexlab{a}})}]{YakubovichZenkovich}%
  \BibitemOpen
  \bibfield  {author} {\bibinfo {author} {\bibfnamefont {E.~I.}\ \bibnamefont
  {Yakubovich}}\ and\ \bibinfo {author} {\bibfnamefont {D.~A.}\ \bibnamefont
  {Zenkovich}},\ }\href@noop {} {\bibfield  {journal} {\bibinfo  {journal}
  {arXiv preprint physics/0110004}\ } (\bibinfo {year}
  {2001}{\natexlab{a}})}\BibitemShut {NoStop}%
\bibitem [{\citenamefont {Kuznetsov}(2002)}]{Kuznetsov2002}%
  \BibitemOpen
  \bibfield  {author} {\bibinfo {author} {\bibfnamefont {E.~A.}\ \bibnamefont
  {Kuznetsov}},\ }\href@noop {} {\bibfield  {journal} {\bibinfo  {journal}
  {JETP Lett.}\ }\textbf {\bibinfo {volume} {76}},\ \bibinfo {pages} {346}
  (\bibinfo {year} {2002})}\BibitemShut {NoStop}%
\bibitem [{\citenamefont {Frisch}\ and\ \citenamefont
  {Villone}(2014)}]{frisch2014cauchy}%
  \BibitemOpen
  \bibfield  {author} {\bibinfo {author} {\bibfnamefont {U.}~\bibnamefont
  {Frisch}}\ and\ \bibinfo {author} {\bibfnamefont {B.}~\bibnamefont
  {Villone}},\ }\href@noop {} {\bibfield  {journal} {\bibinfo  {journal} {Eur.
  Phys. J. H}\ }\textbf {\bibinfo {volume} {39}},\ \bibinfo {pages} {325}
  (\bibinfo {year} {2014})}\BibitemShut {NoStop}%
\bibitem [{\citenamefont {Kuznetsov}\ and\ \citenamefont
  {Ruban}(1998)}]{KuznetsovRuban}%
  \BibitemOpen
  \bibfield  {author} {\bibinfo {author} {\bibfnamefont {E.~A.}\ \bibnamefont
  {Kuznetsov}}\ and\ \bibinfo {author} {\bibfnamefont {V.~P.}\ \bibnamefont
  {Ruban}},\ }\href@noop {} {\bibfield  {journal} {\bibinfo  {journal} {JETP
  Lett.}\ }\textbf {\bibinfo {volume} {67}},\ \bibinfo {pages} {1076} (\bibinfo
  {year} {1998})}\BibitemShut {NoStop}%
\bibitem [{\citenamefont {Kuznetsov}\ and\ \citenamefont
  {Ruban}(2000{\natexlab{a}})}]{KuznetsovRuban2000}%
  \BibitemOpen
  \bibfield  {author} {\bibinfo {author} {\bibfnamefont {E.~A.}\ \bibnamefont
  {Kuznetsov}}\ and\ \bibinfo {author} {\bibfnamefont {V.~P.}\ \bibnamefont
  {Ruban}},\ }\href@noop {} {\bibfield  {journal} {\bibinfo  {journal} {JETP}\
  }\textbf {\bibinfo {volume} {91}},\ \bibinfo {pages} {775} (\bibinfo {year}
  {2000}{\natexlab{a}})}\BibitemShut {NoStop}%
\bibitem [{\citenamefont {Arnold}(1992)}]{arnold}%
  \BibitemOpen
  \bibfield  {author} {\bibinfo {author} {\bibfnamefont {V.~I.}\ \bibnamefont
  {Arnold}},\ }\href@noop {} {\emph {\bibinfo {title} {{Catastrophe theory}}}}\
  (\bibinfo  {publisher} {Springer},\ \bibinfo {year} {1992})\BibitemShut
  {NoStop}%
\bibitem [{\citenamefont {Yakubovich}\ and\ \citenamefont
  {Zenkovich}(2001{\natexlab{b}})}]{yakubovich2001matrix}%
  \BibitemOpen
  \bibfield  {author} {\bibinfo {author} {\bibfnamefont {E.~I.}\ \bibnamefont
  {Yakubovich}}\ and\ \bibinfo {author} {\bibfnamefont {D.~A.}\ \bibnamefont
  {Zenkovich}},\ }\href@noop {} {\bibfield  {journal} {\bibinfo  {journal} {J.
  Fluid Mech.}\ }\textbf {\bibinfo {volume} {443}},\ \bibinfo {pages} {167}
  (\bibinfo {year} {2001}{\natexlab{b}})}\BibitemShut {NoStop}%
\bibitem [{\citenamefont {Kuznetsov}(2006)}]{Kuznetsov2006}%
  \BibitemOpen
  \bibfield  {author} {\bibinfo {author} {\bibfnamefont {E.~A.}\ \bibnamefont
  {Kuznetsov}},\ }\href@noop {} {\bibfield  {journal} {\bibinfo  {journal}
  {JNMP}\ }\textbf {\bibinfo {volume} {13}},\ \bibinfo {pages} {64} (\bibinfo
  {year} {2006})}\BibitemShut {NoStop}%
\bibitem [{\citenamefont {Brachet}\ \emph {et~al.}(1988)\citenamefont
  {Brachet}, \citenamefont {Meneguzzi}, \citenamefont {Politano},\ and\
  \citenamefont {Sulem}}]{Sulem}%
  \BibitemOpen
  \bibfield  {author} {\bibinfo {author} {\bibfnamefont {M.~E.}\ \bibnamefont
  {Brachet}}, \bibinfo {author} {\bibfnamefont {M.}~\bibnamefont {Meneguzzi}},
  \bibinfo {author} {\bibfnamefont {H.}~\bibnamefont {Politano}},\ and\
  \bibinfo {author} {\bibfnamefont {P.~L.}\ \bibnamefont {Sulem}},\ }\href@noop
  {} {\bibfield  {journal} {\bibinfo  {journal} {J. Fluid Mech.}\ }\textbf
  {\bibinfo {volume} {194}},\ \bibinfo {pages} {333} (\bibinfo {year}
  {1988})}\BibitemShut {NoStop}%
\bibitem [{\citenamefont {Weiss}(1991)}]{weiss}%
  \BibitemOpen
  \bibfield  {author} {\bibinfo {author} {\bibfnamefont {J.}~\bibnamefont
  {Weiss}},\ }\href@noop {} {\bibfield  {journal} {\bibinfo  {journal} {Physica
  D: Nonlinear Phenomena}\ }\textbf {\bibinfo {volume} {48}},\ \bibinfo {pages}
  {273} (\bibinfo {year} {1991})}\BibitemShut {NoStop}%
\bibitem [{\citenamefont {Kuznetsov}\ \emph {et~al.}(2007)\citenamefont
  {Kuznetsov}, \citenamefont {Naulin}, \citenamefont {Nielsen},\ and\
  \citenamefont {Rasmussen}}]{kuznetsov2007effects}%
  \BibitemOpen
  \bibfield  {author} {\bibinfo {author} {\bibfnamefont {E.~A.}\ \bibnamefont
  {Kuznetsov}}, \bibinfo {author} {\bibfnamefont {V.}~\bibnamefont {Naulin}},
  \bibinfo {author} {\bibfnamefont {A.~H.}\ \bibnamefont {Nielsen}},\ and\
  \bibinfo {author} {\bibfnamefont {J.~J.}\ \bibnamefont {Rasmussen}},\
  }\href@noop {} {\bibfield  {journal} {\bibinfo  {journal} {Phys. Fluids}\
  }\textbf {\bibinfo {volume} {19}},\ \bibinfo {pages} {105110} (\bibinfo
  {year} {2007})}\BibitemShut {NoStop}%
\bibitem [{\citenamefont {Hasimoto}(1972)}]{hasimoto}%
  \BibitemOpen
  \bibfield  {author} {\bibinfo {author} {\bibfnamefont {H.}~\bibnamefont
  {Hasimoto}},\ }\href@noop {} {\bibfield  {journal} {\bibinfo  {journal} {J.
  Fluid Mech.}\ }\textbf {\bibinfo {volume} {51}},\ \bibinfo {pages} {477}
  (\bibinfo {year} {1972})}\BibitemShut {NoStop}%
\bibitem [{\citenamefont {Zakharov}\ and\ \citenamefont
  {Takhtadzhyan}(1979)}]{ZakharovTakhtadzhyan}%
  \BibitemOpen
  \bibfield  {author} {\bibinfo {author} {\bibfnamefont {V.~E.}\ \bibnamefont
  {Zakharov}}\ and\ \bibinfo {author} {\bibfnamefont {L.~A.}\ \bibnamefont
  {Takhtadzhyan}},\ }\href@noop {} {\bibfield  {journal} {\bibinfo  {journal}
  {Theor. Math. Phys.}\ }\textbf {\bibinfo {volume} {38}},\ \bibinfo {pages}
  {17} (\bibinfo {year} {1979})}\BibitemShut {NoStop}%
\bibitem [{\citenamefont {Chae}(2008)}]{chae2008incompressible}%
  \BibitemOpen
  \bibfield  {author} {\bibinfo {author} {\bibfnamefont {D.}~\bibnamefont
  {Chae}},\ }\bibinfo {title} {{Incompressible Euler Equations: the blow-up
  problem and related results. In: Handbook of Differential Equations:
  Evolutionary Equation (C.M. Dafermos and M. Pokorny, Eds.), Vol. 4}},\ in\
  \href@noop {} {\emph {\bibinfo {booktitle} {{Handbook of Differential
  Equations: Evolutionary Equations}}}}\ (\bibinfo  {publisher} {Elsevier},\
  \bibinfo {year} {2008})\ pp.\ \bibinfo {pages} {1--55}\BibitemShut {NoStop}%
\bibitem [{\citenamefont {Gibbon}(2008)}]{gibbon2008three}%
  \BibitemOpen
  \bibfield  {author} {\bibinfo {author} {\bibfnamefont {J.~D.}\ \bibnamefont
  {Gibbon}},\ }\href@noop {} {\bibfield  {journal} {\bibinfo  {journal}
  {Physica D}\ }\textbf {\bibinfo {volume} {237}},\ \bibinfo {pages} {1894}
  (\bibinfo {year} {2008})}\BibitemShut {NoStop}%
\bibitem [{\citenamefont {Agafontsev}\ \emph {et~al.}(2015)\citenamefont
  {Agafontsev}, \citenamefont {Kuznetsov},\ and\ \citenamefont
  {Mailybaev}}]{agafontsev2015}%
  \BibitemOpen
  \bibfield  {author} {\bibinfo {author} {\bibfnamefont {D.~S.}\ \bibnamefont
  {Agafontsev}}, \bibinfo {author} {\bibfnamefont {E.~A.}\ \bibnamefont
  {Kuznetsov}},\ and\ \bibinfo {author} {\bibfnamefont {A.~A.}\ \bibnamefont
  {Mailybaev}},\ }\href@noop {} {\bibfield  {journal} {\bibinfo  {journal}
  {Phys. Fluids}\ }\textbf {\bibinfo {volume} {27}},\ \bibinfo {pages} {085102}
  (\bibinfo {year} {2015})}\BibitemShut {NoStop}%
\bibitem [{\citenamefont {Agafontsev}\ \emph {et~al.}(2016)\citenamefont
  {Agafontsev}, \citenamefont {Kuznetsov},\ and\ \citenamefont
  {Mailybaev}}]{agafontsev2016development}%
  \BibitemOpen
  \bibfield  {author} {\bibinfo {author} {\bibfnamefont {D.~S.}\ \bibnamefont
  {Agafontsev}}, \bibinfo {author} {\bibfnamefont {E.~A.}\ \bibnamefont
  {Kuznetsov}},\ and\ \bibinfo {author} {\bibfnamefont {A.~A.}\ \bibnamefont
  {Mailybaev}},\ }\href@noop {} {\bibfield  {journal} {\bibinfo  {journal}
  {JETP letters}\ }\textbf {\bibinfo {volume} {104}},\ \bibinfo {pages} {775}
  (\bibinfo {year} {2016})}\BibitemShut {NoStop}%
\bibitem [{\citenamefont {Agafontsev}\ \emph {et~al.}(2017)\citenamefont
  {Agafontsev}, \citenamefont {Kuznetsov},\ and\ \citenamefont
  {Mailybaev}}]{agafontsev2016asymptotic}%
  \BibitemOpen
  \bibfield  {author} {\bibinfo {author} {\bibfnamefont {D.~S.}\ \bibnamefont
  {Agafontsev}}, \bibinfo {author} {\bibfnamefont {E.~A.}\ \bibnamefont
  {Kuznetsov}},\ and\ \bibinfo {author} {\bibfnamefont {A.~A.}\ \bibnamefont
  {Mailybaev}},\ }\href@noop {} {\bibfield  {journal} {\bibinfo  {journal} {J.
  Fluid Mech.}\ }\textbf {\bibinfo {volume} {813}},\ \bibinfo {pages} {R1}
  (\bibinfo {year} {2017})}\BibitemShut {NoStop}%
\bibitem [{\citenamefont {Brachet}\ \emph
  {et~al.}(1992{\natexlab{a}})\citenamefont {Brachet}, \citenamefont
  {Meneguzzi}, \citenamefont {Vincent}, \citenamefont {Politano},\ and\
  \citenamefont {Sulem}}]{Brachet}%
  \BibitemOpen
  \bibfield  {author} {\bibinfo {author} {\bibfnamefont {M.~E.}\ \bibnamefont
  {Brachet}}, \bibinfo {author} {\bibfnamefont {M.}~\bibnamefont {Meneguzzi}},
  \bibinfo {author} {\bibfnamefont {A.}~\bibnamefont {Vincent}}, \bibinfo
  {author} {\bibfnamefont {H.}~\bibnamefont {Politano}},\ and\ \bibinfo
  {author} {\bibfnamefont {P.~L.}\ \bibnamefont {Sulem}},\ }\href@noop {}
  {\bibfield  {journal} {\bibinfo  {journal} {Phys. Fluids A}\ }\textbf
  {\bibinfo {volume} {4}},\ \bibinfo {pages} {2845} (\bibinfo {year}
  {1992}{\natexlab{a}})}\BibitemShut {NoStop}%
\bibitem [{\citenamefont {Agafontsev}\ \emph {et~al.}(2019)\citenamefont
  {Agafontsev}, \citenamefont {Kuznetsov},\ and\ \citenamefont
  {Mailybaev}}]{AgafontsevKuznetsovMailybaev2019}%
  \BibitemOpen
  \bibfield  {author} {\bibinfo {author} {\bibfnamefont {D.~S.}\ \bibnamefont
  {Agafontsev}}, \bibinfo {author} {\bibfnamefont {E.~A.}\ \bibnamefont
  {Kuznetsov}},\ and\ \bibinfo {author} {\bibfnamefont {A.~A.}\ \bibnamefont
  {Mailybaev}},\ }\href@noop {} {\bibfield  {journal} {\bibinfo  {journal}
  {JETP Lett.}\ }\textbf {\bibinfo {volume} {110}},\ \bibinfo {pages} {121}
  (\bibinfo {year} {2019})}\BibitemShut {NoStop}%
\bibitem [{\citenamefont {Kuznetsov}\ \emph {et~al.}(2004)\citenamefont
  {Kuznetsov}, \citenamefont {Passot},\ and\ \citenamefont {Sulem}}]{MHD}%
  \BibitemOpen
  \bibfield  {author} {\bibinfo {author} {\bibfnamefont {E.~A.}\ \bibnamefont
  {Kuznetsov}}, \bibinfo {author} {\bibfnamefont {T.}~\bibnamefont {Passot}},\
  and\ \bibinfo {author} {\bibfnamefont {P.~L.}\ \bibnamefont {Sulem}},\
  }\href@noop {} {\bibfield  {journal} {\bibinfo  {journal} {Phys. Plasmas}\
  }\textbf {\bibinfo {volume} {11}},\ \bibinfo {pages} {1410} (\bibinfo {year}
  {2004})}\BibitemShut {NoStop}%
\bibitem [{\citenamefont {Kuznetsov}(2008)}]{Kuznetsov2008}%
  \BibitemOpen
  \bibfield  {author} {\bibinfo {author} {\bibfnamefont {E.~A.}\ \bibnamefont
  {Kuznetsov}},\ }\href@noop {} {\bibfield  {journal} {\bibinfo  {journal} {J.
  Fluid Mech.}\ }\textbf {\bibinfo {volume} {600}},\ \bibinfo {pages} {167}
  (\bibinfo {year} {2008})}\BibitemShut {NoStop}%
\bibitem [{\citenamefont {Shandarin}\ and\ \citenamefont
  {Zeldovich}(1989)}]{shandarin1989large}%
  \BibitemOpen
  \bibfield  {author} {\bibinfo {author} {\bibfnamefont {S.~F.}\ \bibnamefont
  {Shandarin}}\ and\ \bibinfo {author} {\bibfnamefont {Y.~B.}\ \bibnamefont
  {Zeldovich}},\ }\href@noop {} {\bibfield  {journal} {\bibinfo  {journal}
  {Rev. Mod. Phys.}\ }\textbf {\bibinfo {volume} {61}},\ \bibinfo {pages} {185}
  (\bibinfo {year} {1989})}\BibitemShut {NoStop}%
\bibitem [{\citenamefont {Gurbatov}\ \emph {et~al.}(2012)\citenamefont
  {Gurbatov}, \citenamefont {Saichev},\ and\ \citenamefont
  {Shandarin}}]{GurbatovSaichevShandarin}%
  \BibitemOpen
  \bibfield  {author} {\bibinfo {author} {\bibfnamefont {S.~N.}\ \bibnamefont
  {Gurbatov}}, \bibinfo {author} {\bibfnamefont {A.~I.}\ \bibnamefont
  {Saichev}},\ and\ \bibinfo {author} {\bibfnamefont {S.~F.}\ \bibnamefont
  {Shandarin}},\ }\href@noop {} {\bibfield  {journal} {\bibinfo  {journal}
  {Physics-Uspekhi}\ }\textbf {\bibinfo {volume} {55}},\ \bibinfo {pages} {223}
  (\bibinfo {year} {2012})}\BibitemShut {NoStop}%
\bibitem [{\citenamefont {Kudryavtsev}\ \emph {et~al.}(2013)\citenamefont
  {Kudryavtsev}, \citenamefont {Kuznetsov},\ and\ \citenamefont
  {Sereshchenko}}]{kudryavtsev2013statistical}%
  \BibitemOpen
  \bibfield  {author} {\bibinfo {author} {\bibfnamefont {A.~N.}\ \bibnamefont
  {Kudryavtsev}}, \bibinfo {author} {\bibfnamefont {E.~A.}\ \bibnamefont
  {Kuznetsov}},\ and\ \bibinfo {author} {\bibfnamefont {E.~V.}\ \bibnamefont
  {Sereshchenko}},\ }\href@noop {} {\bibfield  {journal} {\bibinfo  {journal}
  {JETP Lett.}\ }\textbf {\bibinfo {volume} {96}},\ \bibinfo {pages} {699}
  (\bibinfo {year} {2013})}\BibitemShut {NoStop}%
\bibitem [{\citenamefont {Kuznetsov}\ and\ \citenamefont
  {Sereshchenko}(2015)}]{kuznetsov2015anisotropic}%
  \BibitemOpen
  \bibfield  {author} {\bibinfo {author} {\bibfnamefont {E.~A.}\ \bibnamefont
  {Kuznetsov}}\ and\ \bibinfo {author} {\bibfnamefont {E.~V.}\ \bibnamefont
  {Sereshchenko}},\ }\href@noop {} {\bibfield  {journal} {\bibinfo  {journal}
  {JETP Lett.}\ }\textbf {\bibinfo {volume} {102}},\ \bibinfo {pages} {760–}
  (\bibinfo {year} {2015})}\BibitemShut {NoStop}%
\bibitem [{\citenamefont {Kuznetsov}\ and\ \citenamefont
  {Sereshchenko}(2017)}]{KuznetsovSereshchenko2017}%
  \BibitemOpen
  \bibfield  {author} {\bibinfo {author} {\bibfnamefont {E.~A.}\ \bibnamefont
  {Kuznetsov}}\ and\ \bibinfo {author} {\bibfnamefont {E.~V.}\ \bibnamefont
  {Sereshchenko}},\ }\href@noop {} {\bibfield  {journal} {\bibinfo  {journal}
  {JETP Lett.}\ }\textbf {\bibinfo {volume} {105}},\ \bibinfo {pages} {83}
  (\bibinfo {year} {2017})}\BibitemShut {NoStop}%
\bibitem [{\citenamefont {Kuznetsov}(2017)}]{Kuznetsov-2016}%
  \BibitemOpen
  \bibfield  {author} {\bibinfo {author} {\bibfnamefont {E.~A.}\ \bibnamefont
  {Kuznetsov}},\ }in\ \href@noop {} {\emph {\bibinfo {booktitle} {Nonlinear
  waves 2016}}},\ \bibinfo {editor} {edited by\ \bibinfo {editor}
  {\bibfnamefont {A.~M.}\ \bibnamefont {Sergeev}}\ and\ \bibinfo {editor}
  {\bibfnamefont {A.~V.}\ \bibnamefont {Slunyaev}}}\ (\bibinfo  {publisher}
  {IAP RAS, Nizhny Novgorod},\ \bibinfo {year} {2017})\ p.\ \bibinfo {pages}
  {304}\BibitemShut {NoStop}%
\bibitem [{\citenamefont {Kuznetsov}(2019)}]{Kuznetsov-2018}%
  \BibitemOpen
  \bibfield  {author} {\bibinfo {author} {\bibfnamefont {E.~A.}\ \bibnamefont
  {Kuznetsov}},\ }in\ \href@noop {} {\emph {\bibinfo {booktitle} {Nonlinear
  waves 2018}}},\ \bibinfo {editor} {edited by\ \bibinfo {editor}
  {\bibfnamefont {A.~G.}\ \bibnamefont {Litvak}}\ and\ \bibinfo {editor}
  {\bibfnamefont {A.~V.}\ \bibnamefont {Slunyaev}}}\ (\bibinfo  {publisher}
  {IAP RAS, Nizhny Novgorod},\ \bibinfo {year} {2019})\ p.\ \bibinfo {pages}
  {238}\BibitemShut {NoStop}%
\bibitem [{\citenamefont {Salmon}(1988)}]{salmon}%
  \BibitemOpen
  \bibfield  {author} {\bibinfo {author} {\bibfnamefont {R.}~\bibnamefont
  {Salmon}},\ }\href@noop {} {\bibfield  {journal} {\bibinfo  {journal} {Annu.
  Rev. Fluid Mech.}\ }\textbf {\bibinfo {volume} {20}},\ \bibinfo {pages} {225}
  (\bibinfo {year} {1988})}\BibitemShut {NoStop}%
\bibitem [{\citenamefont {Kuznetsov}\ and\ \citenamefont
  {Ruban}(2000{\natexlab{b}})}]{KuznetsovRubanPRE}%
  \BibitemOpen
  \bibfield  {author} {\bibinfo {author} {\bibfnamefont {E.~A.}\ \bibnamefont
  {Kuznetsov}}\ and\ \bibinfo {author} {\bibfnamefont {V.~P.}\ \bibnamefont
  {Ruban}},\ }\href@noop {} {\bibfield  {journal} {\bibinfo  {journal} {Phys.
  Rev. E}\ }\textbf {\bibinfo {volume} {61}},\ \bibinfo {pages} {831} (\bibinfo
  {year} {2000}{\natexlab{b}})}\BibitemShut {NoStop}%
\bibitem [{\citenamefont {Agafontsev}\ \emph {et~al.}(2018)\citenamefont
  {Agafontsev}, \citenamefont {Kuznetsov},\ and\ \citenamefont
  {Mailybaev}}]{agafontsev2017universal}%
  \BibitemOpen
  \bibfield  {author} {\bibinfo {author} {\bibfnamefont {D.~S.}\ \bibnamefont
  {Agafontsev}}, \bibinfo {author} {\bibfnamefont {E.~A.}\ \bibnamefont
  {Kuznetsov}},\ and\ \bibinfo {author} {\bibfnamefont {A.~A.}\ \bibnamefont
  {Mailybaev}},\ }\href@noop {} {\bibfield  {journal} {\bibinfo  {journal}
  {Phys. Fluids}\ }\textbf {\bibinfo {volume} {30}},\ \bibinfo {pages} {095104}
  (\bibinfo {year} {2018})}\BibitemShut {NoStop}%
\bibitem [{\citenamefont {Frisch}(1999)}]{frisch1999turbulence}%
  \BibitemOpen
  \bibfield  {author} {\bibinfo {author} {\bibfnamefont {U.}~\bibnamefont
  {Frisch}},\ }\href@noop {} {\emph {\bibinfo {title} {{Turbulence: the legacy
  of A.N.~Kolmogorov}}}}\ (\bibinfo  {publisher} {Cambridge University Press},\
  \bibinfo {year} {1999})\BibitemShut {NoStop}%
\bibitem [{\citenamefont {Orlandi}\ and\ \citenamefont
  {Pirozzoli}(2010)}]{orlandipirozzoli2010}%
  \BibitemOpen
  \bibfield  {author} {\bibinfo {author} {\bibfnamefont {P.}~\bibnamefont
  {Orlandi}}\ and\ \bibinfo {author} {\bibfnamefont {S.}~\bibnamefont
  {Pirozzoli}},\ }\href@noop {} {\bibfield  {journal} {\bibinfo  {journal}
  {Theor. Comput. Fluid Dyn.}\ }\textbf {\bibinfo {volume} {24}},\ \bibinfo
  {pages} {247} (\bibinfo {year} {2010})}\BibitemShut {NoStop}%
\bibitem [{\citenamefont {Holm}\ and\ \citenamefont
  {Kerr}(2002)}]{holm2002transient}%
  \BibitemOpen
  \bibfield  {author} {\bibinfo {author} {\bibfnamefont {D.~D.}\ \bibnamefont
  {Holm}}\ and\ \bibinfo {author} {\bibfnamefont {R.~M.}\ \bibnamefont
  {Kerr}},\ }\href@noop {} {\bibfield  {journal} {\bibinfo  {journal} {Phys.
  Rev. Lett.}\ }\textbf {\bibinfo {volume} {88}},\ \bibinfo {pages} {244501}
  (\bibinfo {year} {2002})}\BibitemShut {NoStop}%
\bibitem [{\citenamefont {Cichowlas}\ \emph {et~al.}(2005)\citenamefont
  {Cichowlas}, \citenamefont {Bona{\"\i}ti}, \citenamefont {Debbasch},\ and\
  \citenamefont {Brachet}}]{cichowlas2005effective}%
  \BibitemOpen
  \bibfield  {author} {\bibinfo {author} {\bibfnamefont {C.}~\bibnamefont
  {Cichowlas}}, \bibinfo {author} {\bibfnamefont {P.}~\bibnamefont
  {Bona{\"\i}ti}}, \bibinfo {author} {\bibfnamefont {F.}~\bibnamefont
  {Debbasch}},\ and\ \bibinfo {author} {\bibfnamefont {M.}~\bibnamefont
  {Brachet}},\ }\href@noop {} {\bibfield  {journal} {\bibinfo  {journal} {Phys.
  Rev. Lett.}\ }\textbf {\bibinfo {volume} {95}},\ \bibinfo {pages} {264502}
  (\bibinfo {year} {2005})}\BibitemShut {NoStop}%
\bibitem [{\citenamefont {Holm}\ and\ \citenamefont {Kerr}(2007)}]{holm2007}%
  \BibitemOpen
  \bibfield  {author} {\bibinfo {author} {\bibfnamefont {D.~D.}\ \bibnamefont
  {Holm}}\ and\ \bibinfo {author} {\bibfnamefont {R.~M.}\ \bibnamefont
  {Kerr}},\ }\href@noop {} {\bibfield  {journal} {\bibinfo  {journal} {Phys.
  Fluids}\ }\textbf {\bibinfo {volume} {19}},\ \bibinfo {pages} {025101}
  (\bibinfo {year} {2007})}\BibitemShut {NoStop}%
\bibitem [{\citenamefont {Brachet}\ \emph
  {et~al.}(1992{\natexlab{b}})\citenamefont {Brachet}, \citenamefont
  {Meneguzzi}, \citenamefont {Vincent}, \citenamefont {Politano},\ and\
  \citenamefont {Sulem}}]{brachet1992numerical}%
  \BibitemOpen
  \bibfield  {author} {\bibinfo {author} {\bibfnamefont {M.~E.}\ \bibnamefont
  {Brachet}}, \bibinfo {author} {\bibfnamefont {M.}~\bibnamefont {Meneguzzi}},
  \bibinfo {author} {\bibfnamefont {A.}~\bibnamefont {Vincent}}, \bibinfo
  {author} {\bibfnamefont {H.}~\bibnamefont {Politano}},\ and\ \bibinfo
  {author} {\bibfnamefont {P.~L.}\ \bibnamefont {Sulem}},\ }\href@noop {}
  {\bibfield  {journal} {\bibinfo  {journal} {Phys. Fluids A}\ }\textbf
  {\bibinfo {volume} {4}},\ \bibinfo {pages} {2845} (\bibinfo {year}
  {1992}{\natexlab{b}})}\BibitemShut {NoStop}%
\bibitem [{\citenamefont {Ishihara}\ \emph {et~al.}(2009)\citenamefont
  {Ishihara}, \citenamefont {Gotoh},\ and\ \citenamefont
  {Kaneda}}]{ishihara2009study}%
  \BibitemOpen
  \bibfield  {author} {\bibinfo {author} {\bibfnamefont {T.}~\bibnamefont
  {Ishihara}}, \bibinfo {author} {\bibfnamefont {T.}~\bibnamefont {Gotoh}},\
  and\ \bibinfo {author} {\bibfnamefont {Y.}~\bibnamefont {Kaneda}},\
  }\href@noop {} {\bibfield  {journal} {\bibinfo  {journal} {Annu. Rev. Fluid
  Mech.}\ }\textbf {\bibinfo {volume} {41}},\ \bibinfo {pages} {165} (\bibinfo
  {year} {2009})}\BibitemShut {NoStop}%
\bibitem [{\citenamefont {Gotoh}\ \emph {et~al.}(2002)\citenamefont {Gotoh},
  \citenamefont {Fukayama},\ and\ \citenamefont {Nakano}}]{gotoh2002velocity}%
  \BibitemOpen
  \bibfield  {author} {\bibinfo {author} {\bibfnamefont {T.}~\bibnamefont
  {Gotoh}}, \bibinfo {author} {\bibfnamefont {D.}~\bibnamefont {Fukayama}},\
  and\ \bibinfo {author} {\bibfnamefont {T.}~\bibnamefont {Nakano}},\
  }\href@noop {} {\bibfield  {journal} {\bibinfo  {journal} {Phys. Fluids}\
  }\textbf {\bibinfo {volume} {14}},\ \bibinfo {pages} {1065} (\bibinfo {year}
  {2002})}\BibitemShut {NoStop}%
\bibitem [{\citenamefont {Zybin}\ and\ \citenamefont
  {Sirota}(2015)}]{zybin2015stretching}%
  \BibitemOpen
  \bibfield  {author} {\bibinfo {author} {\bibfnamefont {K.~P.}\ \bibnamefont
  {Zybin}}\ and\ \bibinfo {author} {\bibfnamefont {V.~A.}\ \bibnamefont
  {Sirota}},\ }\href@noop {} {\bibfield  {journal} {\bibinfo  {journal} {Phys.
  Usp.}\ }\textbf {\bibinfo {volume} {58}},\ \bibinfo {pages} {556–}
  (\bibinfo {year} {2015})}\BibitemShut {NoStop}%
\bibitem [{\citenamefont {Kraichnan}(1967)}]{kraichnan}%
  \BibitemOpen
  \bibfield  {author} {\bibinfo {author} {\bibfnamefont {R.~H.}\ \bibnamefont
  {Kraichnan}},\ }\href@noop {} {\bibfield  {journal} {\bibinfo  {journal}
  {Phys. Fluids}\ }\textbf {\bibinfo {volume} {10}},\ \bibinfo {pages} {1417}
  (\bibinfo {year} {1967})}\BibitemShut {NoStop}%
\bibitem [{\citenamefont {Boffetta}\ and\ \citenamefont
  {Ecke}(2012)}]{boffetta}%
  \BibitemOpen
  \bibfield  {author} {\bibinfo {author} {\bibfnamefont {G.}~\bibnamefont
  {Boffetta}}\ and\ \bibinfo {author} {\bibfnamefont {R.~E.}\ \bibnamefont
  {Ecke}},\ }\href@noop {} {\bibfield  {journal} {\bibinfo  {journal} {Annu.
  Rev. Fluid Mech.}\ }\textbf {\bibinfo {volume} {44}},\ \bibinfo {pages} {427}
  (\bibinfo {year} {2012})}\BibitemShut {NoStop}%
\bibitem [{\citenamefont {Lilly}(1971)}]{lilly}%
  \BibitemOpen
  \bibfield  {author} {\bibinfo {author} {\bibfnamefont {D.~K.}\ \bibnamefont
  {Lilly}},\ }\href@noop {} {\bibfield  {journal} {\bibinfo  {journal} {J.
  Fluid Mech.}\ }\textbf {\bibinfo {volume} {45}},\ \bibinfo {pages} {395}
  (\bibinfo {year} {1971})}\BibitemShut {NoStop}%
\bibitem [{\citenamefont {Saffman}(1971)}]{saffman}%
  \BibitemOpen
  \bibfield  {author} {\bibinfo {author} {\bibfnamefont {P.~G.}\ \bibnamefont
  {Saffman}},\ }\href@noop {} {\bibfield  {journal} {\bibinfo  {journal} {Stud.
  Appl. Math.}\ }\textbf {\bibinfo {volume} {50}},\ \bibinfo {pages} {377}
  (\bibinfo {year} {1971})}\BibitemShut {NoStop}%
\bibitem [{\citenamefont {Kadomtsev}\ and\ \citenamefont
  {Petviashvili}(1973)}]{KP}%
  \BibitemOpen
  \bibfield  {author} {\bibinfo {author} {\bibfnamefont {B.~B.}\ \bibnamefont
  {Kadomtsev}}\ and\ \bibinfo {author} {\bibfnamefont {V.~I.}\ \bibnamefont
  {Petviashvili}},\ }in\ \href@noop {} {\emph {\bibinfo {booktitle} {Dokl.
  Akad. Nauk SSSR}}},\ Vol.\ \bibinfo {volume} {208}\ (\bibinfo {organization}
  {Russian Academy of Sciences},\ \bibinfo {year} {1973})\ pp.\ \bibinfo
  {pages} {794--796}\BibitemShut {NoStop}%
\bibitem [{\citenamefont {Kuznetsov}(2004)}]{K-04}%
  \BibitemOpen
  \bibfield  {author} {\bibinfo {author} {\bibfnamefont {E.~A.}\ \bibnamefont
  {Kuznetsov}},\ }\href@noop {} {\bibfield  {journal} {\bibinfo  {journal}
  {JETP Lett.}\ }\textbf {\bibinfo {volume} {80}},\ \bibinfo {pages} {83}
  (\bibinfo {year} {2004})}\BibitemShut {NoStop}%
\bibitem [{\citenamefont {Wolibner}(1933)}]{wolibner}%
  \BibitemOpen
  \bibfield  {author} {\bibinfo {author} {\bibfnamefont {W.}~\bibnamefont
  {Wolibner}},\ }\href@noop {} {\bibfield  {journal} {\bibinfo  {journal}
  {Mathematische Zeitschrift}\ }\textbf {\bibinfo {volume} {37}},\ \bibinfo
  {pages} {698} (\bibinfo {year} {1933})}\BibitemShut {NoStop}%
\bibitem [{\citenamefont {Kuznetsov}\ \emph {et~al.}(2010)\citenamefont
  {Kuznetsov}, \citenamefont {Naulin}, \citenamefont {Nielsen},\ and\
  \citenamefont {Rasmussen}}]{KNNR-10}%
  \BibitemOpen
  \bibfield  {author} {\bibinfo {author} {\bibfnamefont {E.~A.}\ \bibnamefont
  {Kuznetsov}}, \bibinfo {author} {\bibfnamefont {V.}~\bibnamefont {Naulin}},
  \bibinfo {author} {\bibfnamefont {A.~H.}\ \bibnamefont {Nielsen}},\ and\
  \bibinfo {author} {\bibfnamefont {J.~J.}\ \bibnamefont {Rasmussen}},\
  }\href@noop {} {\bibfield  {journal} {\bibinfo  {journal} {Theor. Comput.
  Fluid Dyn.}\ }\textbf {\bibinfo {volume} {24}},\ \bibinfo {pages} {253}
  (\bibinfo {year} {2010})}\BibitemShut {NoStop}%
\bibitem [{\citenamefont {Kuznetsov}\ and\ \citenamefont
  {Sereshchenko}(2019)}]{KuznetsovSereshchenko2019}%
  \BibitemOpen
  \bibfield  {author} {\bibinfo {author} {\bibfnamefont {E.~A.}\ \bibnamefont
  {Kuznetsov}}\ and\ \bibinfo {author} {\bibfnamefont {E.~V.}\ \bibnamefont
  {Sereshchenko}},\ }\href@noop {} {\bibfield  {journal} {\bibinfo  {journal}
  {JETP Lett.}\ }\textbf {\bibinfo {volume} {109}},\ \bibinfo {pages} {239}
  (\bibinfo {year} {2019})}\BibitemShut {NoStop}%
\bibitem [{\citenamefont {Falkovich}\ and\ \citenamefont
  {Lebedev}(2011)}]{FalkovichLebedev2011}%
  \BibitemOpen
  \bibfield  {author} {\bibinfo {author} {\bibfnamefont {G.}~\bibnamefont
  {Falkovich}}\ and\ \bibinfo {author} {\bibfnamefont {V.}~\bibnamefont
  {Lebedev}},\ }\href@noop {} {\bibfield  {journal} {\bibinfo  {journal} {Phys.
  Rev. E}\ }\textbf {\bibinfo {volume} {83}},\ \bibinfo {pages} {045301}
  (\bibinfo {year} {2011})}\BibitemShut {NoStop}%
\bibitem [{\citenamefont {Parker}(1963)}]{parker1963kinematical}%
  \BibitemOpen
  \bibfield  {author} {\bibinfo {author} {\bibfnamefont {E.~N.}\ \bibnamefont
  {Parker}},\ }\href@noop {} {\bibfield  {journal} {\bibinfo  {journal}
  {Astrophys. J.}\ }\textbf {\bibinfo {volume} {138}},\ \bibinfo {pages} {552}
  (\bibinfo {year} {1963})}\BibitemShut {NoStop}%
\bibitem [{\citenamefont {Kuznetsov}\ and\ \citenamefont
  {Mikhailov}(2020)}]{KuznetsovMikhailov2020}%
  \BibitemOpen
  \bibfield  {author} {\bibinfo {author} {\bibfnamefont {E.~A.}\ \bibnamefont
  {Kuznetsov}}\ and\ \bibinfo {author} {\bibfnamefont {E.~A.}\ \bibnamefont
  {Mikhailov}},\ }\href@noop {} {\bibfield  {journal} {\bibinfo  {journal}
  {JETP}\ }\textbf {\bibinfo {volume} {131}},\ \bibinfo {pages} {496} (\bibinfo
  {year} {2020})}\BibitemShut {NoStop}%
\end{thebibliography}
\end{document}